%% file: fKPi.tex
\documentclass[11pt, prd, aps, showpacs, nofootinbib, superscriptaddress]{revtex4-1}

\usepackage{geometry} 
\usepackage{amsfonts,amsmath,amssymb,amscd}
\usepackage{adjustbox}
\usepackage{graphicx}
\usepackage{epstopdf}
\usepackage[latin9]{inputenc}
\usepackage{adjustbox}
\usepackage[colorlinks=true,backref=false, linktocpage=true,citecolor=blue,urlcolor=blue,linkcolor=blue,pdfpagemode=UseOutlines]{hyperref}
\usepackage{cleveref}

\Crefname{equation}{Eq.}{Eqs.}
\crefname{equation}{eq.}{eqs.}

\usepackage{tabularx}
\newcommand{\re}{\operatorname{Re}}

\newcommand{\Tr}{\operatorname{Tr}}
\newcommand{\be}{\begin{equation}}
\newcommand{\ee}{\end{equation}}
\newcommand{\bea}{\begin{eqnarray}}
\newcommand{\eea}{\end{eqnarray}}

\newcommand{\mubar}{\overline{\mu}}
\newcommand{\epsbar}{\overline{\epsilon}}
\newcommand{\Dsw}{D_{\mathrm{sw}}}
\newcommand{\Qsw}{Q_{\mathrm{sw}}}
\newcommand{\Qhat}{\hat{Q}}
\newcommand{\What}{\hat{W}}
\newcommand{\Qp}{Q_{+}}
\newcommand{\Qm}{Q_{-}}
\newcommand{\Qpm}{Q_{\pm}}
\newcommand{\mutilde}{\tilde{\mu}}

\newcommand{\Bern}{Institute for Theoretical Physics, Albert Einstein Center for Fundamental Physics,\\University of Bern, Sidlerstrasse 5, CH-3012 Bern, Switzerland}
\newcommand{\hiskp}{HISKP (Theory), Rheinische Friedrich-Wilhelms-Universit{\"a}t Bonn,\\Nussallee 14-16, 53115 Bonn, Germany}
\newcommand{\hpca}{High Performance Computing and Analytics Lab, Rheinische Friedrich-Wilhelms-Universit{\"a}t Bonn,\\ Friedrich-Hirzebruch-Allee 8, 53115 Bonn, Germany}
\newcommand{\itwm}{Fraunhofer Institute for Industrial Mathematics (ITWM),\\Fraunhofer-Platz 1, 67663 Kaiserslautern, Germany}
\newcommand{\CyprusU}{Department of Physics, University of Cyprus, 20537 Nicosia, Cyprus}
\newcommand{\CyprusI}{Computation-based Science and Technology Research Center, The Cyprus Institute,\\20 Konstantinou Kavafi Street, 2121 Nicosia, Cyprus}
\newcommand{\Jena}{University of Jena, Institute for Theoretical Physics, Max-Wien-Platz 1, D-07743 Jena, Germany}
\newcommand{\Parma}{Dipartimento di Scienze Matematiche, Fisiche e Informatiche, Universit{\`a} di Parma  and  INFN, Gruppo Collegato  di  Parma, \\Parco  Area  delle Scienze  7/a (Campus), I-43124  Parma Italy}
\newcommand{\Romadue}{Dipartimento di Fisica, Universit{\`a} di Roma ``Tor Vergata" and INFN, Sezione di Tor Vergata,\\Via della Ricerca Scientifica 1, I-00133 Roma, Italy}
\newcommand{\Romatre}{Dipartimento di Fisica, Universit{\`a} Roma Tre and INFN, Sezione di Roma Tre,\\Via della Vasca Navale 84, I-00146 Rome, Italy}
\newcommand{\RomatreINFN}{Istituto Nazionale di Fisica Nucleare, Sezione di Roma Tre,\\Via della Vasca Navale 84, I-00146 Rome, Italy}

\begin{document}

\title{Ratio of kaon and pion leptonic decay constants\\[2mm] with $N_f = 2 + 1 + 1$ Wilson-clover twisted-mass fermions}

\author{C.~Alexandrou}\affiliation{\CyprusU}\affiliation{\CyprusI}
\author{S.~Bacchio}\affiliation{\CyprusI}
\author{G.~Bergner}\affiliation{\Jena}
\author{P.~Dimopoulos}\affiliation{\Parma}
\author{J.~Finkenrath}\affiliation{\CyprusI}
\author{R.~Frezzotti}\affiliation{\Romadue} 
\author{M.~Garofalo}\affiliation{\Romatre}\affiliation{\hiskp}
\author{B.~Kostrzewa}\affiliation{\hpca}
\author{G.~Koutsou}\affiliation{\CyprusI}
\author{P.~Labus}\affiliation{\itwm}
\author{F.~Sanfilippo}\affiliation{\RomatreINFN}
\author{S.~Simula}\affiliation{\RomatreINFN}
\author{M.~Ueding}\affiliation{\hiskp}
\author{C.~Urbach}\affiliation{\hiskp}
\author{U.~Wenger}\affiliation{\Bern}

\begin{abstract}
%\centerline{\large for the Extended Twisted Mass Collaboration}
\centerline{\includegraphics[height=6cm]{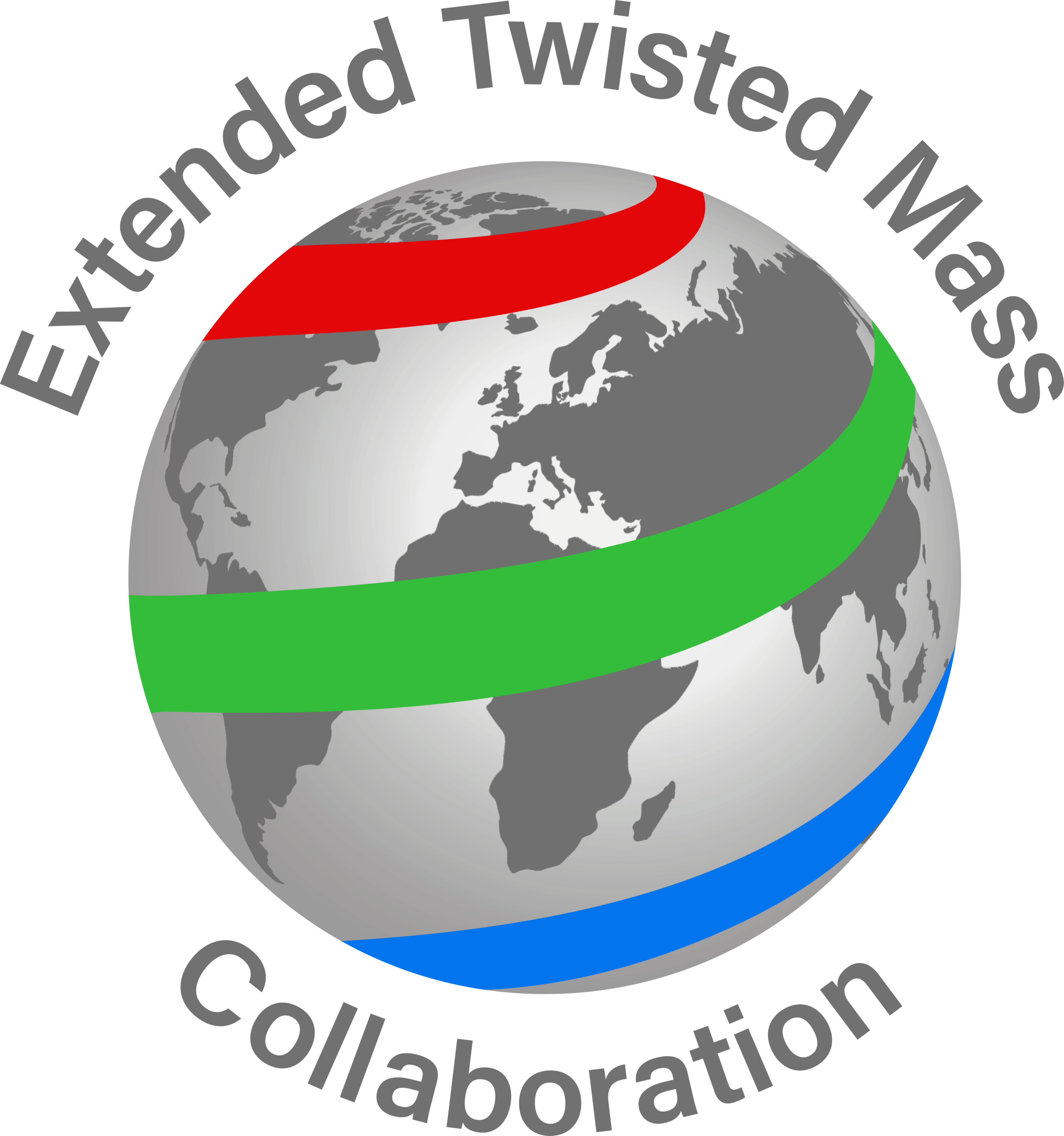}}

\vspace{1cm}
We present a determination of the ratio of kaon and pion leptonic decay constants in isosymmetric QCD (isoQCD), $f_K / f_\pi$, making use of the gauge ensembles produced by the Extended Twisted Mass Collaboration (ETMC) with $N_f = 2 + 1 + 1$ flavors of Wilson-clover twisted-mass quarks, including configurations close to the physical point for all dynamical flavors. The simulations are carried out at three values of the lattice spacing ranging from $\sim 0.068$ to $\sim 0.092$ fm with linear lattice size up to $L \sim 5.5$~fm. The scale is set by the PDG value of the pion decay constant, $f_\pi^{isoQCD} = 130.4~(2)$ MeV, at the isoQCD pion point, $M_\pi^{isoQCD} = 135.0~(2)$ MeV, obtaining for the gradient-flow (GF) scales the values $w_0 = 0.17383~(63)$ fm, $\sqrt{t_0} = 0.14436~(61)$ fm and $t_0 / w_0 = 0.11969~(62)$ fm.
The data are analyzed within the framework of SU(2) Chiral Perturbation Theory (ChPT) without resorting to the use of renormalized quark masses.
At the isoQCD kaon point $M_K^{isoQCD} = 494.2~(4)$ MeV we get $(f_K / f_\pi)^{isoQCD} = 1.1995~(44)$, where the error includes both statistical and systematic uncertainties.
Implications for the Cabibbo-Kobayashi-Maskawa (CKM) matrix element $|V_{us}|$ and for the first-row CKM unitarity are discussed.
\end{abstract}

\maketitle

%\listoftodos

\newpage

%%%%%%%%%%%%%%%%%%%%%%%%%%%%%%%%%%%%%%%%%%%%%%%%%%
\section{Introduction}
\label{sec:intro}
%%%%%%%%%%%%%%%%%%%%%%%%%%%%%%%%%%%%%%%%%%%%%%%%%%

The leptonic decay constants of charged pseudoscalar (P) mesons are the crucial hadronic ingredients necessary for obtaining precise information on the Cabibbo-Kobayashi-Maskawa (CKM) matrix elements describing the weak mixings among quark flavors~\cite{Cabibbo:1963yz,Kobayashi:1973fv}.
Within the Standard Model (SM) the unitarity of the CKM matrix imposes important constraints on various sums of squares of matrix elements and, therefore, any violation of such constraints would imply the presence of physics beyond the SM.
The way the CKM entries can be determined is based on the knowledge of the experimental leptonic decay rates and of the corresponding theoretical calculations.
In particular, both the charged kaon and pion leptonic decay widths into muons are known experimentally with a good precision~\cite{Zyla:2020zbs}, obtaining for their ratio the value
\be
    \label{eq:Gamma_KPi_exp}
    \frac{\Gamma(K \to \mu \nu_\mu [\gamma])}{\Gamma(\pi \to \mu \nu_\mu [\gamma])} = 1.3367 ~ (2)_\pi ~ (29)_K ~ [29] ~ ,
\ee
where $[\gamma]$ stands for the contribution of virtual and real photons.
On the theoretical side, within the SM the above ratio is given by
\be
      \label{eq:Gamma_KPi}
      \frac{\Gamma(K \to \mu \nu_\mu [\gamma])}{\Gamma(\pi \to \mu \nu_\mu [\gamma])} = \left| \frac{V_{us}}{V_{ud}} \frac{f_K}{f_\pi} \right|^2 
         \frac{M_\pi^3}{M_K^3} \left( \frac{M_K^2 - m_\mu^2}{M_\pi^2 - m_\mu^2} \right)^2 
         \left( 1 + \delta R_{K \pi} \right) ,
 \ee
where $V_{ud}$ and $V_{us}$ are the relevant CKM entries, $M_{\pi{(K)}}$ is the charged pion(kaon) mass, $m_\mu$ is the muon mass and $\delta R_{K \pi}$ represents 
the isospin breaking (IB) corrections due both to the mass difference ($m_d - m_u$) between the light $u$- and $d$-quarks and to the quark electric charges.
Finally, in Eq.~(\ref{eq:Gamma_KPi}) $f_K / f_\pi$ is the ratio of kaon and pion leptonic decay constants defined in isosymmetric QCD (isoQCD), i.e.~with $m_u = m_d$ and zero quark electric charges.

Recently~\cite{Giusti:2017dwk,DiCarlo:2019thl} the IB correction $\delta R_{K \pi}$ has been determined using a non-perturbative approach, based on first principles, through QCD+QED simulations on the lattice, obtaining $\delta R_{K \pi} = -0.0126 ~ (14)$.
From Eq.~(\ref{eq:Gamma_KPi_exp}) one has
\be
    \label{eq:ratioVf}
    \left| \frac{V_{us}}{V_{ud}} \right| \frac{f_K}{f_\pi} = 0.27683 ~ (29)_{\mathrm{exp}} ~ (20)_{\mathrm{th}} = 0.27683 ~ (35) ~ , ~
\ee
which corresponds to an accuracy of $\simeq 0.13 \%$.
As well known~\cite{Gasser:2010wz}, the IB correction $\delta R_{K \pi}$ and the isoQCD ratio $f_K / f_\pi$ separately depend on the prescription used to define what is meant by isoQCD, while only the product $(f_K / f_\pi) \sqrt{1 + \delta R_{K \pi}}$ is independent on such prescription.
The hadronic prescription adopted in Refs.~\cite{Giusti:2017dwk,DiCarlo:2019thl} corresponds to
\bea
      \label{eq:MPi_isoQCD}
      M_\pi^{isoQCD} & = & 135.0 ~ (2) ~ {\rm MeV} ~ , ~ \\[2mm]
      \label{eq:MK_isoQCD}
      M_K^{isoQCD} & = & 494.2 ~ (4) ~ {\rm MeV} ~ , ~ \\[2mm]
     \label{eq:fPi_isoQCD}
      f_\pi^{isoQCD} & = & 130.4 ~ (2) ~ {\rm MeV} ~ ,
\eea
while the quantity ($m_d - m_u$) is obtained from the difference between the experimental charged and neutral kaon masses.
The physical pion and kaon masses~(\ref{eq:MPi_isoQCD}-\ref{eq:MK_isoQCD}) are consistent with those recommended by FLAG-3~\cite{Aoki:2016frl}, and the pion decay constant (\ref{eq:fPi_isoQCD}), derived according to Ref.~\cite{Patrignani:2016xqp} adopting the value of the CKM entry $|V_{ud}|$ from Ref.~\cite{Hardy:2016vhg}, is used to set the lattice scale\footnote{\label{footnote:GRS}In Ref.~\cite{DiCarlo:2019thl} it has been shown that within the precision of the lattice simulations the prescription given by Eqs.~(\ref{eq:MPi_isoQCD}-\ref{eq:fPi_isoQCD}) is equivalent to the Gasser-Rusetsky-Scimemi (GRS) scheme~\cite{Gasser:2003hk}, where the renormalized quark masses and coupling constant in a given short-distance scheme (viz.~the $\overline{\rm MS}$ scheme) and at a given scale (viz.~2 GeV) are equal in the full QCD+QED and isoQCD theories. For completeness we mention that in the charm sector the $D_s$-meson mass $M_{D_S}^{isoQCD}$ was chosen to be equal to its experimental value $M_{D_s^+} =1969.0 ~ (1.4)$ MeV~\cite{Zyla:2020zbs}.}.

In this work we present our determination of the leptonic decay constant ratio $f_K / f_\pi$ at the physical isoQCD point given by Eqs.~(\ref{eq:MPi_isoQCD}-\ref{eq:fPi_isoQCD}), evaluated using the ETMC gauge ensembles produced with $N_f = 2+1+1$ flavors of Wilson Clover twisted-mass quarks, including configurations close to the physical point for all dynamical flavors~\cite{Alexandrou:2018egz,Dimopoulos:2020eqd}.
The lattice data will be analyzed within the framework of SU(2) Chiral Perturbation Theory (ChPT) without making use of renormalized quark masses\footnote{An 
analysis of the kaon and pion masses and decay constants in terms of renormalized quark masses is ongoing and will be presented in a forthcoming ETMC publication.}.
By means of the pion data we determine the gradient-flow (GF) scales $w_0$~\cite{Borsanyi:2012zs}, $\sqrt{t_0}$~\cite{Luscher:2010iy} and $t_0 / w_0$ adopting the physical value~(\ref{eq:fPi_isoQCD}) at the pion point~(\ref{eq:MPi_isoQCD}) to set the lattice scale, obtaining
\bea
    \label{eq:w0_intro}
     w_0 & = & 0.17383~ (63) ~ {\rm fm} ~ , ~ \\[2mm]
    \label{eq:t0_intro}
     \sqrt{t_0} & = & 0.14436 ~ (61) ~ {\rm fm} ~ , ~ \\[2mm]
    \label{eq:t0w0_intro}
     t_0 / w_0 & = & 0.11969 ~ (62) ~ {\rm fm} ~ , ~
\eea
where the error includes both statistical and systematic uncertainties.
Our findings (\ref{eq:w0_intro}-\ref{eq:t0_intro}) are a little larger than the MILC results~\cite{Bazavov:2015yea} $w_0 = 0.1714_{-12}^{+15}$ fm and $\sqrt{t_0} = 0.1416_{-5}^{+8}$ fm as well as the HPQCD results~\cite{Dowdall:2013rya} $w_0 = 0.1715~(9)$ fm and $\sqrt{t_0} = 0.1420~(8)$ fm, both obtained using the hadronic value (\ref{eq:fPi_isoQCD}) to set the lattice scale. 
Within $\simeq 1.5$ standard deviations our result (\ref{eq:w0_intro}) is consistent with the recent, precise BMW determination $w_0 = 0.17236~(70)$ fm, obtained in Ref.~\cite{Borsanyi:2020mff} using the $\Omega^-$-baryon mass to set the lattice scale.
Furthermore, the differences with the recent results $w_0 = 0.1709(11)$ fm and $\sqrt{t_0} = 0.1422(14)$ fm, obtained in Ref.~\cite{Miller:2020evg} using the $\Omega^-$-baryon mass to set the lattice scale, are within $\sim 2$ and $\sim 1.5$ standard deviations, respectively.

As for the ratio $f_K / f_\pi$ we determine its value at the physical isoQCD point~(\ref{eq:MPi_isoQCD}-\ref{eq:fPi_isoQCD}) and in the continuum and infinite volume limits, obtaining
\be
     \label{eq:fKPi_intro}
     \left( \frac{f_K}{f_\pi} \right)^{isoQCD} = 1.1995 ~ (44) ~ , ~
\ee
where again the error includes both statistical and systematic uncertainties.

The IB correction $\delta R_{K \pi} = -0.0126 ~ (14)$, determined in Refs.~\cite{Giusti:2017dwk,DiCarlo:2019thl} and adopted in Eqs.~(\ref{eq:Gamma_KPi}-\ref{eq:ratioVf}), stems from the sum of a QED and a strong IB terms, which are both prescription dependent as well as their sum and the isoQCD value (\ref{eq:fKPi_intro}). 
Within the GRS prescription (see footnote~\ref{footnote:GRS}) they are equal respectively to $-0.0062 ~ (12)$ and $-0.0064 ~ (7)$.
Thus, for the ratio of kaon and pion leptonic decay constant including strong IB effects (which we remind is prescription dependent) we get 
\be
    \label{eq:fKPi_IB_intro}
     \frac{f_{K^+}}{f_{\pi^+}} = 1.1957 ~ (44) ~ .
\ee

For comparison, the $N_f = 2+1+1$ determinations, entering the FLAG-4 average~\cite{Aoki:2019cca}, yield the value $(f_K^+ / f_ \pi^+) = 1.1932 ~ (19)$~\cite{Dowdall:2013rya,Carrasco:2014poa,Bazavov:2017lyh}, which is well consistent with our result~(\ref{eq:fKPi_IB_intro}).
Once corrected for the strong IB effects obtained in Refs.~\cite{Dowdall:2013rya,Carrasco:2014poa,Bazavov:2017lyh}, the FLAG-4 average becomes $(f_K / f_ \pi)^{isoQCD} = 1.1966 ~ (18)$, which agrees with our finding~(\ref{eq:fKPi_intro}). 

Taking the updated value $|V_{ud}| = 0.97370~(14)$ from super-allowed nuclear beta decays~\cite{Zyla:2020zbs,Seng:2018yzq}, Eqs.~(\ref{eq:ratioVf}) and (\ref{eq:fKPi_intro}) yield the following value for the CKM element $|V_{us}|$:
\be  
    \label{eq:Vus_intro}
    |V_{us}| = 0.22472 ~ (24)_{\mathrm{exp}} ~ (84)_{\mathrm{th}} = 0.22472 ~ (87) ~ , ~
\ee
which is nicely consistent with the latest estimate $|V_{us}| = 0.2252~(5)$ from leptonic modes provided by the PDG~\cite{Zyla:2020zbs}.
Correspondingly, using $|V_{ub}| = 0.00382~(24)$~\cite{Zyla:2020zbs} the first-row CKM unitarity  becomes
\be
    \label{eq:unitarity_intro}
    |V_{ud}|^2 + |V_{us}|^2 + |V_{ub}|^2 = 0.99861 ~ (48) ~ ,
\ee
which would imply a $\simeq 3 \sigma$ tension with unitarity from leptonic modes.

The plan of the paper is as follows.

In Section~\ref{sec:ETMC} some details of the ETMC gauge ensembles and of the simulations are illustrated, while a more complete description is provided in Appendix~\ref{sec:appA}.
For each gauge ensemble the pion mass and decay constant are extracted from the relevant two-point correlation functions using a single exponential fit in the appropriate regions of large time distances.
Alternatively, in Appendix~\ref{sec:appB} the extraction of the ground-state properties is performed through the multiple exponential procedure of Ref.~\cite{Romiti:2019qim}.
For one gauge ensemble (cA211.12.48), because of  a small deviation from maximal twist, the mass and the decay constant are corrected as described in Appendix~\ref{sec:appC}.
In Section~\ref{sec:SU2} the SU(2) ChPT predictions at next-to-leading order (NLO) for the pion decay constant $f_\pi$, including finite volume effects (FVEs), are presented.
For the ensembles cB211.25.XX, sharing the same light-quark mass and lattice spacing and differing only for the lattice size $L$, the FVEs are investigated using both the NLO and the resummed NNLO formulae of Ref.~\cite{Colangelo:2005gd}.
In Section~\ref{sec:w0}, adopting the physical value~(\ref{eq:fPi_isoQCD}) at the pion point~(\ref{eq:MPi_isoQCD}), we perform two determinations of the GF scale $w_0$ using the data for either $f_\pi$ or the quantity $X_\pi \equiv (f_\pi M_\pi^4)^{1/5}$, which is found to be less affected by statistical and systematic errors.
The two determinations of $w_0$ agree very nicely, but the one based on the quantity $X_\pi$ turns out to be more precise by a factor of $\approx 2.5$.
In the same way the other two GF scales $\sqrt{t_0}$ and $t_0 / w_0$ are determined in Appendix~\ref{sec:appD}, where our calculations of the relative GF scales $w_0 / a$, $\sqrt{t_0} / a$ and $t_0 / (w_0 a)$ at the physical pion point are also described.
In Section~\ref{sec:fKPi} we analyze the data for the decay constant ratio $f_K / f_\pi$ using SU(2) ChPT.
In Section~\ref{sec:CKM} the implications for $V_{us}$ and the first-row CKM unitarity are discussed.
Finally, our conclusions are collected in Section~\ref{sec:conclusions} .

%%%%%%%%%%%%%%%%%%%%%%%%%%%%%%%%%%%%%%%%%%%%%%%%%%
\section{ETMC ensembles}
\label{sec:ETMC}
%%%%%%%%%%%%%%%%%%%%%%%%%%%%%%%%%%%%%%%%%%%%%%%%%%

In this work we make use of the gauge ensembles produced recently by ETMC in isoQCD with $N_f = 2 + 1 + 1$ flavors of Wilson-clover twisted-mass quarks and described in Refs.~\cite{Alexandrou:2018egz,Dimopoulos:2020eqd}.
The gluon action is the improved Iwasaki one~\cite{Iwasaki:1985we}, while the fermionic action includes a Clover term~\cite{Sheikholeslami:1985ij} with a coefficient fixed by its estimate in one-loop tadpole boosted perturbation theory~\cite{Aoki:1998qd}.
Its inclusion turns out to be very beneficial for reducing cutoff effects, in particular on the neutral pion mass, thereby making numerically stable simulations close to the physical pion point~\cite{Alexandrou:2018egz}.

The Wilson mass counterterms of the two degenerate light-quarks as well as of the strange and charm quarks are chosen in order to guarantee automatic ${\cal{O}}(a)$-improvement~\cite{Frezzotti:2003ni,Frezzotti:2005gi}.
The masses of the strange and charm sea quarks are tuned to their physical values for each ensemble~\cite{Alexandrou:2018egz,Dimopoulos:2020eqd}.
For the valence strange and charm sectors, a mixed action setup employing Osterwalder-Seiler fermions~\cite{Osterwalder:1977pc}, with the same critical mass as determined in the unitary setup, has been adopted in order to avoid any undesired strange-charm quark mixing (through cutoff effects) and to preserve the automatic ${\cal{O}}(a)$-improvement of physical observables~\cite{Frezzotti:2004wz}.

Some properties of the ETMC ensembles, which are relevant for this work, are collected in Table~\ref{tab:simudetails}, while the simulation setup is described in detail in \Cref{sec:appA}.
With respect to Ref.~\cite{Dimopoulos:2020eqd} two other dedicated gauge ensembles, cB211.25.24 and cB211.25.32, have been produced for the investigation of finite volume effects (FVEs).
\begin{table}[!htb]
\begin{center}
\begin{tabular}{||c|c|c|c|c||c|c|c||c||}
\hline
 ensemble & ~$\beta$~ & ~$V / a^4$~ & ~$a ~ \mbox{(fm)}$~ & ~$a \mu_\ell$~ & ~$M_\pi~ \mbox{(MeV)}$~ & ~$L~ \mbox{(fm)}$~ & ~$M_\pi L$~ & ~confs~\\
\hline
cA211.53.24& $1.726$ & $24^3 \times ~48$ & $~0.0947~(4)~$ & $~0.00530~$ & $~346.4~(1.6)~$ & $2.27$ & $3.99$ & $~628$~\\
cA211.40.24&               & $24^3 \times ~48$ &                              & $~0.00400~$ & $~301.6~(2.1)~$ & $2.27$ & $3.47$ & $~662$~\\
cA211.30.32&               & $32^3 \times ~64$ &                              & $~0.00300~$ & $~261.1~(1.1)~$ & $3.03$ & $4.01$ & $1237$~\\
cA211.12.48&               & $48^3 \times ~96$ &                              & $~0.00120~$ & $~167.1~(0.8)~$ & $4.55$ & $3.85$ & $~322$~\\
\hline
cB211.25.24& $1.778$ & $24^3 \times ~48$ & $~0.0816~(3)~$ & $~0.00250~$ & $~259.2~(3.0)~$  & $1.96$ & $2.57$ & $~500$~\\
cB211.25.32&               & $32^3 \times ~64$ &                              & $~0.00250~$ & $~253.3~(1.4)~$  & $2.61$ & $3.35$ & $~400$~\\
cB211.25.48&               & $48^3 \times ~96$ &                              & $~0.00250~$ & $~253.0~(1.0)~$  & $3.92$ & $5.02$ & $~314$~\\
cB211.14.64&               & $64^3 \times 128$ &                              & $~0.00140~$ & $~189.8~(0.7)~$  & $5.22$ & $5.02$ & $~437$~\\
cB211.072.64&               & $64^3 \times 128$ &                              & $~0.00072~$ & $~136.8~(0.6)~$  & $5.22$ & $3.62$ & $~374$~\\
\hline
cC211.06.80& $1.836$ & $80^3 \times 160$ & $~0.0694~(3)~$ & $~0.00060~$ & $~134.2~(0.5)~$ & $5.55$ & $3.78$ & $~401$~\\
\hline
\end{tabular}
\end{center}
\caption{\it Summary of the simulated light-quark bare mass, $a \mu_\ell = a \mu_u = a \mu_d$, of the pion mass $M_\pi$, of the lattice size $L$ and of the product $M_\pi L$ for the various ETMC gauge ensembles used in this work. The values of the lattice spacing $a$ in the fourth column, estimated in Appendix~\ref{sec:appD_sect2} using the relative GF scale $w_0 / a$ of Table~\ref{tab:relGF}, and the values of $M_\pi$ and $L$ in the sixth and seventh columns correspond to the absolute scale $w_0 = 0.17383~(63)$ fm (see Eq.~(\ref{eq:w0_intro})). In the last column the number of gauge configurations analyzed for each ensemble is presented.}
\label{tab:simudetails}
\end{table} 

Note that in the case of the ensembles cB211.072.64 and cC211.06.80, corresponding respectively to a lattice spacing equal to $a \approx 0.082$ fm and $a \approx 0.069$ fm, the pion mass is simulated quite close to the physical isoQCD value (\ref{eq:MPi_isoQCD}).

For each ensemble we compute the pion correlator given by
 \be
    C_\pi(t) = \frac{1}{L^3} \sum\limits_{x, z} \left\langle 0 \right| P_5 (x) P_5^\dag (z) \left| 0 \right\rangle \delta_{t, (t_x  - t_z )} ~ ,
    \label{eq:P5P5}
 \ee
where $P_5 (x) = \overline{q}_\ell(x) \gamma_5 q_\ell(x)$ is a local interpolating pion field.
The Wilson parameters of the two mass-degenerate valence quarks are always chosen to have opposite values. 
In this way the squared pion mass differs from its continuum counterpart only by terms of ${\cal{O}}(a^2 \mu_\ell)$~\cite{Frezzotti:2003ni, Frezzotti:2005gi}.

At large time distances one has
 \be
    C_\pi(t)_{ ~ \overrightarrow{t  \gg a, ~ (T - t) \gg a} ~ } \frac{\mathcal{Z}_\pi}{2M_\pi} \left[ e^{ - M_\pi  t}  + e^{ - M_\pi (T - t)} \right] ~ ,
    \label{eq:larget}
 \ee
so that the pion mass $M_\pi$ and the matrix element $\mathcal{Z}_\pi = | \langle\pi | \overline{q}_\ell \gamma_5 q_\ell | 0 \rangle|^2$ can be extracted from the exponential fit given in the r.h.s.~of Eq.~(\ref{eq:larget}). 

For maximally twisted fermions the value of $\mathcal{Z}_\pi$ determines the pion decay constant $f_\pi$ without the need of the knowledge of any renormalization constant~\cite{Frezzotti:2000nk,Frezzotti:2003ni}, namely
 \be
    af_\pi = 2 a \mu_\ell \frac{\sqrt{a^4 \mathcal{Z}_\pi}}{aM_\pi ~ \mbox{sinh}(aM_\pi)} ~ .
    \label{eq:decayPS}
 \ee
 
\begin{figure}[htb!]
\begin{center}
\includegraphics[scale=0.75]{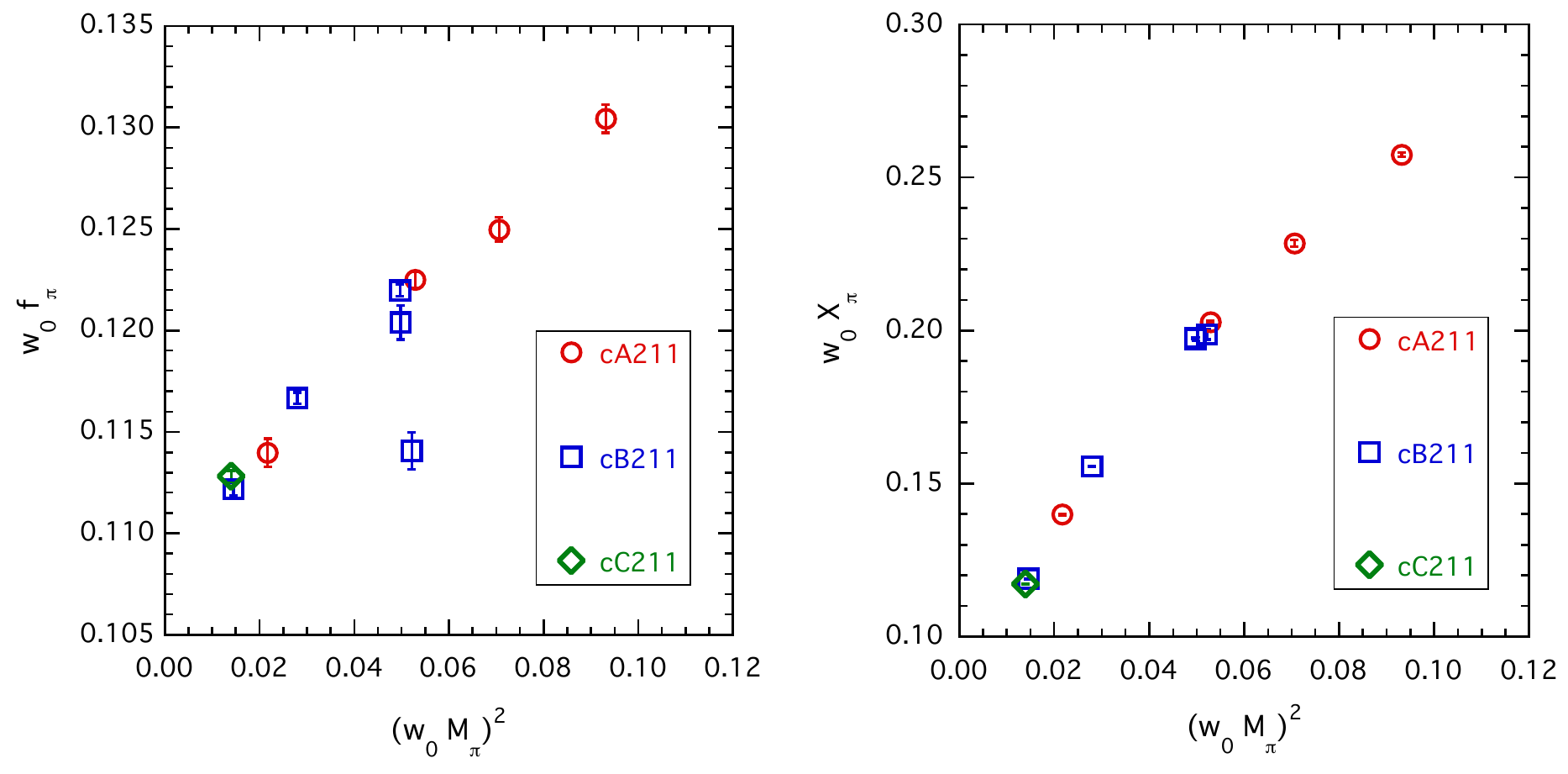}
\end{center}
\vspace{-0.75cm}
\caption{\it \small Values of the pion decay constant $w_0 f_\pi$ (left panel) and of the quantity $w_0 X_\pi = w_0 (f_\pi M_\pi^4)^{1/5}$ (right panel) versus the squared pion mass $(w_0 M_\pi)^2$ in units of the GF scale $w_0$. For the ensemble cA211.12.48 the corrected value of $f_\pi$ given by Eq.~(\ref{eq:decay_PS_corrected}) is considered.}
\label{fig:pion_data}
\end{figure}

The time intervals $[t_{min}, t_{max}]$ adopted for the fit (\ref{eq:larget}) of the pion correlation function (\ref{eq:P5P5}) as well as the extracted values of the pion mass and decay constant in lattice units are collected in Table~\ref{tab:plateaux}.
As anticipated in the Introduction, in this work we will make also use of the data for the quantity $X_\pi$ defined as
\be
     \label{eq:XPi}
     X_\pi \equiv \left( f_\pi M_\pi^4 \right)^{1/5} ~ , ~
\ee
which turns out to be less affected by lattice artifacts (see below Fig.~\ref{fig:pion_data} and later Section~\ref{sec:XPi_FVE}).
The values of $X_\pi$ in lattice units are shown in the last column of Table~\ref{tab:plateaux}.
The statistical errors of the lattice data lie in the range $0.1 \div 1.1 \%$ for the pion mass, in the range $0.2 \div 0.8 \%$ for the pion decay constant and in the range $0.1 \div 0.9 \%$ for the quantity $X_\pi$.
We stress that in the case of the four ensembles cA211.12.48,  cB211.14.64, cB211.072.64 and cC211.06.80 (which correspond to $M_\pi \lesssim 190$ MeV) the statistical errors of $aX_\pi$ turn out to be less than half of those of $af_\pi$.
\begin{table}[!hbt]
\begin{center}	
\begin{tabular}{||c|c|c||c||c|c|c||}
\hline
ensemble& $\beta$ & $V / a^4$ & $[t_{\rm min} / a, \, t_{\rm max} / a]$ & $a M_\pi$ & $a f_\pi$& $aX_\pi$\\
\hline
cA211.53.24& $1.726$ & $24^3 \times ~48$ & $[13, \, 22]$ & $~0.16626~(51)~$ & $~0.07106~(36)~$ & $~0.14027~(41)~$\\
cA211.40.24&               & $24^3 \times ~48$ & $[13, \, 22]$ & $~0.14477~(70)~$ & $~0.06809~(30)~$ & $~0.12450~(44)~$ \\
cA211.30.32&               & $32^3 \times ~64$ & $[13, \, 28]$ & $~0.12530~(16)~$ & $~0.06674~(15)~$ & $~0.11047~(12)~$ \\
cA211.12.48&               & $48^3 \times ~96$ & $[13, \, 40]$ & $~0.08022~(18)~$ & $~0.06133~(33)~$ & $~0.07621~(10)~$ \\
\hline
cB211.25.24& $1.778$ & $24^3 \times ~48$ & $[14, \, 22]$ & $~0.10720~(118)~$ & $~0.05355~(42)~$ & $~0.09331~(79)~$ \\
cB211.25.32&               & $32^3 \times ~64$ & $[14, \, 28]$ & $~0.10475~(45)~$   & $~0.05652~(38)~$ & $~0.09259~(26)~$ \\
cB211.25.48&               & $48^3 \times ~96$ & $[14, \, 42]$ & $~0.10465~(14)~$   & $~0.05726~(12)~$ & $~0.09276~(10)~$ \\
cB211.14.64&               & $64^3 \times 128$ & $[14, \, 56]$ & $~0.07848~(10)~$   & $~0.05477~(12)~$ & $~0.07303~~(6)~$ \\
cB211.072.64&               & $64^3 \times 128$ & $[14, \, 56]$ & $~0.05659~~(8)~$   & $~0.05267~(14)~$ & $~0.05578~~(5)~$ \\
\hline
cC211.06.80& $1.836$ & $80^3 \times 160$ & $[15, \, 70]$ & $~0.04720~~(7)~$ & $~0.04504~(10)~$ & $~0.04676~~(5)~$ \\
\hline
\end{tabular}
\end{center}
\caption{\it The time intervals $[t_{min}, t_{max}]$ adopted in the fit (\ref{eq:larget}) of the pion correlation function (\ref{eq:P5P5}) together with the extracted values of the pion mass $M_\pi$, the decay constant $f_\pi$ and the quantity $X_\pi$, given by Eq.~(\ref{eq:XPi}), in lattice units. Errors are statistical only.}
\label{tab:plateaux}
\end{table} 

An alternative way to extract the pion mass and decay constant is the ODE procedure of Ref.~\cite{Romiti:2019qim}.
The results obtained by applying this method to the pion correlation function (\ref{eq:P5P5}) are collected in Appendix~\ref{sec:appB} and found to be totally consistent with the findings of the single exponential fit (\ref{eq:larget}) of Table~\ref{tab:plateaux}.

In the case of the ensemble cA211.12.48 due to a small deviation from maximal twist a correction needs to be applied.
According to Appendix~\ref{sec:appC} the squared pion mass is left uncorrected, while for the pion decay constant $f_\pi$ we use the following formula
\be
      f_\pi |_{corrected} = f_\pi ~ K_\ell ~ , ~
      \label{eq:decay_PS_corrected}
\ee
where
\be
     K_\ell \equiv \sqrt{1 + (Z_A ~ m_{PCAC} / \mu_\ell)^2} ~ ,
     \label{eq:Kell}
\ee
$m_{PCAC}$ is the bare untwisted PCAC mass, $Z_A$ is the renormalization constant of the axial current and $\mu_\ell$ is the bare twisted mass of the light valence quarks.
For the ensemble cA211.12.48 one has $Z_A \approx 0.75$ and $m_{PCAC} / \mu_\ell \simeq -0.21 ~ (5)$~\cite{Dimopoulos:2020eqd}.

The statistical accuracy of the correlator (\ref{eq:P5P5}) is significantly improved by using the so-called ``one-end" stochastic method~\cite{McNeile:2006bz}, which includes spatial stochastic sources at a single time slice randomly chosen.
Statistical errors are evaluated using the jackknife procedure.

The results obtained for the pion decay constant $w_0 f_\pi$ and for the quantity $w_0 X_\pi$ (see Eq.~(\ref{eq:XPi})), are shown in Fig.~\ref{fig:pion_data} for all the gauge ensembles.
By comparing the results corresponding to the ensembles cB211.25.XX the FVEs are clearly visible in the case of $f_\pi$, while they are almost absent in the case of $X_\pi$.
Moreover, also discretization effects in $X_\pi$ turn out to be smaller than those present in $f_\pi$.

%%%%%%%%%%%%%%%%%%%%%%%%%%%%%%%%%%%%%%%%%%%%%%%%%%
\section{The pion decay constant $f_\pi$ within SU(2) ChPT}
\label{sec:SU2}
%%%%%%%%%%%%%%%%%%%%%%%%%%%%%%%%%%%%%%%%%%%%%%%%%%

Within SU(2) ChPT~\cite{Gasser:1983yg} the pion decay constant $f_\pi$ is given at NLO by
\be
    f_\pi = f \left[ 1 - 2 \xi_\ell \mbox{log}(\xi_\ell) + 2 A_1 \xi_\ell \right] ~ , ~
   \label{eq:fPi}
\ee
where 
\be
      \xi_\ell \equiv \frac{2B \, m_\ell }{(4 \pi f)^2} ~ 
      \label{eq:xi_ell}
 \ee
with $m_\ell = m_u = m_d$ being the renormalized light-quark mass.
In Eqs.~(\ref{eq:fPi}-\ref{eq:xi_ell}) $B$ and $f$ are the LO SU(2) ChPT low-energy constants (LECs), while the coefficient $A_1$ is related to the NLO LEC $\bar{\ell}_4^{phys}$ by 
\be
    \bar{\ell}_4^{phys} = A_1 + 2 ~ \mbox{log}\left( \frac{4\pi f}{M_\pi^{isoQCD}} \right) ~ .
    \label{eq:l4}
\ee

For the squared pion mass one has at NLO
\be
    M_\pi^2 = 2 B m_\ell \left[ 1 + \xi_\ell \mbox{log}(\xi_\ell) + C_1 \xi_\ell  \right] ~ , ~ 
    \label{eq:MPi2}
\ee
where the coefficient $C_1$ is related to the NLO LEC $\bar{\ell}_3^{phys}$ by
\be
    \bar{\ell}_3^{phys} = - C_1 + 2 ~ \mbox{log}\left( \frac{4\pi f}{M_\pi^{isoQCD}}\right) ~ .
    \label{eq:l3}
\ee

%%%%%%%%%%%%%%%%%%%%%%%%%%%%%%%%%%%%%%%%%%%%%%%%%%
\subsection{Finite volume effects within NLO SU(2) ChPT}
\label{sec:FVE}
%%%%%%%%%%%%%%%%%%%%%%%%%%%%%%%%%%%%%%%%%%%%%%%%%%

The structure of FVEs on the pion decay constant can be studied using SU(2) ChPT~\cite{Gasser:1983yg}.
At NLO FVEs come entirely from the discretized sum over periodic momenta of the loop contributions.
For a finite spatial volume $V = L^3$ one has
\be
    f_\pi(L) = f_\pi(L \to \infty) \left[1 + \Delta_{FVE}^\pi(L) \right] ~ , ~
    \label{eq:fPi_FVE}
\ee
where $f_\pi(L \to \infty)$ is given by Eq.~(\ref{eq:fPi}).
The correction term $\Delta_{FVE}^\pi(L)$ can be obtained from the chiral log in Eq.~(\ref{eq:fPi}) via the following replacement 
\be
    \xi_\ell \mbox{log}(\xi_\ell)  \to \xi_\ell ~ \widetilde{g}_1(\lambda) ~ , ~
    \label{eq:A_FVE}
\ee
where $\lambda \equiv \sqrt{2B m_\ell} L = \sqrt{\xi_\ell} ~ 4\pi f L$ and
\be
    \widetilde{g}_1(\lambda) = 4 \sum_{n=1}^\infty \frac{m(n)}{\sqrt{n} \lambda} K_1\left( \sqrt{n} \lambda \right) ~
    \label{eq:g1T}
\ee
with $K_1$ being a Bessel function of the second kind and $m(n)$ the multiplicities of a three-dimensional vector $\vec{n}$ having integer norm $n$ (i.e.~$m(n) = \{6, 12, 8, 6, ...\}$).
At sufficiently large values of $\lambda$ the Bessel function can be replaced by its asymptotic expansion, which leads to
\be
    \widetilde{g}_1(\lambda) \simeq 4 \sqrt{\frac{\pi}{2}} \sum_{n=1}^\infty \frac{m(n)}{(\sqrt{n} \lambda)^{3/2}} e^{- \sqrt{n} \lambda} ~ . ~
    \label{eq:g1T_asymptotic}
\ee
Thus, within NLO SU(2) ChPT the quantity $\Delta_{FVE}^\pi(L)$ is given  by
\be
      \Delta_{FVE}^\pi(L) = - 2 \xi_\ell ~ \widetilde{g}_1(\lambda) ~ . ~
      \label{eq:Delta_pi}
\ee

In the case of the squared pion mass one gets
\be
    M_\pi^2(L) = M_\pi^2(L \to \infty) \left[1 - \frac{1}{4} \Delta_{FVE}^\pi(L) \right]^2 ~ , ~
    \label{eq:MPi2_FVE}
\ee
where $M_\pi^2(L \to \infty)$ is given by Eq.~(\ref{eq:MPi2}).

%%%%%%%%%%%%%%%%%%%%%%%%%%%%%%%%%%%%%%%%%%%%%%%%%%
\subsection{FVEs for the ensembles cB211.25.XX}
\label{sec:B25XX}
%%%%%%%%%%%%%%%%%%%%%%%%%%%%%%%%%%%%%%%%%%%%%%%%%%

In this Section we study the FVEs on the pion mass and decay constant corresponding to the three ensembles cB211.25.XX of Table~\ref{tab:simudetails}, which share the same light-quark mass and lattice spacing but differ only for the lattice size $L$.
We consider SU(2) ChPT both at NLO, i.e.~the Gasser-Leutwyler (GL) formulae (\ref{eq:fPi_FVE}) and (\ref{eq:MPi2_FVE}), and at NNLO + resummation, i.e.~the Colangelo-D\"urr-Haefeli (CDH) formulae~\cite{Colangelo:2005gd}.
The latter ones read as
\bea
    \label{eq:fPi_CDH}
    f_\pi(L) & = & f_\pi(\infty) \left\{ 1 - 2 \xi_\pi \widetilde{g}_1(M_\pi L) + 2 \xi_\pi^2 \left[ C_{f_\pi}^{(1)} \widetilde{g}_1(M_\pi L) + 
                          C_{f_\pi}^{(2)} \widetilde{g}_2(M_\pi L)  + S_{f_\pi}^{(4)} \right] \right\} ~ , ~ \\[2mm]
    \label{eq:MPi_CDH}
    M_\pi(L) & = & M_\pi(\infty) \left\{ 1 + \frac{1}{2} \xi_\pi \widetilde{g}_1(M_\pi L) - \xi_\pi^2 \left[ C_{M_\pi}^{(1)} \widetilde{g}_1(M_\pi L) + 
                            C_{M_\pi}^{(2)} \widetilde{g}_2(M_\pi L)  + S_{M_\pi}^{(4)} \right] \right\}  ~ , ~ \quad                
\eea
where $\widetilde{g}_1$ is defined in Eq.~(\ref{eq:g1T}), while
\be
    \widetilde{g}_2(\lambda) \equiv 4 \sum_{n=1}^\infty \frac{m(n)}{\sqrt{n} \lambda} \frac{K_2\left( \sqrt{n} \lambda \right)}{\sqrt{n} \lambda} ~
    \label{eq:g2T}
\ee
and
\bea
      \label{eq:Cf1}
      C_{f_\pi}^{(1)} & = & -\frac{7}{9} + 2 \overline{\ell}_1 + \frac{4}{3} \overline{\ell}_2 - 3 \overline{\ell}_4 ~ , ~ \\[2mm]
      \label{eq:Cf2}
      C_{f_\pi}^{(2)} & = & \frac{112}{9} - \frac{8}{3} \overline{\ell}_1 - \frac{32}{3} \overline{\ell}_2 ~ , ~ \\[2mm]
       \label{eq:CM1}
      C_{M_\pi}^{(1)} & = & -\frac{55}{18} + 4 \overline{\ell}_1 + \frac{8}{3} \overline{\ell}_2 - \frac{5}{2} \overline{\ell}_3 - 2 \overline{\ell}_4 ~ , ~ \\[2mm]
      \label{eq:CM2}
      C_{M_\pi}^{(2)} & = & C_{f_\pi}^{(2)} = \frac{112}{9} - \frac{8}{3} \overline{\ell}_1 - \frac{32}{3} \overline{\ell}_2 ~ 
\eea
with $\overline{\ell}_i$ being NLO LECs that have a logarithmic pion mass dependence
\be
     \overline{\ell}_i = \overline{\ell}_i^{phys} + 2 \mbox{log}\left( \frac{M_\pi^{isoQCD}}{M_\pi}\right) ~ . ~
\ee
Finally, in Eqs.~(\ref{eq:fPi_CDH})-(\ref{eq:MPi_CDH}) the NNLO terms $S_{f_\pi}^{(4)}$ and $S_{M_\pi}^{(4)}$ are defined in the Appendix A of Ref.~\cite{Colangelo:2005gd}, but useful approximate analytic formulae are given by~\cite{Colangelo:2005gd}
\bea
      \label{eq:Sf4}
      S_{f_\pi}^{(4)} & = & \left( \frac{4}{3} s_0 - \frac{13}{6} s_1 \right) \widetilde{g}_1(M_\pi L) - \left( \frac{40}{3} s_0 - 4 s_1 - \frac{8}{3} s_2 -
                                       \frac{13}{3} s_3 \right) \widetilde{g}_2(M_\pi L) ~ , ~ \quad  \\[2mm]
      \label{eq:SM4}
      S_{M_\pi}^{(4)} & = & \frac{13}{3} s_0 ~ \widetilde{g}_1(M_\pi L) - \left( \frac{40}{3} s_0 + \frac{32}{3} s_1 + \frac{26}{3} s_2 \right) 
                                         \widetilde{g}_2(M_\pi L) ~
\eea
with
\be
      s_0 = 2 - \frac{\pi}{2} ~ , ~ \qquad
      s_1 = \frac{\pi}{4} - \frac{1}{2} ~ , ~ \qquad
      s_2 = \frac{1}{2} - \frac{\pi}{8} ~ , ~ \qquad
      s_3 = \frac{3 \pi}{16} - \frac{1}{2} ~ .
\ee

The expansion variable $\xi_\pi$ is defined as~\cite{Colangelo:2005gd}
\be
   \label{eq:csipi_CDH}
    \xi_\pi \equiv \frac{M_\pi^2}{(4 \pi f_\pi)^2} ~ . ~
\ee
Different choices of the expansion variable are possible: one can replace $f_\pi$ with the LO LEC $f$ and/or replace $M_\pi^2$ with $2 B m_\ell$ (and correspondingly $M_\pi L$ with $\sqrt{2 B m_\ell} L$ in the arguments of the functions $\widetilde{g}_1$ and $ \widetilde{g}_2$). 
At NLO (i.e., for the GL formula) the above changes are equivalent, since any difference represents a NNLO effect.
Instead, in the CDH formula additional terms appear at NNLO, which can be found in Ref.~\cite{Frezzotti:2008dr}.
Here we consider only the alternative definition
\be
   \label{eq:csipi_alt}
    \xi_\pi \to \frac{M_\pi^2}{(4 \pi f)^2} ~ , ~
\ee
which requires the addition to the r.h.s~of Eq.~(\ref{eq:fPi_CDH}) of the term $f_\pi(\infty) \left\{  8 \xi_\pi^2 \overline{\ell}_4 \widetilde{g}_1(M_\pi L) \right\}$ and to the r.h.s.~of Eq.~(\ref{eq:MPi_CDH}) of the term $M_\pi(\infty) \left\{  - 2 \xi_\pi^2 \overline{\ell}_4 \widetilde{g}_1(M_\pi L) \right\}$.

The GL formula corresponds to Eqs.~(\ref{eq:fPi_CDH})-(\ref{eq:MPi_CDH}) with all $C$'s and $S$'s set equal to zero.
The CDH formula requires the knowledge of the values of the four NLO LECs $\overline{\ell}_i^{phys}$ with $i = 1, ...~4$.

\begin{figure}[htb!]
\begin{center}
\includegraphics[scale=0.75]{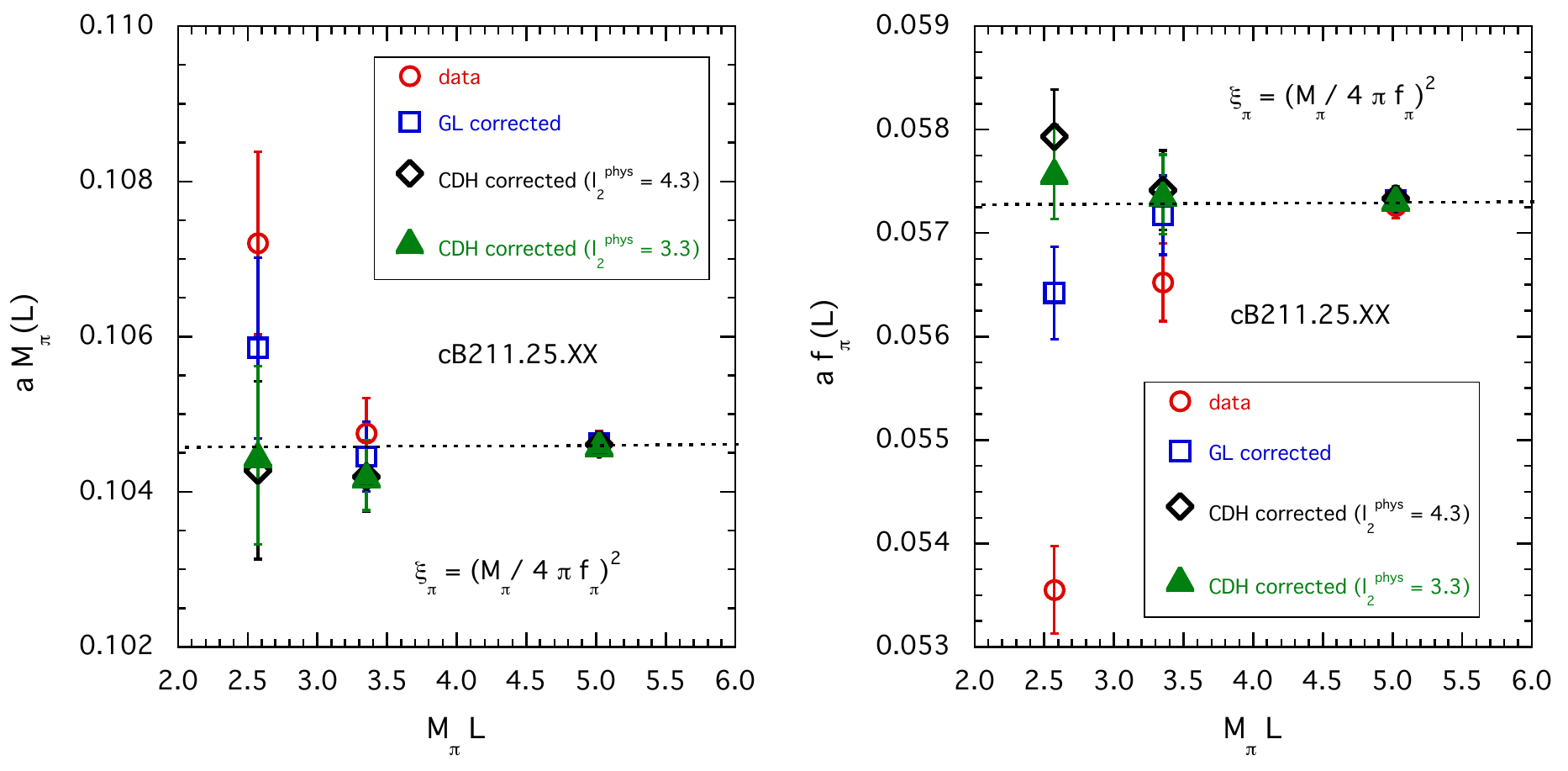}
\end{center}
\vspace{-0.75cm}
\caption{\it \small Values of the pion mass (left panel) and pion decay constant (right panel) in lattice units for the three ensembles cB211.25.XX of Table~\ref{tab:simudetails}. The red circles represent the data versus $M_\pi L$. The expansion variable $\xi_\pi$ is given by Eq.~(\ref{eq:csipi_CDH}). The blue squares correspond to the data corrected by the GL formula, while black diamonds represent the data corrected by the CDH formula, adopting for the NLO LECs the values $\overline{\ell}_1^{phys} = -0.4$, $\overline{\ell}_2^{phys} = 4.3$, $\overline{\ell}_3^{phys} = 3.2$ and $\overline{\ell}_4^{phys} = 4.4$. The green triangles correspond to the CDH correction assuming $\overline{\ell}_2^{phys} = 3.3$. The horizontal dotted lines are the the values of the pion mass and decay constant in the infinite volume limit.}
\label{fig:FVE0}
\end{figure}

\begin{figure}[htb!]
\begin{center}
\includegraphics[scale=0.75]{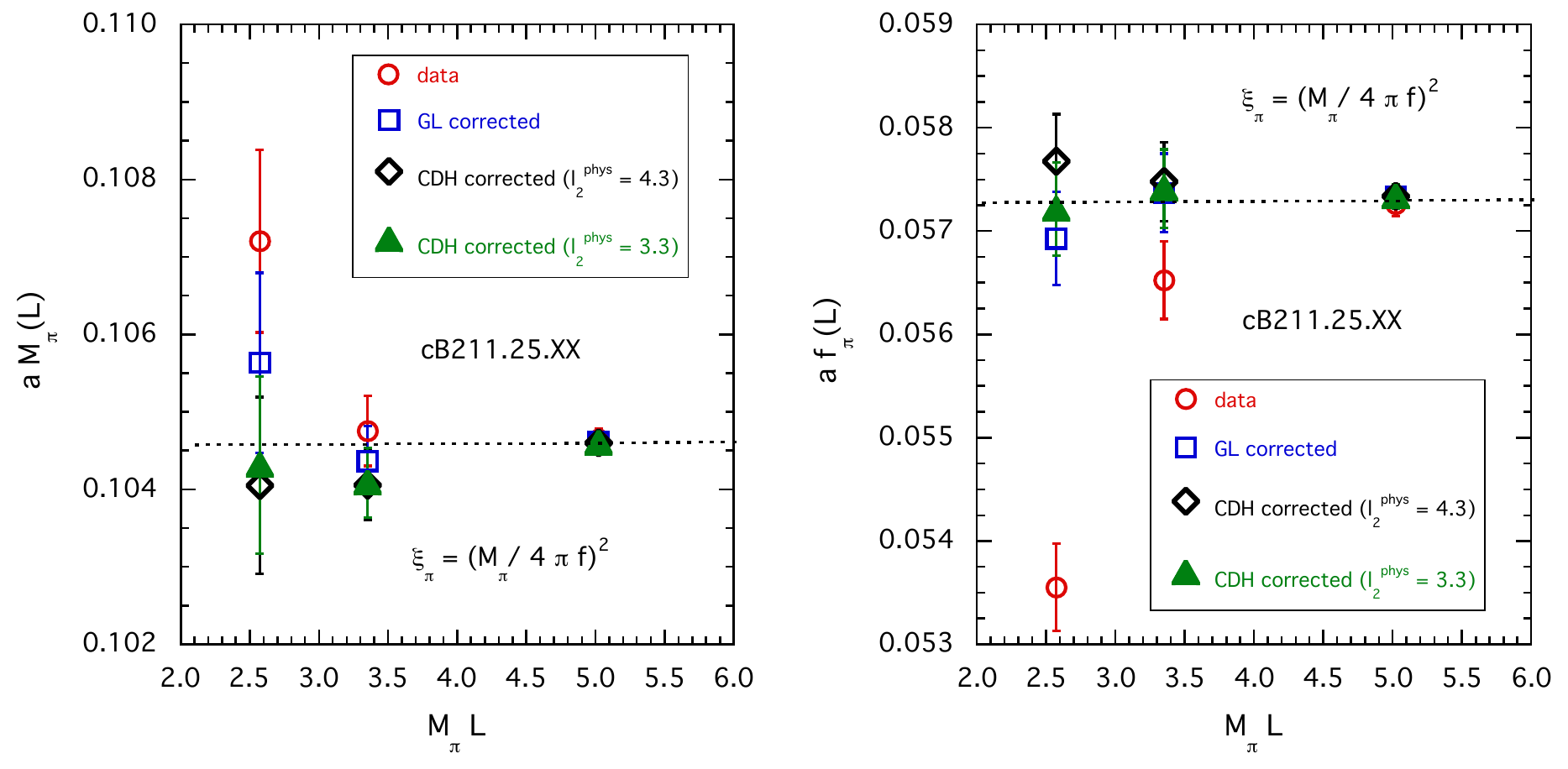}
\end{center}
\vspace{-0.75cm}
\caption{\it \small The same as in Fig.~\ref{fig:FVE0}, but adopting the alternative definition (\ref{eq:csipi_alt}) for the expansion variable $\xi_\pi$ and assuming $f = 122.5$ MeV and $a = 0.080$ fm.}
\label{fig:FVE1}
\end{figure}

In Figs.~\ref{fig:FVE0} and \ref{fig:FVE1} we compare the FVEs on the pion mass and decay constant for the three ensembles cB211.25.XX of Table~\ref{tab:simudetails}, evaluated using the GL and CDH formulae and assuming respectively the two definitions (\ref{eq:csipi_CDH}) and (\ref{eq:csipi_alt}) for the expansion variable $\xi_\pi$.

In the case of CDH formula we adopt the following values of the NLO LECs:  $\overline{\ell}_1^{phys} = -0.4$, $\overline{\ell}_2^{phys} = 4.3$, $\overline{\ell}_3^{phys} = 3.2$ and $\overline{\ell}_4^{phys} = 4.4$ (see Ref.~\cite{Frezzotti:2008dr}).
The CDH results depend on such a choice and the sensitivity to the specific value of $\overline{\ell}_2^{phys}$ is illustrated in both figures by the green triangles.

It can be seen that the GL formula applied to both the pion mass and decay constant works quite well for $M_\pi L \gtrsim 3$, particularly in the case of the definition (\ref{eq:csipi_alt}) of the expansion variable $\xi_\pi$.
The above condition is satisfied by all ETMC ensembles of Table~\ref{tab:simudetails} except the ensemble cB211.25.24.

%%%%%%%%%%%%%%%%%%%%%%%%%%%%%%%%%%%%%%%%%%%%%%%%%%
\subsection{FVEs for the quantity $X_\pi$}
\label{sec:XPi_FVE}
%%%%%%%%%%%%%%%%%%%%%%%%%%%%%%%%%%%%%%%%%%%%%%%%%%

\begin{figure}[htb!]
\begin{center}
\includegraphics[scale=0.70]{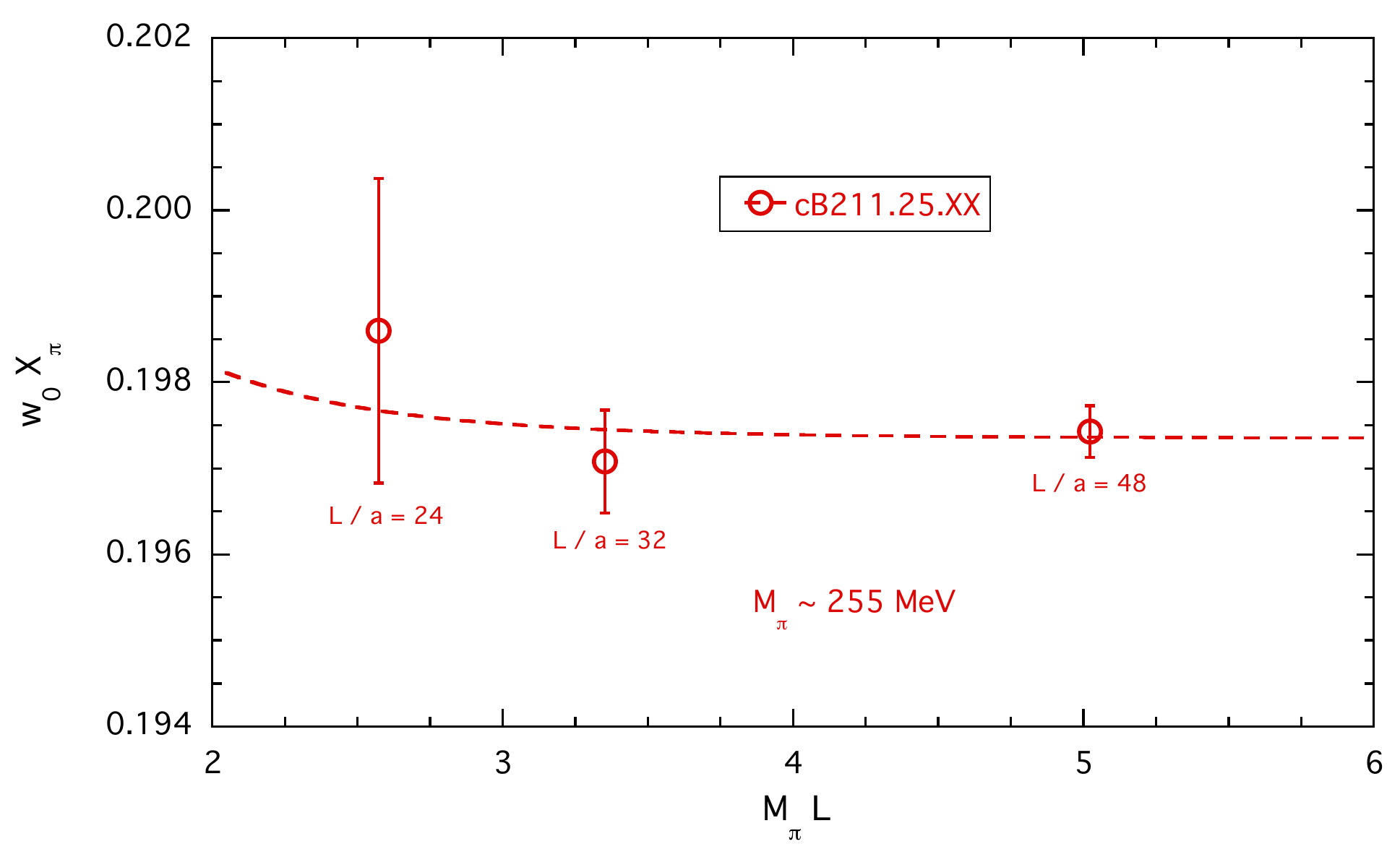}
\end{center}
\vspace{-1.0cm}
\caption{\it \small Values of $w_0 X_\pi$ versus $M_\pi L$ for the three ensembles cB211.25.XX differing only for the lattice size L. The dashed line indicates the simple exponential fit of the form $A \left[ 1 + B e^{- M_\pi L} / (M_\pi L)^{3/2} \right]$.}
\label{fig:cB25XX}
\end{figure}

The interesting feature of the quantity $X_\pi$, given by Eq.~(\ref{eq:XPi}), is the absence of NLO chiral logs in its SU(2) ChPT expansion (see Eqs.~(\ref{eq:fPi}) and (\ref{eq:MPi2})) when expressed in terms of quark masses.
This implies the absence of FVEs at NLO, which in turn is also the origin of the small FVEs observed in the right panel of Fig.~\ref{fig:pion_data}.
This point is better elucidated in Fig.~\ref{fig:cB25XX}, where the results corresponding to the three ensembles cB211.25.XX differing only in the lattice size $L$ are shown.

%%%%%%%%%%%%%%%%%%%%%%%%%%%%%%%%%%%%%%%%%%%%%%%%%%
\section{Determination of the GF scale $w_0$ from the pion data}
\label{sec:w0}
%%%%%%%%%%%%%%%%%%%%%%%%%%%%%%%%%%%%%%%%%%%%%%%%%%

Let's now apply the SU(2) ChPT predictions for interpolating the pion data to the physical pion mass and for extrapolating them to the continuum and infinite volume limits.
The goal is to determine the GF scale $w_0$ adopting the physical value (\ref{eq:fPi_isoQCD}) at the pion point~(\ref{eq:MPi_isoQCD}) without resorting to the use of the renormalized light-quark mass.
In the next two subsections we separately analyze the pion decay constant $f_\pi$ and the quantity $X_\pi$, respectively.

%%%%%%%%%%%%%%%%%%%%%%%%%%%%%%%%%%%%%%%%%%%%%%%%%%
\subsection{Determination of $w_0$ using the data for $f_\pi$}
\label{sec:fPi}
%%%%%%%%%%%%%%%%%%%%%%%%%%%%%%%%%%%%%%%%%%%%%%%%%%

Using the simulated values $a M_\pi$ and $a f_\pi$ in lattice units we evaluate the expansion variable $\xi_\pi$, defined (from now on) as
\be
    \xi_\pi \equiv \frac{(a M_\pi)^2}{(4 \pi a f_\pi)^2} = \frac{M_\pi^2}{(4 \pi f_\pi)^2} ~ , ~
    \label{eq:xiPi}
\ee
which depends on neither $w_0$ nor $w_0 / a$.
Then, for each gauge ensemble we calculate the FVE correction $\Delta_{FVE}^\pi(L)$ as
\be
    \Delta_{FVE}^\pi(L) = -2 \xi_\pi ~ \widetilde{g}_1(M_\pi L) ~
    \label{eq:FVE_NLO}
\ee
and we re-express the quantity $\xi_\ell$ (see Eq.~(\ref{eq:xi_ell})) in terms of the pion mass in the infinite volume limit (see Eq.~(\ref{eq:MPi2_FVE}))
\be
   \xi_\ell \to \xi \equiv \frac{M_\pi^2(L \to \infty)}{(4\pi f)^2} = \frac{(w_0 M_\pi)^2}{(4\pi w_0 f)^2} ~ 
                                   \frac{1}{\left[ 1 - \frac{1}{4} \Delta_{FVE}^\pi(L) \right]^2} ~ , ~ 
   \label{eq:xi_ell_meson}
\ee
where only the knowledge of $w_0 / a$ is required to calculate the pion mass in units of $w_0$ and the free parameter becomes $w_0 f$.

We correct the data of the pion decay constant $w_0 f_\pi(L)$ for FVEs (see Eq.~(\ref{eq:fPi_FVE})), namely
\be
     w_0 f_\pi(L \to \infty) = \frac{w_0 f_\pi(L)}{1 + \Delta_{FVE}^\pi(L)} ~ . ~
     \label{eq:fpi_infty}
\ee
Analogously, for the pion mass $w_0 M_\pi(L)$ one has
\be
     w_0 M_\pi(L \to \infty) = \frac{w_0 M_\pi(L)}{1 - \frac{1}{4} \Delta_{FVE}^\pi(L)} ~ . ~
     \label{eq:Mpi_infty}
\ee

The data for $w_0 f_\pi(L \to \infty)$ are fitted in terms of the variable $\xi$ (see Eq.~(\ref{eq:xi_ell_meson})) using the following functional form
\be
     w_0 f_\pi(L \to \infty) = w_0 f \left[ 1 - 2 \xi \mbox{log}(\xi) + 2 A_1 \xi + A_2 \xi^2 + \frac{a^2}{w_0^2} \left( D_0 + D_1 \xi \right) \right]
     \label{eq:fpi_NNLO}
\ee
where with respect to a pure NLO ansatz we have added a possible higher-order term quadratic in $\xi$ as well as discretization effects proportional to $a^2$ and $a^2 M_\pi^2$.

The free parameters appearing in Eq.~(\ref{eq:fpi_NNLO}) are $w_0 f$, $A_1$, $A_2$, $D_0$, $D_1$ and their values are obtained from a standard $\chi^2$-minimization.
From the value of $w_0 f$ the GF scale $w_0$ can be determined as follows.
Let us consider the physical value of the variable (\ref{eq:xiPi}), namely
\be
    \xi_\pi^{isoQCD} \equiv \left[ \frac{M_\pi^{isoQCD}}{4 \pi f_\pi^{isoQCD}} \right]^2 = 0.006785 ~ (29) ~ . ~
    \label{eq:xiPi_isoQCD}
\ee
Using Eq.~(\ref{eq:fpi_NNLO}) in the continuum limit the physical value of the variable (\ref{eq:xi_ell_meson}), namely $\xi^{isoQCD} = (M_\pi^{isoQCD} / 4 \pi f)^2$, can be obtained by solving the relation
\be
    \xi_\pi^{isoQCD} = \frac{\xi^{isoQCD}}{\left[ 1 - 2 \xi^{isoQCD} \mbox{log}(\xi^{isoQCD}) + 2 A_1 \xi^{isoQCD} + A_2 (\xi^{isoQCD})^2 \right]^2} ~ .
\ee
In this way the value of the LEC $f$ in physical units is given by $f = M_\pi^{isoQCD} / (4 \pi \sqrt{\xi^{isoQCD}})$ and, therefore, $w_0$ can be determined using the value of $w_0 f$.

\begin{figure}[htb!]
\begin{center}
\includegraphics[scale=0.75]{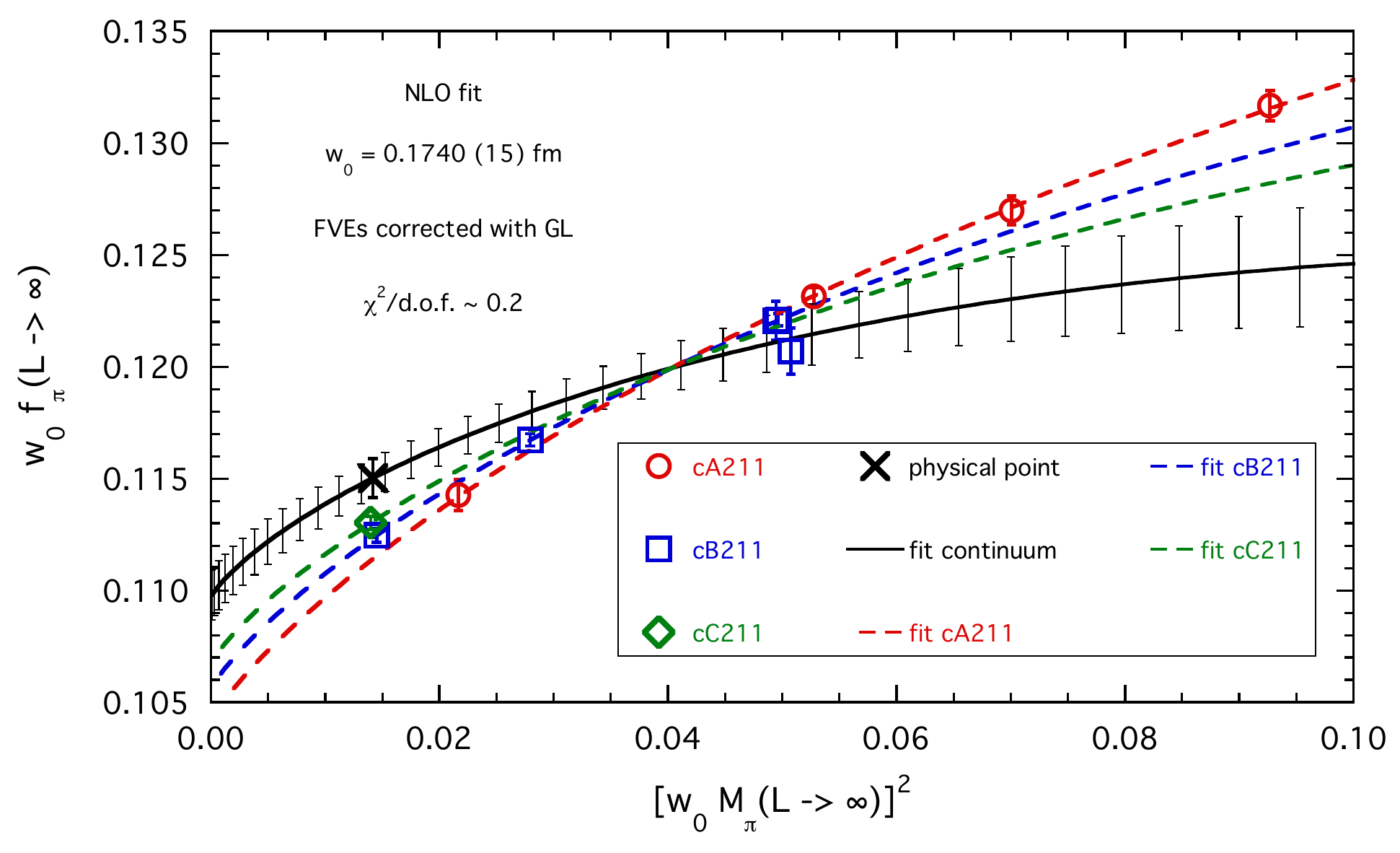}
\end{center}
\vspace{-0.75cm}
\caption{\it \small Values of the pion decay constant $w_0 f_\pi$ corrected for FVEs according to Eq.~(\ref{eq:fpi_infty}) (open markers) and compared to the results of the NLO ChPT fit corresponding to $A_2 = 0$ in Eq.~(\ref{eq:fpi_NNLO}) applied to all data points ($M_\pi \lesssim 350$ MeV). The solid line represents the results of the fit in the continuum limit, while the dashed lines correspond to the fit evaluated at each value of $\beta$. The cross represents the result at the physical pion point~(\ref{eq:MPi_isoQCD}) corresponding to the value $w_0 = 0.1740\,(15)$ fm, obtained as described in the text.}
\label{fig:fPi_NLO}
\end{figure}

We start by considering a pure NLO fit, i.e.~$A_2 = 0$, including only the discretization effect proportional to $a^2$, i.e.~$D_1 = 0$ in Eq.~(\ref{eq:fpi_NNLO}), and we apply it to all pion data up to $M_\pi \simeq 350$ MeV.
The discretization coefficient $D_0$ turns out to be quite small, $D_0 = -0.05\,(4)$, and the corresponding $\chi^2/({\rm d.o.f.})$ is equal to $\chi^2/({\rm d.o.f.}) \simeq 1.5$ for 10 data points and 3 parameters.
For the GF scale $w_0$ we get $w_0 = 0.1712\,(14)$ fm, which exhibits a $\simeq 0.8 \%$ accuracy.
However, a drastic improvement in the quality of the fit is obtained by including the discretization term proportional to $a^2 M_\pi^2$, i.e.~$D_1 \neq 0$.
This leads to $\chi^2/({\rm d.o.f.}) \simeq 0.2$, obtaining for $w_0$ the value
\be
     \label{eq:w0_NLO}
     w_0 = 0.1740 ~ (15) ~ {\rm fm} ~ 
\ee
with $f = 124.4\,(6)$ MeV and $\bar{\ell}_4^{phys} = 3.24\,(29)$ (see Eq.~(\ref{eq:l4})).
The quality of the above fit is illustrated in Fig.~\ref{fig:fPi_NLO}.
The result (\ref{eq:w0_NLO}) is confirmed by a NLO fit without the discretization effects proportional to $a^2 M_\pi^2$ (i.e.~$D_1 = 0$), but limited to pion masses below $\simeq 190$ MeV (4 data points and 3 parameters).
In this case one gets $w_0 = 0.1736\,(15)$ fm, $f = 122.8\,(4)$ MeV, $\bar{\ell}_4^{phys} = 4.06\,(18)$ and $\chi^2/({\rm d.o.f.}) \simeq 0.1$. 

In order to investigate systematic effects we include the quadratic term proportional to $A_2$, obtaining $w_0 = 0.1737\,(16)$ fm, $f = 124.3\,(7)$ MeV, $\bar{\ell}_4^{phys} = 3.26\,(30)$ and $\chi^2/({\rm d.o.f.}) \simeq 0.2$, and we check also the impact of FVEs by multiplying the correction $\Delta_{FVE}^\pi(L)$ of Eq.~(\ref{eq:FVE_NLO}) by a factor $\kappa_{FVE}$ used as a further free parameter in the NLO fit.
The factor $\kappa_{FVE}$ turns out to be consistent with unity, $\kappa_{FVE} = 1.20\,(18)$, and we get $w_0 = 0.1743\,(16)$ fm, $f = 124.5\,(6)$ MeV, $\bar{\ell}_4^{phys} = 3.18\,(30)$ and $\chi^2/({\rm d.o.f.}) \simeq 0.1$.

After averaging the above results our determinations of $w_0$, $f$ and $\bar{\ell}_4^{phys}$ based on the analysis of $f_\pi$ are
\bea
     \label{eq:w0_fPi}
     w_0 & = & 0.17390 ~ (157)_{stat+fit} ~ (30)_{syst} ~ [160] ~{\rm fm} ~ , ~ \\[2mm]
     \label{eq:f_fPi}
     f & = & 124.0 ~ (6)_{stat+fit} ~ (7)_{syst} ~ [9] ~{\rm MeV} ~ , ~ \\[2mm]
     \label{eq:l4_fPi}
     \bar{\ell}_4^{phys} & = & 3.44 ~ (27)_{stat+fit} ~ (36)_{syst} ~ [45] ~ , ~ 
\eea
where $()_{stat+fit}$ incorporates the uncertainties induced by both the statistical errors and the fitting procedure itself, $()_{syst}$ corresponds to the uncertainty related to chiral interpolation, discretization and finite-volume effects, while the last error is their sum in quadrature. 
More precisely, the various systematic uncertainties are estimated by considering the results obtained with $A_2 =0$ or $A_2 \neq 0$ in the case of the chiral extrapolation, with $D_1 \neq 0$ or $D_1 = 0$ (but limited to $M_\pi < 190$ MeV) for the discretization effects and with $\kappa_{FVE} = 1$ or $\kappa_{FVE} \neq 1$ for the FVEs.

%%%%%%%%%%%%%%%%%%%%%%%%%%%%%%%%%%%%%%%%%%%%%%%%%%
\subsection{Determination of the GF scale $w_0$ using the data for $X_\pi$}
\label{sec:XPi}
%%%%%%%%%%%%%%%%%%%%%%%%%%%%%%%%%%%%%%%%%%%%%%%%%%

In this Section we illustrate the results of the analysis of the lattice data for the quantity $w_0 X_\pi$ adopting the following fitting function
\bea
    \label{eq:XPi_fit}
    w_0 X_\pi & = & (w_0 f)  \left\{ (4 \pi)^4 \xi^2 \left[ 1 - 2 \xi \mbox{log}(\xi) + 2 A_1 \xi + A_2^\prime \xi^2  + a^2 \left( D_0^\prime + D_1^\prime \xi \right) \right] \right\}^{1/5} ~ 
                               \nonumber \\[2mm]
                     & \cdot & \left( 1 + F_{FVE} ~ \xi^2 e^{-M_\pi L} / (M_\pi L)^{3/2} \right) ~ , ~
\eea
where the variable $\xi$ is defined by Eq.~(\ref{eq:xi_ell_meson}), given in terms of the pion mass corrected for the FVEs using the GL formula (\ref{eq:FVE_NLO}), and the coefficient $A_1$ is related to the LEC $\bar{\ell}_4^{phys}$ by Eq.~(\ref{eq:l4}).
In Eq.~(\ref{eq:XPi_fit}) we have taken into account that the FVEs on $X_\pi$ start only at NNLO, i.e.~at order ${\cal{O}}(\xi^2)$.
Their impact is obtained by including ($F_{FVE} \neq 0$) or by excluding ($F_{FVE} = 0$) the higher order FVEs.
Moreover, the NLO chiral log is present only because we employ meson masses and it would disappear if the light-quark mass would be instead considered (in this case the linear coefficient $A_1$ provides directly the difference $\bar{\ell}_4^{phys} - \bar{\ell}_3^{phys}$).

\begin{figure}[htb!]
\begin{center}
\includegraphics[scale=0.80]{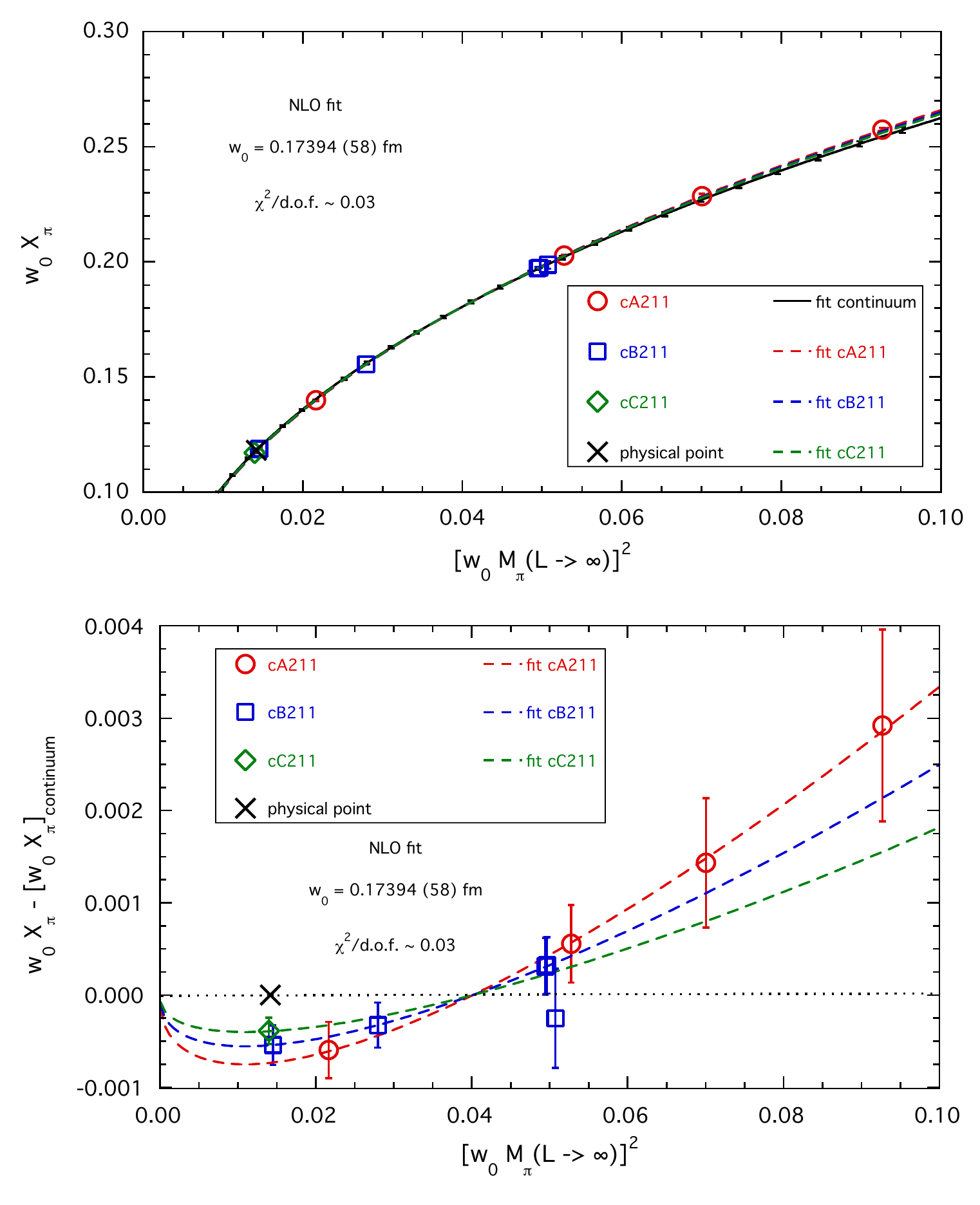}
\end{center}
\vspace{-0.75cm}
\caption{\it \small Top panel: values of the quantity $w_0 X_\pi$ (open markers) compared to the results of the NLO fit corresponding to $A_2^\prime = F_{FVE} = 0$ in Eq.~(\ref{eq:XPi_fit}) applied to all data points ($M_\pi \lesssim 350$ MeV). The solid line represents the results of the fit in the continuum limit, while the dashed lines correspond to the fit evaluated at each value of $\beta$. The cross represents the result at the physical pion point~(\ref{eq:MPi_isoQCD}) corresponding to the value $w_0 = 0.17394\,(58)$ fm. Bottom panel: the quantity $w_0 X_\pi$ after subtraction of its extrapolation to the continuum limit. The discretization terms proportional both to $a^2$ and to $a^2 \xi$ present in Eq.~(\ref{eq:XPi_fit}) are clearly visible.}
\label{fig:XPi_NLO}
\end{figure}

We have performed several fits similar to those adopted in Section~\ref{sec:fPi} and the corresponding results are collected in Table~\ref{tab:XPi}.
\begin{table}[hbt!]
\begin{center}
{\small
\begin{tabular}{||c|c|c|c||c||c|c||c||}
\hline
~$A_2^\prime \neq 0$~ & ~$D_1^\prime \neq 0$~ & ~$F_{FVE} \neq 0$~ & ~range of $M_\pi$~ & ~$w_0$~(fm)~ & ~$f$~(MeV)~ & ~$\bar{\ell}_4^{phys}$~ & ~$\chi^2 / \mbox{(d.o.f.)}$~ \\
\hline \hline
no   & no  & no   & ~$<$ 350 MeV~ & ~0.17213~(47)~ & ~122.4~(0.7)~ & ~4.23~~(9)~ & ~0.26~ \\ \hline
\hline
no   & yes & no   & ~$<$ 350 MeV~ & ~0.17394~(58)~ & ~124.4~(1.2)~ & ~3.24~(29)~ & ~0.03~ \\ \hline
no   & no   & no   & ~$<$ 190 MeV~ & ~0.17343~(53)~ & ~122.8~(1.0)~ & ~4.04~(16)~ & ~0.05~ \\ \hline
yes & yes & no   & ~$<$ 350 MeV~ & ~0.17378~(56)~ & ~124.3~(1.3)~ & ~3.27~(30)~ & ~0.04~ \\ \hline
no   & yes & yes & ~$<$ 350 MeV~ & ~0.17415~(61)~ & ~124.6~(1.3)~ & ~3.15~(35)~ & ~0.02~ \\ \hline
\hline
\end{tabular}
}
\end{center}
\vspace{-0.25cm}
\caption{\it \small Results for $w_0$ obtained by fitting the lattice data for $w_0 X_\pi$ using Eq.~(\ref{eq:XPi_fit}) and adopting the isoQCD values~(\ref{eq:MPi_isoQCD}) and~(\ref{eq:fPi_isoQCD}) for fixing the lattice scale at the physical pion point.}
\label{tab:XPi}
\end{table}
The quality of the NLO fit with $D_1^\prime \neq 0$ is illustrated in Fig.~\ref{fig:XPi_NLO}, where it is also clearly visible the presence of discretization effects proportional to $a^2 M_\pi^2$, as already observed in the case of $w_0 f_\pi$ (see Fig.~\ref{fig:fPi_NLO}).
We stress that for both quantities, $w_0 f_\pi$ and $w_0 X_\pi$, the inclusion of a discretization term proportional to $a^2 M_\pi^2$ leads to higher values of $w_0$.
This result is reassuringly confirmed also by a NLO fit without such a discretization term (i.e.~$D_1^\prime = 0$), but limited to pion masses below $\simeq 190$ MeV (see the fourth row of Table~\ref{tab:XPi}).

By averaging the last four results of Table~\ref{tab:XPi} one has
\bea
   \label{eq:w0_XPi}
    w_0 & = & 0.17383 ~ (57)_{\rm stat+fit} ~ (26)_{\rm syst} ~ [63] ~ \mbox{fm} ~ , ~ \\[2mm]
   \label{eq:f_XPi}
    f & = & 124.0 ~ (1.2)_{\rm stat+fit} ~ (0.7)_{\rm syst} ~ [1.4] ~ \mbox{MeV} ~ , ~ \\[2mm]
   \label{eq:l4_XPi}
    \bar{\ell}_4^{phys} & = & 3.43 ~ (28)_{\rm stat+fit} ~ (36)_{\rm syst} ~ [46] ~ , ~
\eea
which nicely agree with the corresponding results obtained by the analysis of $f_\pi$ given by Eqs.~(\ref{eq:w0_fPi}-\ref{eq:l4_fPi}).
Note that the determination of $w_0$ obtained using $X_\pi$ is more precise than the one from $f_\pi$ by a factor equal to $\approx 2.5$.

Our result~(\ref{eq:w0_XPi}) is slightly larger than both the MILC result $w_0 = 0.1714_{-12}^{+15}$ fm from Ref.~\cite{Bazavov:2015yea} and the HPQCD result $w_0 = 0.1715~(9)$ fm from Ref.~\cite{Dowdall:2013rya}, obtained using the value~(\ref{eq:fPi_isoQCD}) to set the lattice scale. 
Within $\simeq 1.5$ standard deviations it is consistent with the recent, precise BMW determination $w_0 = 0.17236~(70)$, obtained in Ref.~\cite{Borsanyi:2020mff} using the $\Omega^-$-baryon mass to set the lattice scale.
Furthermore, the difference with the recent result $w_0 = 0.1709(11)$ fm, obtained in Ref.~\cite{Miller:2020evg} using the $\Omega^-$-baryon mass to set the lattice scale, is within $\sim 2$ standard deviations.

In Appendix~\ref{sec:appD_sect2} the procedure used in this Section to determine the GF scale $w_0$ is repeated in the case of the scales $\sqrt{t_0}$ and $t_0 / w_0$, obtaining
\bea
   \label{eq:t0s}
    \sqrt{t_0} & = & 0.14436 ~ (54)_{\rm stat+fit} ~ (30)_{\rm syst} ~ [61] ~ \mbox{fm} ~ , ~ \\[2mm]
   \label{eq:f_t0s}
    f & = & 124.1 ~ (1.2)_{\rm stat+fit} ~ (0.7)_{\rm syst} ~ [1.4] ~ \mbox{MeV} ~ , ~ \\[2mm]
   \label{eq:l4_t0s}
    \bar{\ell}_4^{phys} & = & 3.37 ~ (27)_{\rm stat+fit} ~ (38)_{\rm syst} ~ [47] ~
\eea
and
\bea
   \label{eq:t0w0}
    t_0 / w_0 & = & 0.11969 ~ (52)_{\rm stat+fit} ~ (33)_{\rm syst} ~ [62] ~ \mbox{fm} ~ , ~ \\[2mm]
   \label{eq:f_t0w0}
    f & = & 124.2 ~ (1.4)_{\rm stat+fit} ~ (0.8)_{\rm syst} ~ [1.6] ~ \mbox{MeV} ~ , ~ \\[2mm]
   \label{eq:l4_t0w0}
    \bar{\ell}_4^{phys} & = & 3.31 ~ (27)_{\rm stat+fit} ~ (40)_{\rm syst} ~ [48] ~ . ~
\eea
Our finding (\ref{eq:t0s}) is larger than the MILC result $\sqrt{t_0} = 0.1416_{-5}^{+8}$ fm from Ref.~\cite{Bazavov:2015yea} and the HPQCD result $\sqrt{t_0} = 0.1420~(8)$ fm from Ref.~\cite{Dowdall:2013rya}, while within $\simeq 1.5$ standard deviations it is consistent with the recent result $\sqrt{t_0} = 0.1422(14)$ fm from Ref.~\cite{Miller:2020evg}.

The values of the lattice spacing corresponding to the three GF scales are collected in Table~\ref{tab:spacings} of Appendix~\ref{sec:appD_sect2}.

%%%%%%%%%%%%%%%%%%%%%%%%%%%%%%%%%%%%%%%%%%%%%%%%%%
\section{SU(2) ChPT analysis of $f_K / f_\pi$}
\label{sec:fKPi}
%%%%%%%%%%%%%%%%%%%%%%%%%%%%%%%%%%%%%%%%%%%%%%%%%%

The kaon correlator
 \be
    C_K(t) = \frac{1}{L^3} \sum\limits_{x, z} \left\langle 0 \right| \overline{q}_s(x) \gamma_5 q_\ell(x) \overline{q}_\ell(z) \gamma_5 q_s(z) 
                   \left| 0 \right\rangle \delta_{t, (t_x  - t_z )} ~
    \label{eq:P5P5_K}
 \ee
has been evaluated for three values of the (valence) strange bare quark mass $a \mu_s$ at each value of $\beta$, namely: $a \mu_s = \{ 0.0176, 0.0200, 0.0264\}$ for the ensembles cA211, $a \mu_s = \{ 0.0148, 0.0185, 0.0222\}$ for the ensembles cB211 and $a \mu_s = \{ 0.0128, 0.0161, 0.0193\}$ for the ensemble cC211.06.80.

At large time distances one has
 \be
    C_K(t)_{ ~ \overrightarrow{t  \gg a, ~ (T - t) \gg a} ~ } \frac{\mathcal{Z}_K}{2M_K} \left[ e^{ - M_K  t}  + e^{ - M_K (T - t)} \right] ~ ,
    \label{eq:larget_K}
 \ee
which allows the extraction of the kaon mass $M_K$ and the matrix element $\mathcal{Z}_K = | \langle K | \overline{q}_s\gamma_5 q_\ell | 0 \rangle|^2$ from the exponential fit given in the r.h.s.~of Eq.~(\ref{eq:larget_K}). 
The kaon decay constant $f_K$ is given by
 \be
    af_K = \left( a \mu_\ell + a \mu_s \right) \frac{\sqrt{a^4 \mathcal{Z}_K}}{aM_K ~ \mbox{sinh}(aM_K)} ~ 
    \label{eq:decayK}
 \ee
and, using the pion data (\ref{eq:decayPS}) for $f_\pi$, the ratio $f_K / f_\pi$ is evaluated at each simulated strange bare quark mass.
The time intervals $[t_{min}, t_{max}]$ adopted for the fit (\ref{eq:larget_K}) of the kaon correlation function (\ref{eq:P5P5_K}) are the same as those used for the case of the pion correlator, collected in Table~\ref{tab:plateaux}.

As in the case of the pion data (see Section~\ref{sec:ETMC}), due to a small deviation from maximal twist, a correction should be applied to observables of the ensemble cA211.12.48.
We use the following formula (see Appendix~\ref{sec:appB})
\be
      f_K |_{corrected} \simeq f_K \cdot K_f
      \label{eq:decay_K_corrected}
\ee
with
\be
    K_f = \frac{1}{\mbox{cos}[(\theta_s + \theta_\ell)/2]} ~ ,
    \label{eq:mtm_correction}
\ee
where, we remind,
\be
     \frac{1}{\mbox{cos}(\theta_i)} \equiv K_i = \sqrt{1 + (Z_A ~ m_{PCAC} / \mu_i)^2} ~ ,
\ee
$m_{PCAC}$ is the bare untwisted PCAC mass, $Z_A$ is the renormalization constant of the axial current and $\mu_i$ is the bare twisted mass of the valence quarks.
In the degenerate case $m_s = m_\ell$ one gets $K_f = K_\ell$, i.e.~Eq.~(\ref{eq:Kell}), while for $m_s >> m_\ell$ one has $K_f \simeq 1/ \mbox{cos}(\theta_\ell / 2)$.

Since the LECs of the SU(2) ChPT depend on the value of the (renormalized) strange quark mass $m_s$, we need to interpolate the ratio $f_K / f_\pi$ at an approximately fixed value of $m_s$.
To this end we take advantage of the fact that the meson mass combination $2M_K^2 - M_\pi^2$ is proportional to $m_s$ at LO in ChPT.
Thus, for each gauge ensemble, adopting a simple quadratic spline, the lattice data for $f_K / f_\pi$ are interpolated at a reference kaon mass given by
\be
    M_K^{ref} \equiv \sqrt{ \left( M_K^{isoQCD} \right)^2 + \frac{M_\pi^2 - \left( M_\pi^{isoQCD} \right)^2}{2} } ~
     \label{eq:MKref}
\ee
with $M_\pi^{isoQCD}$ and $M_K^{isoQCD}$ chosen as in Eqs.~(\ref{eq:MPi_isoQCD}) and~(\ref{eq:MK_isoQCD}), respectively. 
The physical units for $M_\pi$ (and consequently for $M_K^{ref}$) are obtained by using the results for the lattice spacing given in Table~\ref{tab:spacings} of Appendix~\ref{sec:appD_sect2} for each choice of the GF scale.
In what follows we make use of our determination (\ref{eq:w0_XPi}) of the GF scale $w_0$.
In this way the renormalized strange quark mass $m_s^{ref}$ corresponding to $M_K^{ref}$ is kept close to its physical value.

\begin{figure}[htb!]
\begin{center}
\includegraphics[scale=0.75]{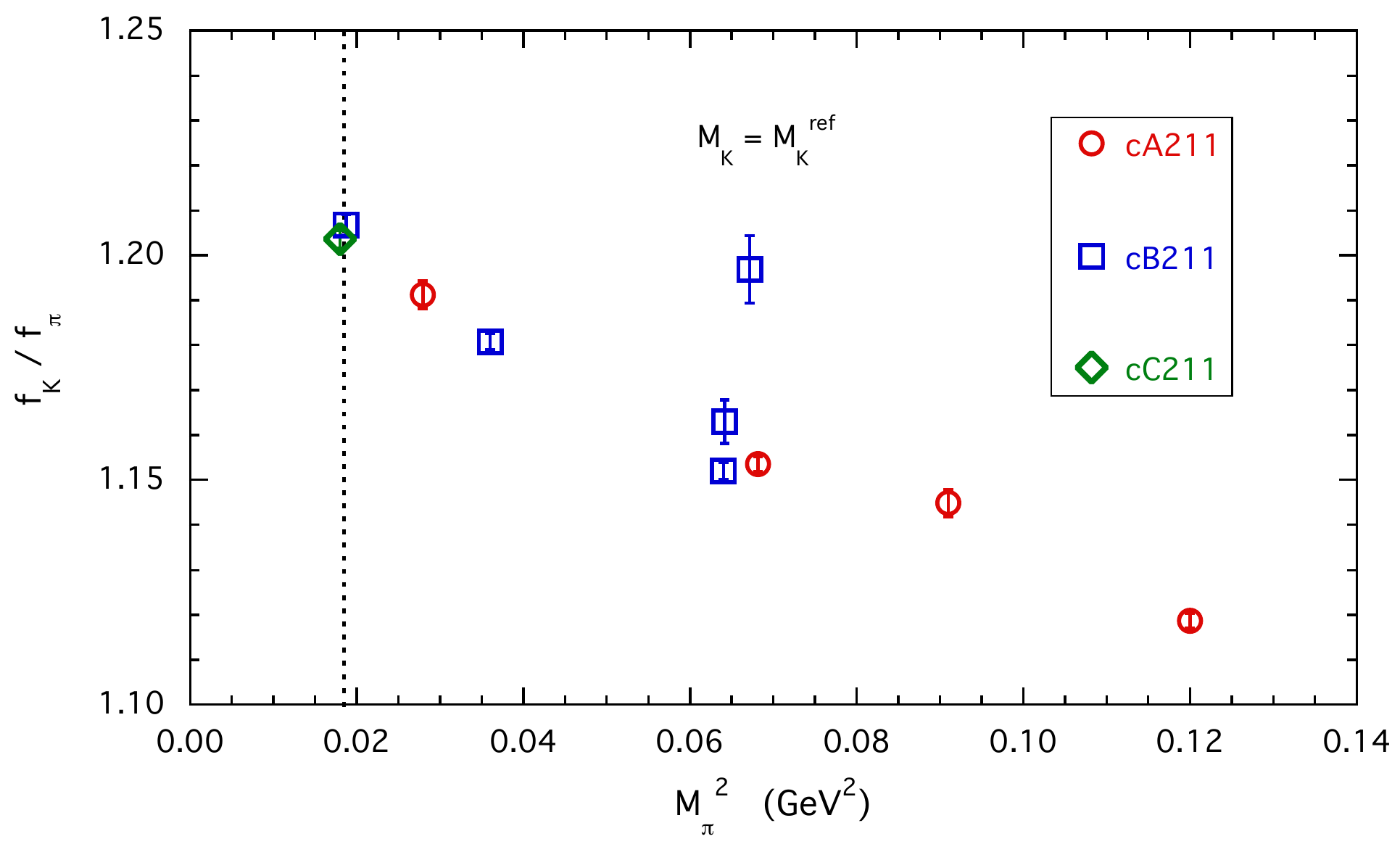}
\end{center}
\vspace{-0.75cm}
\caption{\it \small Values of the ratio $f_K / f_\pi$ interpolated at the kaon reference mass (\ref{eq:MKref}) versus the squared pion mass. The vertical dotted line indicates the location of the physical isoQCD point (\ref{eq:MPi_isoQCD}). For the ensemble cA211.12.48 the corrected value of the ratio $f_K / f_\pi$, obtained using Eqs.~(\ref{eq:decay_PS_corrected}) and (\ref{eq:decay_K_corrected}), is considered.}
\label{fig:fKPi_data}
\end{figure}

The results obtained for the ratio $f_K / f_\pi$ interpolated at the kaon reference mass (\ref{eq:MKref}) are shown in Fig.~\ref{fig:fKPi_data} for all the ETMC gauge ensembles.
The statistical errors of the data lie in the range $0.1 \div 0.6 \%$.

We now apply the correction for FVEs using the GL formula and the expansion variable $\overline{\xi}_\pi$ defined as
\be
    \overline{\xi}_\pi \equiv \frac{M_\pi^2(L)}{(4 \pi f)^2} ~ , ~
    \label{eq:csi_fKPi}
\ee
where $f$ is fixed at the value given by Eq.~(\ref{eq:f_XPi}).
For the pion and kaon decay constants the NLO FVE corrections are respectively given by~\cite{Colangelo:2005gd} 
\bea
    \label{eq:FVE_pion}
   \Delta_{FVE}^\pi(L) & = & -2 \overline{\xi}_\pi~ \widetilde{g}_1(M_\pi L) ~ \\[2mm]
    \label{eq:FVE_kaon}
    \Delta_{FVE}^K(L) & = & -\frac{3}{4} \overline{\xi}_\pi ~ \widetilde{g}_1(M_\pi L) ~ , ~
\eea
so that the overall FVE correction for $f_K / f_\pi$ is given by
\be
     \frac{f_K}{f_\pi}(L \to \infty) = \frac{f_K}{f_\pi}(L) \left[ 1 - \frac{5}{4} \overline{\xi}_\pi ~ \widetilde{g}_1(M_\pi L) \right] ~ . ~
     \label{eq:fKPi_FVE}
\ee

Finally, in terms of the variable $\xi$, defined in Eq.~(\ref{eq:xi_ell_meson}), the data for $(f_K / f_\pi)(L \to \infty)$ are fitted using the following ansatz
\be
     \frac{f_K}{f_\pi}(L \to \infty) = R_0 \left[ 1 + \frac{5}{4} \xi \mbox{log}(\xi) + R_1 \xi + R_2 \xi^2 + \frac{a^2}{w_0^2} \left( \widetilde{D}_0 + \widetilde{D}_1 \xi \right) \right] 
     \label{eq:fKPi_fit}
\ee
where with respect to the well-known SU(2) ChPT prediction at NLO a quadratic term in $\xi$ as well as discretization effects proportional to $a^2$ and $a^2 M_\pi^2$ have been added.

The free parameters appearing in Eq.~(\ref{eq:fKPi_fit}) are $R_0$, $R_1$, $R_2$, $\widetilde{D}_0$, $\widetilde{D}_1$ and their values are obtained by a straightforward $\chi^2$-minimization procedure.
We have performed several fits based on Eq.~(\ref{eq:fKPi_fit}) and the results for the ratio $(f_K / f_\pi)^{isoQCD}$ at the physical pion point~(\ref{eq:MPi_isoQCD}) are collected in Table~\ref{tab:fKPi}.
\begin{table}[hbt!]
\begin{center}
\begin{tabular}{||c|c|c||c||c||}
\hline
~$R_2 \neq 0$~ & ~$\widetilde{D}_1 \neq 0$~ & ~range of $M_\pi$~ & ~$(f_K / f_\pi)^{isoQCD}$~ & ~$\chi^2 / \mbox{(d.o.f.)}$~ \\
\hline \hline
no   & no  & ~$<$ 350 MeV~ & ~1.1995~(35)~ & ~0.53~ \\ \hline
no   & yes & ~$<$ 350 MeV~ & ~1.1984~(54)~ & ~0.58~ \\ \hline
no   & no  & ~$<$ 190 MeV~ & ~1.2005~(48)~ & ~1.40~ \\ \hline
yes & no  & ~$<$ 350 MeV~ & ~1.1998~(32)~ & ~0.37~ \\ \hline
\hline
\end{tabular}
\end{center}
\vspace{-0.25cm}
\caption{\it \small Results for the decay constant ratio $(f_K / f_\pi)^{isoQCD}$ at the physical isoQCD point, given by Eqs.~(\ref{eq:MPi_isoQCD}) and~(\ref{eq:MK_isoQCD}), obtained using the fitting function~(\ref{eq:fKPi_fit}).}
\label{tab:fKPi}
\end{table}

\begin{figure}[htb!]
\begin{center}
\includegraphics[scale=0.75]{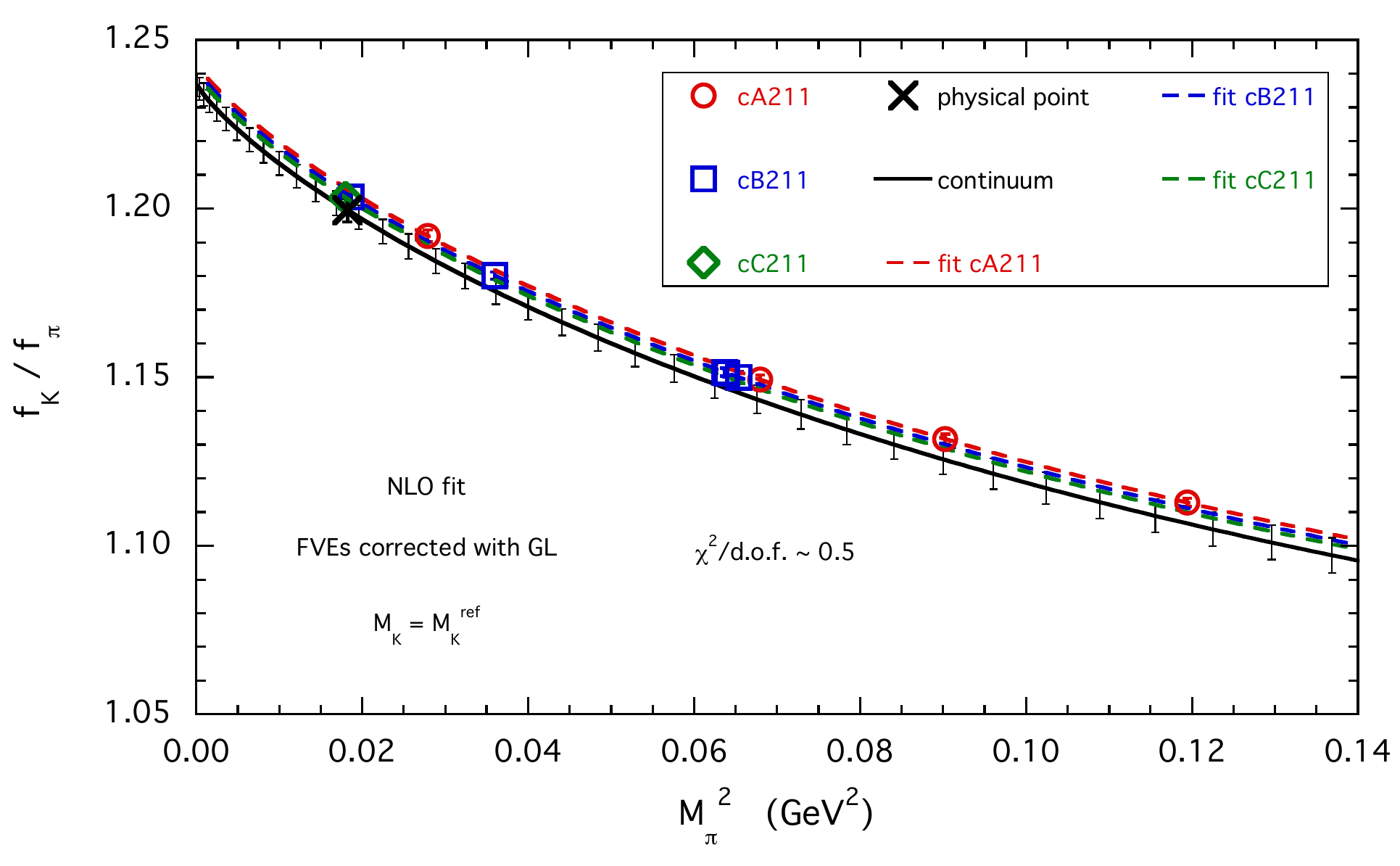}
\end{center}
\vspace{-0.75cm}
\caption{\it \small Values of the ratio $f_K / f_\pi$ corrected for FVEs according to Eq.~(\ref{eq:fKPi_FVE}) (open markers) compared to the results of the NLO fit corresponding to $R_2 = \widetilde{D}_1 = 0$ in Eq.~(\ref{eq:fKPi_fit}) applied to all pion masses ($M_\pi \lesssim 350$ MeV). The solid line represents the results of the fit in the continuum limit, while the dashed lines correspond to the fit evaluated at each value of $\beta$. The cross represents the result at the physical pion point~(\ref{eq:MPi_isoQCD}).}
\label{fig:fKPi_NLO}
\end{figure}

The quality of the NLO fit with $R_2 = \widetilde{D}_1 = 0$ is illustrated in Fig.~\ref{fig:fKPi_NLO}.
It can be seen that FVEs are properly taken care of and that discretization effects are quite small.
As a check of the impact of FVEs we multiply the GL correction in Eq.~(\ref{eq:fKPi_FVE}) by a factor $\kappa_{FVE}$, which is treated as a further free parameter in the NLO fit.
The factor $\kappa_{FVE}$ turns out to be consistent with unity, $\kappa_{FVE} = 1.19\,(24)$, and the NLO result $(f_K / f_\pi)^{isoQCD} = 1.1995~(35)$ is reassuringly confirmed.

Putting together all the various results we obtain
\be
     \label{eq:fKPi_isoQCD}
     \left( \frac{f_K}{f_\pi} \right)^{isoCQD} = 1.1995 ~ (44)_{stat+fit} ~ (7)_{syst} ~ [44] ~ , ~
\ee
where we remind that $()_{stat+fit}$ incorporates the uncertainties induced by both the statistical errors and the fitting procedure itself.
Adopting the results of the ODE procedure (see Appendix~\ref{sec:appB}) for the extraction of the pion and kaon masses and decay constants the analysis of the ratio $f_K / f_\pi$ yields 
\be
     \label{eq:fKPi_ODE}
     \left( \frac{f_K}{f_\pi} \right)^{isoCQD} = 1.1994 ~ (43)_{\rm stat+fit} ~ (7)_{\rm syst} ~ [43]  ~ , ~  
\ee
which compares very well with the finding (\ref{eq:fKPi_isoQCD}).

The present result (\ref{eq:fKPi_isoQCD}) improves drastically the precision of the previous $N_f = 2+1+1$ ETMC determination $(f_K / f_ \pi)^{isoQCD} = 1.188 ~ (15)$~\cite{Carrasco:2014poa} by a factor of $\simeq 3.5$ reaching the level of $\simeq 0.4 \%$. 
For comparison, the $N_f = 2+1+1$ determinations, entering the FLAG-4 average~\cite{Aoki:2019cca} and corrected for strong IB effects, yield a consistent value within the uncertainties, namely $(f_K / f_ \pi)^{isoQCD} = 1.1966 ~ (18)$~\cite{Dowdall:2013rya,Carrasco:2014poa,Bazavov:2017lyh}.
Our finding (\ref{eq:fKPi_isoQCD}) is also in good agreement with the recent determination $(f_K / f_ \pi)^{isoQCD} = 1.1964 ~ (44)$ obtained in Ref.~\cite{Miller:2020xhy} adopting the same isoQCD prescription in a mixed-action approach (domain-wall valence quarks with staggered sea quarks).

%%%%%%%%%%%%%%%%%%%%%%%%%%%%%%%%%%%%%%%%%%%%%%%%%%
\section{Implications for $V_{us}$ and the first-row CKM unitarity}
\label{sec:CKM}
%%%%%%%%%%%%%%%%%%%%%%%%%%%%%%%%%%%%%%%%%%%%%%%%%%

Inserting our isoQCD result (\ref{eq:fKPi_isoQCD}) into Eq.~(\ref{eq:ratioVf}) the ratio of the CKM entries $V_{us}$ and $V_{ud}$ is given by
\be  
   \left|  \frac{V_{us}}{V_{ud}} \right| = 0.23079 ~ (24)_{\mathrm{exp}} ~ (87)_{\mathrm{th}} = 0.23079 ~ (90) ~ . ~
    \label{eq:VusVud}
\ee
Using the value $|V_{ud}| = 0.97370~(14)$ from super-allowed nuclear beta decays~\cite{Zyla:2020zbs,Seng:2018yzq}, which updates the old result $V_{ud} = 0.97420~(21)$ from Ref.~\cite{Hardy:2016vhg}, Eq.~(\ref{eq:ratioVf}) yields the following value for the CKM element $|V_{us}|$:
\be  
    \label{eq:Vus}
    |V_{us}| = 0.22472 ~ (24)_{\mathrm{exp}} ~ (84)_{\mathrm{th}} = 0.22472 ~ (87) ~ , ~
\ee
which is in good agreement with the latest estimate $|V_{us}| = 0.2252~(5)$ from leptonic modes provided by the PDG~\cite{Zyla:2020zbs}.

Using the values $|V_{ub}| = 0.00382~(24)$~\cite{Zyla:2020zbs} and $|V_{ud}| = 0.97370~(14)$~\cite{Zyla:2020zbs,Seng:2018yzq} our result~(\ref{eq:Vus}) implies for the unitarity of the first-row of the CKM matrix the value
\be
     \label{eq:unitarity}
     |V_{ud}|^2 + |V_{us}|^2 + |V_{ub}|^2 =  0.99861~ (48) ~ , ~
\ee
which in turn would imply a $\simeq 3 \sigma$ tension with unitarity from leptonic modes.
Had we used the result $V_{ud} = 0.97420 ~ (21)$ from Ref.~\cite{Hardy:2016vhg} the first-row CKM unitarity would be fulfilled within one standard deviation, i.e.~within a precision of $\simeq 0.5$ permil.

Another source of information on $V_{us}$ is represented by the semileptonic $K_{\ell 3}$ decay.
In this case the relevant hadronic quantity is the vector form factor at zero momentum transfer $f_+(0)$.
From the high-precision experimental data on $K_{\ell 3}$ decays one has $V_{us} f_+(0) = 0.2165 ~ (4)$~\cite{Moulson:2017ive}.

\begin{figure}[htb!]
\begin{center}
\includegraphics[scale=0.75]{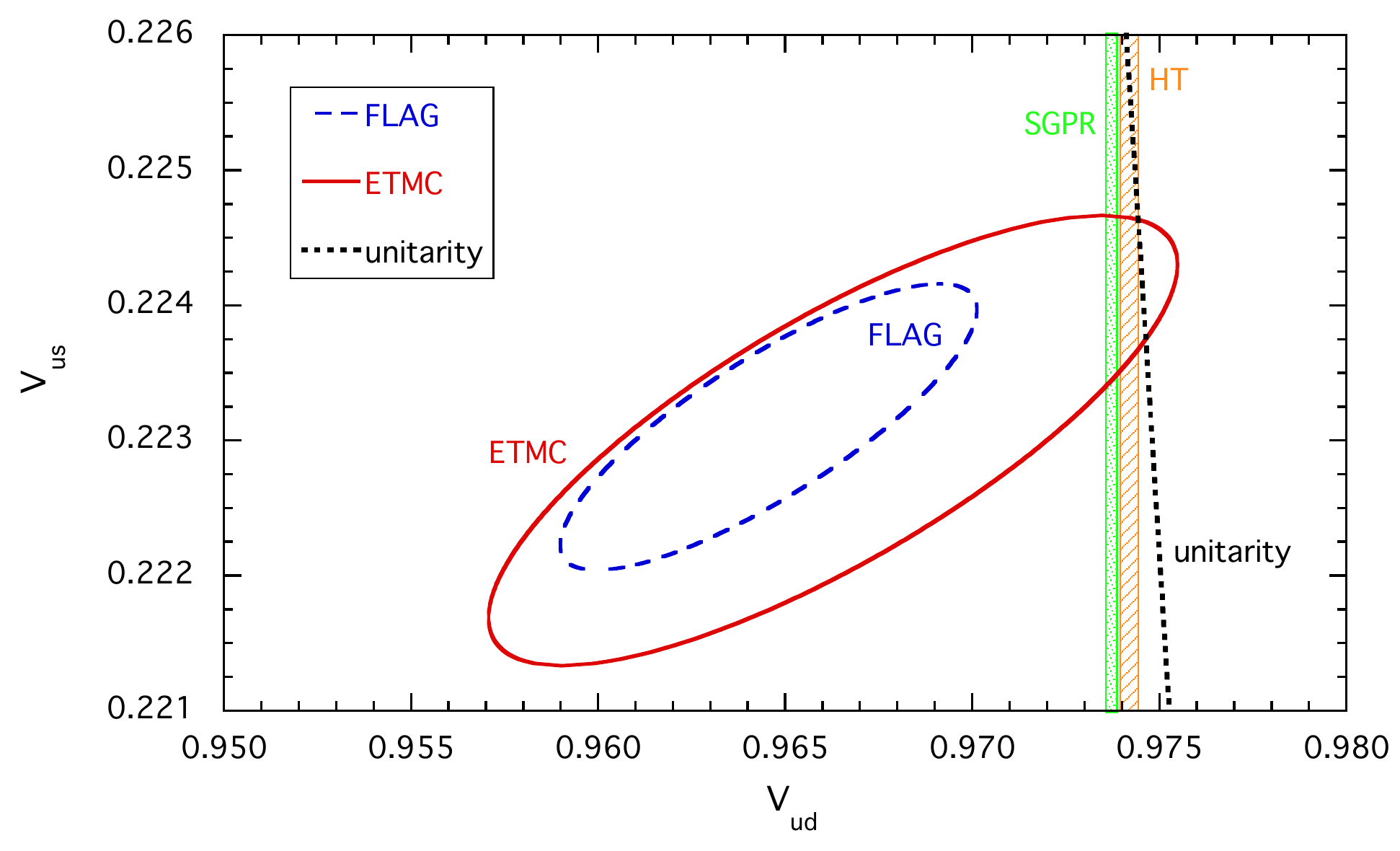}
\end{center}
\vspace{-0.75cm}
\caption{\it \small The plot compares the information for $V_{ud}$ and $V_{us}$ obtained in the FLAG-4 review for $N_f = 2 + 1 + 1$~\cite{Aoki:2019cca} and in this work by using Eq.~(\ref{eq:VusVud}) and the semileptonic result of Ref.~\cite{Carrasco:2016kpy}. The determinations of $V_{ud}$ obtained from superallowed nuclear $\beta$ transitions obtained in Refs.~\cite{Hardy:2016vhg,Seng:2018yzq} are also shown as green and orange bands, labelled respectively $HT$ and $SGPR$. The dotted line indicates the correlation between $V_{ud}$ and $V_{us}$ that follows assuming the unitarity of the CKM-matrix. The ellipses represent $68 \%$ likelihood contours.}
\label{fig:ellipses}
\end{figure}

Using the ETMC determination $f_+(0) = 0.9709 ~ (46)$ obtained with Wilson twisted-mass quarks in Ref.~\cite{Carrasco:2016kpy}, one gets the semileptonic result $V_{us} = 0.2230 ~ (11)$ to be compared with the leptonic one given in Eq.~(\ref{eq:Vus}).
The above finding is combined with Eq.~(\ref{eq:VusVud}) to obtain the red ellipse in Fig.~\ref{fig:ellipses}, which represents a $68 \%$ likelihood contour.
For comparison the blue ellipse corresponds to the FLAG-4 contour for $N_f = 2+1+1$~\cite{Aoki:2019cca}, defined by the bands corresponding to $V_{us} = 0.2231 ~ (7)$ and $V_{us} / V_{ud} = 0.2313 ~ (5)$.
The two determinations of $V_{ud}$ obtained in Refs.~\cite{Hardy:2016vhg} and~\cite{Seng:2018yzq} are also shown.
Finally, the dotted line represents the correlation between $V_{us}$ and $V_{ud}$ when the CKM matrix is taken to be unitary.

%%%%%%%%%%%%%%%%%%%%%%%%%%%%%%%%%%%%%%%%%%%%%%%%%%
\section{Conclusions}
\label{sec:conclusions}
%%%%%%%%%%%%%%%%%%%%%%%%%%%%%%%%%%%%%%%%%%%%%%%%%%

We have presented a determination of the ratio of kaon and pion leptonic decay constants in isoQCD, $f_K / f_\pi$, adopting the gauge ensembles produced by ETMC with $N_f = 2 + 1 + 1$ flavors of Wilson-clover twisted-mass quarks, including configurations close to the physical point for all dynamical flavors. 

The simulations are carried out at three values of the lattice spacing ranging from $\sim 0.068$ to $\sim 0.092$ fm with linear lattice size up to $L \sim 5.5$~fm. 
The scale is set using the value the pion decay constant $f_\pi^{isoQCD} = 130.4~(2)$ MeV taken from Ref.~\cite{Patrignani:2016xqp}.
Two observables, $f_\pi$ and $(f_\pi M_\pi^4)^{1/5}$, have been analyzed within the framework of SU(2) ChPT without making use of renormalized quark masses.
The latter quantity is found to be marginally affected by lattice artifacts and provides a precise determination of the GF scales, namely: $w_0 = 0.17383~(63)$ fm, $\sqrt{t_0} = 0.14436~(61)$ fm and $t_0 / w_0 = 0.11969~(62)$ fm.

As for the decay constant ratio $f_K / f_\pi$ we get at the physical isoQCD point, defined by Eqs.~(\ref{eq:MPi_isoQCD}-\ref{eq:fPi_isoQCD}), the result
\be
     \label{eq:fKPi_final}
     \left( \frac{f_K}{f_\pi} \right)^{isoQCD} = 1.1995 ~ (44) ~ , ~
\ee
where the error includes both statistical and systematic uncertainties in quadrature.
Our result (\ref{eq:fKPi_final}) agrees nicely with the recent $N_f = 2+1+1$ determinations, entering the FLAG-4 average~\cite{Aoki:2019cca} and corrected for strong IB effects, namely $(f_K / f_ \pi)^{isoQCD} = 1.1966 ~ (18)$~\cite{Dowdall:2013rya,Carrasco:2014poa,Bazavov:2017lyh}.

Taking the updated value $|V_{ud}| = 0.97370~(14)$ from super-allowed nuclear beta decays~\cite{Zyla:2020zbs,Seng:2018yzq}, Eqs.~(\ref{eq:ratioVf}) and (\ref{eq:fKPi_final}) yield the following value for the CKM element $|V_{us}|$:
\be  
    \label{eq:Vus_final}
    |V_{us}| = 0.22472 ~ (24)_{\mathrm{exp}} ~ (84)_{\mathrm{th}} = 0.22472 ~ (87) ~ , ~
\ee
which is nicely consistent with the latest estimate $|V_{us}| = 0.2252~(5)$ from leptonic modes provided by the PDG~\cite{Zyla:2020zbs}.
Correspondingly, using $|V_{ub}| = 0.00382~(24)$~\cite{Zyla:2020zbs} the first-row CKM unitarity  becomes
\be
    \label{eq:unitarity_final}
    |V_{ud}|^2 + |V_{us}|^2 + |V_{ub}|^2 = 0.99861 ~ (48) ~ ,
\ee
which would imply a $\simeq 3 \sigma$ tension with unitarity from leptonic modes.

\section*{Acknowledgments}

We thank all the ETMC members for a very productive collaboration. We warmly thank G.C.~Rossi for his continuous support and for a careful reading of the manuscript.
We are grateful to B.~Jo{\'o} for his kind support with our refactoring and extension of the QPhiX lattice QCD library.

We acknowledge PRACE (Partnership for Advanced Computing in Europe) for awarding us access to the high-performance computing system Marconi and Marconi100 at CINECA (Consorzio Interuniversitario per il Calcolo Automatico dell'Italia Nord-orientale) under the grants Pra17-4394, Pra20-5171 and Pra22-5171, and CINECA for providing us CPU time under the specific initiative INFN-LQCD123. We also acknowledge PRACE for awarding us access to HAWK, hosted by HLRS, Germany, under the grant with Acid 33037.
The authors gratefully acknowledge the Gauss Centre for Supercomputing e.V.~(www.gauss-centre.eu) for funding the project pr74yo by providing computing time on the GCS Supercomputer SuperMUC at Leibniz Supercomputing Centre (www.lrz.de).
Some of the ensembles for this study were generated on Jureca Booster~\cite{jureca} and Juwels~\cite{JUWELS} at the J{\"u}lich Supercomputing Centre (JSC) and we gratefully acknowledge the computing time granted there by the John von Neumann Institute for Computing (NIC).

The project has received funding from the Horizon 2020 research and innovation program of the European Commission under the Marie Sklodowska-Curie grant agreement No 642069 (HPC-LEAP) and under grant agreement No 765048 (STIMULATE).
The project was funded in part by the NSFC (National Natural Science Foundation of China) and the DFG (Deutsche Forschungsgemeinschaft, German Research Foundation) through the Sino-German Collaborative Research Center grant TRR110 ``Symmetries and the Emergence of Structure in QCD'' (NSFC Grant No. 12070131001, DFG Project-ID 196253076 - TRR 110).

R.F.~acknowledges the University of Rome Tor Vergata for the support granted to the project PLNUGAMMA.
F.S.~and S.S.~are supported by the Italian Ministry of Research (MIUR) under grant PRIN 20172LNEEZ.
F.S.~is supported by INFN under GRANT73/CALAT.
P.D.~acknowledges support form the European Unions Horizon 2020 research and innovation programme under the Marie Sk\l{}odowska-Curie grant agreement No. 813942 (EuroPLEx) and from INFN.~under the research project INFN-QCDLAT.
S.B.~and J.F.~are supported by the H2020 project PRACE6-IP (grant agreement No 82376) and the COMPLEMENTARY/0916/0015 project funded by the Cyprus Research Promotion Foundation.
The authors acknowledge support from project NextQCD, co-funded by the European Regional Development Fund and the Republic of Cyprus through the Research and Innovation Foundation (EXCELLENCE/0918/0129).

\appendix

\input{app_gauge_config_sim_params}

\input{app_ode_procedure}

\input{app_max_twist_corrections}

\input{app_determination_of_gf_scales}

\bibliography{refs_fKPi}

\end{document}

%% file: app_gauge_config_sim_params.tex
%%%%%%%%%%%%%%%%%%%%%%%%%%%%%%%%%%%%%%%%%%%%%%%%%%
\section{Algorithmic details and parameters for the ETMC gauge ensembles}
\label{sec:appA}
%%%%%%%%%%%%%%%%%%%%%%%%%%%%%%%%%%%%%%%%%%%%%%%%%%

In \Cref{subsec:integrator}, we present the algorithmic setup employed for the generation of our ensembles of gauge configurations, while the simulation parameters are given in \Cref{tab:simparams}.

\begin{table}[htb!]
      \begin{adjustbox}{width=\textwidth,center}
        \begin{tabular}{||c|c|c|c|c|c|c|c|c|c||}
        	\hline
        	  ensemble   & $\beta$ & $c_{\mathrm{SW}}$ &   $\kappa$   &     $V / a^4$     & $a \mu_\ell$ & $a \mu_\sigma$ & $a \mu_\delta$ & $\lambda_{\rm min}$ & $\lambda_{\rm max}$ \\ \hline
        	cA211.53.24  & $1.726$ &      $1.74$       & $0.1400645$  & $24^3 \times ~48$ & $~0.00530~$  &   $~0.1408$    &   $~0.1521$    &  $0.0000376$   &     $4.7$      \\
        	cA211.40.24  &         &                   &              & $24^3 \times ~48$ & $~0.00400~$  &                &                &                &                \\
        	cA211.30.32  &         &                   &              & $32^3 \times ~64$ & $~0.00300~$  &                &                &                &                \\
        	cA211.12.48  &         &                   & $0.1400650$  & $48^3 \times ~96$ & $~0.00120~$  &                &                &                &                \\ \hline
        	cB211.25.24  & $1.778$ &      $1.69$       & $0.1394267$  & $24^3 \times ~48$ & $~0.00250~$  &  $~0.1246864$  &  $~0.131052$  &  $0.0000344$   &     $4.3$      \\
        	cB211.25.32  &         &                   &              & $32^3 \times ~64$ & $~0.00250~$  &                &                &                &                \\
        	cB211.25.48  &         &                   &              & $48^3 \times ~96$ & $~0.00250~$  &                &                &                &                \\
        	cB211.14.64  &         &                   &              & $64^3 \times 128$ & $~0.00140~$  &                &                &                &                \\
        	cB211.072.64 &         &                   & $0.1394265$  & $64^3 \times 128$ & $~0.00072~$  &                &                &   $0.00005$    &     $4.7$      \\ \hline
        	cC211.06.80  & $1.836$ &     $~1.6452$     & $0.13875285$ & $80^3 \times 160$ & $~0.00060~$  &  $~0.106586$   &  $~0.107146$   &  $0.0000376$   &     $4.7$      \\ \hline
        \end{tabular}
    \end{adjustbox}
    \caption{Simulation parameters for the ensembles used for this study. Please refer to \Cref{subsec:integrator} for details on the integrator setup.}
    \label{tab:simparams}
\end{table}

\subsection{Integrator Setups}
\label{subsec:integrator}

In the generation of gauge ensembles via the Hybrid Monte Carlo algorithm, the effective lattice action can be represented by a sum over \emph{monomials} corresponding to different contributions to the partition function as defined below.
In the integration of the equations of motion, the forces contributed by the different monomials differ by orders of magnitude, allowing them to be integrated on different time scales accordingly, as detailed in \Cref{tab:simsetup} below.

\subsubsection{Monomial Types}

We define below the different types of monomials that we employ in our effective lattice action to simulate QCD using $N_f=2+1+1$ twisted mass clover fermions.

\begin{description}
    \item[Gauge] [gau($\beta,c_1$)]\\
       \begin{equation}
         \frac{\beta}{3}\sum_x\left(  c_0\sum_{\substack{
               \mu,\nu=1\\1\leq\mu<\nu}}^4\{1-\re\Tr(U^{1\times1}_{x,\mu,\nu})\}\Bigr. 
           \Bigl.\ +\ 
           c_1\sum_{\substack{\mu,\nu=1\\\mu\neq\nu}}^4\{1
           -\re\Tr(U^{1\times2}_{x,\mu,\nu})\}\right)\, ,
       \end{equation}
       with $c_0=(1-8c_1)$, for the Iwasaki action used here~\cite{Iwasaki:1985we}, $c_1=-0.331$.

    \item[Degenerate Determinant] [det($\rho$)]\\
        The action contribution of a degenerate doublet of clover-improved twisted mass quarks is given by
        \begin{equation}
            \label{eq:eosw0}
            \begin{split}
                S[\chi,\bar\chi,U] = \sum_x & \Biggl\{ \bar\chi(x)[1+2\kappa
                c_{SW}T + 2i \kappa\mu\gamma_5\tau^3]\chi(x)  \Bigr. \\
                & -\kappa\bar\chi(x)\sum_{\mu = 1}^4\Bigl[ U_\mu(x)(r-\gamma_\mu)\chi(x+a\hat\mu)\bigr. \\
                & + \Bigl. \bigl. U_\mu^\dagger(x-a\hat\mu)(r+\gamma_\mu)\chi(x-a\hat\mu)\Bigr]
                \Biggr\} \\
                \equiv &\sum_{x,y}\bar\chi(x) M_{xy}\chi(y)\, ,
            \end{split}
        \end{equation}
        in the twisted basis and in the hopping parameter normalisation, where $T$ is the clover term.
        In our simulations we use the conventional value $r=1$.
    
        For convenience, we define $\tilde\mu\equiv2\kappa\mu$ and absorb $2\kappa c_{SW}$ into $T$, defining the two-flavour operator
        \begin{equation}
          \label{eq:eosw1}
          Q\equiv \gamma_5 M = \begin{pmatrix}
            \Qp & \\\
            & \Qm \\
          \end{pmatrix} \;,
        \end{equation}
        and the Hermitian operator $\Qsw = \gamma_5 \Dsw$, where in turn $\Dsw$ is the clover-improved Wilson Dirac operator.
        We then have $\Qpm = \Qsw \pm i\mutilde$, such that $Q_{+}^\dagger = Q_{-}$ and $Q_{+} Q_{-} = \Qsw^2 + \mutilde^2$.
        The contribution to the partition function of the mass-degenerate (light) twisted mass quark doublet is thus given by $\det \left( Q_{+} Q_{-} \right) = \det \left( \Qsw^2 + \mutilde^2 \right)$.    
        
        An even-odd Schur decomposition of the sub-matrices $\Qpm$ then gives
        \begin{equation}
          \label{eq:eosw2}
          \begin{split}
            Q^\pm &= \gamma_5\begin{pmatrix}
              1 + T_{ee} \pm i\tilde\mu\gamma_5 & M_{eo} \\
              M_{oe}    & 1 + T_{oo} \pm i\tilde\mu\gamma_5 \\
            \end{pmatrix} =
            \gamma_5\begin{pmatrix}
              M_{ee}^\pm & M_{eo} \\
              M_{oe}    & M_{oo}^\pm \\
            \end{pmatrix} \\
            & =
            \begin{pmatrix}
              \gamma_5M_{ee}^\pm & 0 \\
              \gamma_5M_{oe}  & 1 \\
            \end{pmatrix}
            \begin{pmatrix}
              1       & (M_{ee}^\pm)^{-1}M_{eo}\\
              0       & \gamma_5(M_{oo}^\pm-M_{oe}(M_{ee}^\pm)^{-1}M_{eo})\\
            \end{pmatrix}\, ,
        \end{split}
        \end{equation}
        from which we obtain $\Qhat_{\pm}$ defined only on the odd sites of the lattice
        \begin{equation}
            \Qhat_{\pm} = \gamma_5 \left( M_{oo}^{\pm} - M_{oe} \left( M_{ee}^{\pm} \right)^{-1} M_{eo} \right) \;.
        \end{equation}
        The light quark determinant can then be reexpressed as
        \begin{equation}
            \det \left( Q_{+} Q_{-} \right) = \det \left( M_{ee}^{+} M_{ee}^{-} \right) \cdot \det \left( \Qhat_{+} \Qhat_{-} \right)\;.
        \end{equation}
        
        In order to implement mass preconditioning, the $\Qhat_\pm$ can be shifted by a constant through the addition of a further twisted mass: $\What_{\pm}(\rho) = \Qhat_{\pm} \pm i\rho$, such that $\What_{+} \What_{-} = \Qhat_{+} \Qhat_{-} + \rho^2$.
        It should be noted that this shift is applied to the even-odd-preconditioned operator, such that the factor $M_{ee}^{\pm}$ remains independent of $\rho$ since its inverse is non-trivial.
        
        In terms of pseudofermion fields, one thus obtains a contribution to the partition function
        \begin{equation}
            \int \mathcal{D}\phi_1^\dagger \; \mathcal{D}\phi_1 \; \exp{\left\lbrace - \phi_1^\dagger \; (\What_{+} \What_{-})^{-1} \; \phi_1 \right\rbrace} \;,
            \label{eq:det_massprec}
        \end{equation}
        which we refer to as the degenerate determinant and a corresponding contribution
        \begin{equation}
            \int \mathcal{D} \phi_2^\dagger \; \mathcal{D}\phi_2 \; \exp{\left\lbrace - \phi_2^\dagger \; \What_{-} \frac{1}{\Qhat_{+} \Qhat_{-}} \What_{+} \; \phi_2 \right\rbrace}\,,
            \label{eq:detratio_compens}
        \end{equation}
        which we refer to as a determinant ratio.
    
    \item[Determinant Ratio] [detrat($\rho_b, \rho_t$)]\\
        \Cref{eq:detratio_compens} generalises to the introduction of multiple shifts $\rho_1, \rho_2, \ldots, \rho_n$ to contributions of the form:
        \begin{equation}
            \int \mathcal{D} \phi_i^\dagger \; \mathcal{D}\phi_i \; \exp{\left\lbrace - \phi_i^\dagger \; \What_{-}(\rho_t) \frac{1}{\What_{+}(\rho_b) \What_{-}(\rho_b)} \What_{+}(\rho_t) \; \phi_i \right\rbrace}\,.
            \label{eq:detratio_general}
        \end{equation}
        
        The pseudofermion fields $\phi_i$ are defined only on the odd sites of the lattice and are generated from a random spinor field $R_i$, sampled from a normalized Gaussian distribution at the beginning of each molecular dynamics trajectory.
        In the case of the determinant, we have $\phi_i = \Qhat_{+} R_i$, while in the case of the determinant ratio we have $\phi_j = \left( \What_{+}(\rho_t) \right)^{-1} \What_{+}(\rho_b) R_j$.
        
        The complete mass-preconditioned contribution with $n$ shifts is thus given by
        \begin{equation}
            \det(\rho_n) \cdot \textrm{detrat}(\rho_{n-1},\rho_n) \cdot \textrm{detrat}(\rho_{n-2},\rho_{n-1}) \cdot \ldots \cdot \textrm{detrat}(0,\rho_1) \;,
        \end{equation}
        where the last factor has the form of \Cref{eq:detratio_compens} with the target twisted quark mass in $\Qhat_{\pm}$.
        In general, the different contributions are integrated on multiple time scales because their contributions to the force differ by orders of magnitude.
        
    \item[Rational Approximation Partial Fraction] [rat($n_\ell,n_k$)]\\
        The Dirac operator for the non-degenerate flavour doublet employed in the strange-charm sector reads
        \begin{equation}
            D_h(\mubar,\epsbar) = \Dsw \cdot 1_f + i\mubar\gamma_5\tau^3_f - \epsbar\tau^1_f \;,
        \end{equation}
        with the property
        \begin{equation}
            D_h^\dagger = \tau^1_f \gamma_5 D_h \gamma_5 \tau^1_f\;.
        \end{equation}
        Equivalently, as used (without the clover term) in Ref.~\cite{Chiarappa:2006ae}, one may write
        \begin{equation}
            D^{'}_h(\mu_\sigma, \mu_\delta) = \Dsw \cdot 1_f + i\mu_\sigma \gamma_5 \tau^1_f + \mu_\delta \tau^3_f\,,
        \end{equation}
        which is related to $D_h$ by $D_h^{'} = (1+i\tau^2_f)D_h(1-i\tau^2_f)/2$ and $(\mu_\sigma, \mu_\delta) \rightarrow (\mubar,-\epsbar)$.
        
        As before, we define $Q_h = \gamma_5 D_h$ and the implementation of even-odd preconditioning translates straightforwardly from the mass-degenerate case, although the construction of $M^h_{ee}$ has to take into account the additional (off-diagonal) flavour structure.
        
        The operator $\Qhat_h$, defined only on the odd sites, has the property $\Qhat_h^\dagger = \tau^1_f \Qhat_h \tau^1_f$ and the non-degenerate quark doublet contributes a factor
        \begin{equation}
            \det \left( Q_h \right) \propto \det \left( \Qhat_h \right)
        \end{equation}
        to the partition function, which we simulate via
        \begin{equation}
            \left[ \det \left(\Qhat_h^2 \right) \right]^{1/2} \approx \det \left( \mathcal{R}^{-1} \right)\;,
        \end{equation}
        where we made use of the shorthand notation $\Qhat_h^2 = \Qhat_h \tau^1_f \Qhat_h \tau^1_f$.
        
        We use a rational approximation of order $N$ (see Refs.~\cite{Kennedy:1998cu,Clark:2006fx,Luscher:2010ae})
        \begin{equation}
            \mathcal{R}\left( \Qhat_h^2 \right) = A\prod^{N}_{i=1} \frac{\Qhat_h^2 + a_{2i}}{\Qhat^2 + a_{2i-1}} \approx \frac{1}{\sqrt{\Qhat^2_h}} \;.
        \end{equation}
        For this, we employ the Zolotarev solution~\cite{zolotarev1877application} for the optimal approximation to $1/\sqrt{y}$, where the coefficients $a_i$ satisfy the property
        \begin{equation}
            a_1 > a_2 > \ldots > a_{2N} > 0\;.
        \end{equation}
        The amplitude $A$, the coefficients $a_i$ and the maximal deviation of the rational approximation $\delta = \max_{y} |1-\sqrt{y}\mathcal{R}(y)|$ are computed analytically at given order $N$ and lower bound $\epsilon <y<1$. These are
        \begin{align}
        a_{i} &= \text{cs}^2\left(i \cdot v,\sqrt{1-\epsilon}\right) &\text{with }\ v = \frac{K\left(\sqrt{1-\epsilon}\right)}{2N+1}\label{eq:zolotarev_shifts}\\
        A &= \frac{2}{1+\sqrt{1-d^2}}\prod_{j=1}^{N}\frac{s_{2j-1}}{s_{2j}}&\text{with }\ s_{i} = \text{sn}^2\left(i \cdot v,\sqrt{1-\epsilon}\right)\\
        \delta &= \frac{d^2}{1+\sqrt{1-d^2}}&\text{with }\ d = (1-\epsilon)^{\frac{2N+1}{2}}\prod_{j=1}^{N}s_{(2j-1)}^2
        \end{align}
        where $\text{sn}(u,k)$ and $\text{cs}(u,k)=\text{cn}(u,k) / \text{sn}(u,k)$ are Jacobi elliptic functions and $K(k)$ is the complete elliptic integral. In all our simulations we use $N=10$ and $\epsilon=\lambda_{\min} / \lambda_{\max}$ where $\lambda_{\min}$ and $\lambda_{\max}$ are respectively the lower and upper bound of the eigenvalues of $\Qhat_h^2$. In order to have all the eigenvalues $\lambda$ in the range $\epsilon <\lambda<1$, we re-scale $\Qhat_h^2$ with $\lambda_{\max}$.
        The values of $\lambda_{\min}$ and $\lambda_{\max}$ per ensemble are given in \Cref{tab:simparams}.
        In the simulation, we explicitly check that the eigenvalues of $\Qhat_h^2$ remain within these bounds.
        
        The factors in the approximation can be grouped
        \begin{equation}
            \mathcal{R}\left( \Qhat^2_h \right) = A r_1^\ell \left( \Qhat^2_h \right) \cdot r_\ell^k \left( \Qhat^2_h \right) \cdot \ldots \;,
        \end{equation}
        where
        \begin{equation}
            r_{n_\ell}^{n_k}\left( \Qhat^2_h \right) = \prod_{i=n_\ell}^{n_k} \frac{ \Qhat^2_h + a_{2i-1} }{ \Qhat^2_h + a_{2i} } = \textrm{rat} (n_\ell,n_k)\;.
        \end{equation}
        
        We perform partial fraction expansions of the terms 
        \begin{equation}
            r_{n_\ell}^{n_k} \left( \Qhat_h^2 \right) = 1 + \sum_{i=n_\ell}^{n_k} \frac{q_i}{\Qhat^2_h + \mu_i^2}\;,
        \end{equation}
        such that the necessary matrix inverse can be calculated efficiently using a multi-shift solver.
        The coefficients $q_i$ are given by
        \begin{equation}
            q_i = \left( a_{2i-1} - a_{2i} \right) \prod_{m=n_\ell, m \neq i}^{n_k} \frac{a_{2m-1} - a_{2i}}{a_{2m}-a_{2i}} \;, i = n_\ell, \ldots, n_k \;.
        \end{equation}
        We can further define $\mu_i = \sqrt{a_{2i}}$ and $\nu_i = \sqrt{a_{2i-1}}$ and express $q_i$ as
        \begin{equation}
            q_i = \left( \nu_i^2 - \mu_i^2 \right) \prod_{m=n_\ell,m \neq i}^{n_k} \frac{\nu_m^2 - \mu_i^2}{\mu_m^2 - \mu_i^2}, \;i = n_\ell,\ldots,n_k\;.
        \end{equation}
        At the beginning of each trajectory, pseudofermion fields are generated as follows: again a random spinor field $R$ is sampled from a Gaussian distribution. Now, we need to compute pseudofermion fields $\phi$ from
        \[                                              
        R^\dagger R = \phi^\dagger \mathcal{R} \phi     
        \]
        and, therefore, we need operators $C^\dagger$ and $C$ with the property
        \[                                                      \mathcal{R}^{-1} = C^\dagger\cdot C\,,\qquad\Rightarrow\ \phi =  C\cdot R\,.
        \]
        $C$ is given by (inspired by twisted mass)              
        \[
        C\ =\ \prod_{i=1}^N \frac{\hat Q_h + i\mu_i}{\hat Q_h + i\nu_i}
        \]                                                      which can again be written as a partial fraction
        \[
        C\ =\ 1 + i\sum_{i=1}^N \frac{r_i}{\hat Q_h + i\nu_i}\,,
        \]                                                      with
        \[                                                      r_i = (\mu_i - \nu_i)\prod_{m=1, m\neq i}^N \frac{\mu_m -\nu_i}{\nu_m - \nu_i}\,,\quad i = 1,...,N\,.
        \]
        The rational approximation $\mathcal{R}$ can be applied to a vector using a multi-mass solver and the partial fraction representation. The same works for $C$: after solving $N$ equations simultaneously for $(\hat Q_h^2 + \nu_i^2)^{-1},\quad i = 1,...,N$, we have to multiply every term with $(\hat Q_h - i\nu_i)$. The hermitian conjugate of $C$ is given by
        \[
        C^\dagger\ =\ 1 - i\sum_{i=1}^N \frac{r_i}{\hat Q_h - i\nu_i}\,,
        \]
        using $\hat Q_h^\dagger = \hat Q_h$. For the acceptance step only the application of $\mathcal{R}$ is needed.
        
    \item[Rational Approximation Correction Factor] [ratcor($n$)]\\
        The rational approximation $\mathcal{R}$ only has a finite precision. This finite precision can be accounted for during the acceptance step in the HMC by estimating~\cite{Luscher:2010ae} $1-|\hat{Q}_h|\mathcal{R}$, if the rational approximation is precise enough. This can be achieved by including a monomial $\det(|\hat{Q}_h|\mathcal{R})$ in the simulation, for which one needs an operator $B$
        \[
        B\cdot B^\dagger = |\hat{Q}_h|\mathcal{R}\,.
        \]
        Following Ref.~\cite{Luscher:2010ae}, $B$ can be written as
        \[
        B = (1 + Z)^{1/4} = \sum_{i=0}^m c_i Z^i = 1 + \frac{1}{4} Z - \frac{3}{32} Z^2 + \frac{7}{128} Z^3 + \ldots
        \]
        with $Z = \hat Q_h^2\mathcal{R}^2 -1$. The series converges rapidly and can, thus, be truncated after a few terms, $m+1$. The convergence can actually be controlled during the simulation and the truncation does not need to be fixed.
        We choose to sum the series until the contribution of the given term to the acceptance Hamiltonian is below the residual precision squared, $r^2$, that we employ for the solution of the linear systems involved in the approximation of $\mathcal{R}$ \emph{in the acceptance step}, such that $\lvert c_m Z_m \phi \rvert^2 < r^2$, which we typically choose to be at the limit of double precision arithmetic.
        
        For this monomial the pseudofermion field is computed from
        \[
        \phi = B\cdot R\,,
        \]
        where $R$ is again a Gaussian random vector, see above.
\end{description}

\subsubsection{Simulation parameters}

In \Cref{tab:simsetup} we list monomials and parameters used per ensemble. The monomials are grouped in various timescales where the one with the highest id is the outermost timescale (with the fewest integration steps) into which the other time scales are nested.
For the various timescales two integrator types are used, either the second order minimal norm (2MN) integrator or its extension with a force gradient (2MNFG), making the latter a fourth-order integrator~\cite{OMELYAN2003272}.
The number of steps per timescale is indicated with $N_s$.

The time evolution operator $\operatorname{exp}\left[ (\delta\tau) H_{\mathrm{MD}} \right]$ for a given MD-Hamiltonian can be decomposed into ``kinetic" and ``potential" parts, $\operatorname{exp}\left[ (\delta\tau) ( T + V ) \right]$.
To a given order $n$ in the time step $\delta\tau$, this can be factorised
\begin{equation}
    \operatorname{exp}\left[ (\delta\tau) ( T + V ) \right] = \prod_{i=1}^{n} \operatorname{exp}[c_i (\delta\tau) T] \operatorname{exp}[d_i (\delta\tau) V] + \mathcal{O}\left[ (\delta\tau)^{n+1} \right] \,.
    \label{eq:symplectic}
\end{equation}
By expanding the left-hand side of \Cref{eq:symplectic} (being mindful of the non-commutativity of $T$ and $V$) and matching the coefficients $c_i$ and $d_i$ of terms of equal order in $\delta\tau$, explicit factorizations can be constructed.
In practice, the expansion of the left-hand side is only formal and one attempts instead to formulate order equations in the coefficients $c_i$ and $d_i$ to eliminate terms which are expensive to compute (stemming from commutators of $T$ and $V$) to satisfy the equality to some approximation.
For 2MN and 2MNFG, these equations can be reformulated in terms of a coefficient $\lambda$.
%It can be shown that $\lambda \approx 0.1931833275$ is optimal for the former
%while $\lambda = 1/6$ is the only possible value for the force-gradient version %of the integrator.

The 2MNFG scheme is now given by setting $\lambda = 1/6$, which cancels out one of the second order commutators $[T,[V,T]]$. Now, the remaining term $[V,[V,T]]$ can  be canceled using the force gradient term.
It turns out that for the 2MN integrator an optimal value for $\lambda$
is larger than $1/6$. 
Namely assuming unity of the second order commutators and neglecting any correlations leads to the optimal value of $\lambda \approx 0.1931833275$.
In the usage of the 2MN integrator with multiple time scales, experience suggests that further deviations from this optimal value improve the acceptance rate, such that we often use schemes with increasing values of $\lambda$ from the innermost to the outermost time scale, as shown in \Cref{tab:simsetup}.

\begin{table}
\begin{minipage}[t]{0.4\textwidth}
\begin{adjustbox}{width=\textwidth,center,valign=t}
	\begin{tabular}{||c|c|c|c|c||}
		\hline
		Id & Type  & $N_s$ & $\lambda$ &                        Monomials                        \\ \hline\hline
		\multicolumn{5}{||c||}{cA211.53.24, 5 timescales, $\tau=1.0$}                                        \\ \hline
		0  &  2MN  &   1   &   0.193   &                    gau($\beta,c_1$)                     \\
		1  &  2MN  &   1   &   0.195   &                    det($0.1$)                    \\ %det($0.10420836$)
		2  &  2MN  &   1   &   0.197   &       detrat($0.02,0.1$), rat($0,5$)       \\ %detrat($0.02016936,0.10420836$)
		3  &  2MN  &   1   &   0.200   &       detrat($0.003,0.02$), rat($6,7$)       \\ %detrat($0.00336156,0.02016936$) 
		4  &  2MN  &   9   &   0.205   &           detrat($0,0.003$), rat($8,9$)            \\ %detrat($0,0.00336156$)
		\hline\hline
		\multicolumn{5}{||c||}{cA211.40.24, 5 timescales, $\tau=1.0$}                                        \\ \hline
		0  &  2MN  &   1   &   0.193   &                    gau($\beta,c_1$)                     \\
		1  &  2MN  &   1   &   0.195   &                    det($0.1$)                    \\ %det($0.10420836$)
		2  &  2MN  &   1   &   0.197   &       detrat($0.02,0.1$), rat($0,5$)       \\ %detrat($0.02016936,0.10420836$)
		3  &  2MN  &   1   &   0.200   &       detrat($0.003,0.02$), rat($6,7$)       \\ %detrat($0.00336156,0.02016936$)
		4  &  2MN  &   9   &   0.205   &           detrat($0,0.003$), rat($8,9$)            \\ %detrat($0,0.00336156$)
		\hline\hline
		\multicolumn{5}{||c||}{cA211.30.32, 5 timescales, $\tau=1.0$}                                        \\ \hline
		0  &  2MN  &   1   &   0.193   &                    gau($\beta,c_1$)                     \\
		1  &  2MN  &   1   &   0.195   &                    det($0.1$)                    \\ %det($0.10420836$) 
		2  &  2MN  &   1   &   0.197   &       detrat($0.02,0.1$), rat($0,5$)       \\ %detrat($0.02016936,0.10420836$)
		3  &  2MN  &   1   &   0.200   &       detrat($0.003,0.02$), rat($6,7$)       \\ %detrat($0.00336156,0.02016936$)
		4  &  2MN  &  12   &   0.205   &           detrat($0,0.003$), rat($8,9$)            \\ %detrat($0,0.00336156$)
		\hline\hline
		\multicolumn{5}{||c||}{cA211.12.48, 6 timescales, $\tau=1.0$}                                        \\ \hline
		0  &  2MN  &   1   &   0.185   &                    gau($\beta,c_1$)                     \\
		1  &  2MN  &   1   &   0.190   &                   det($0.16$)                   \\ %det($0.1573086528$)
		2  &  2MN  &   1   &   0.195   &     detrat($0.03,0.16$), rat($0,2$)     \\ %detrat($0.0312600528,0.1573086528$)
		3  &  2MN  &   1   &   0.200   &     detrat($0.006,0.03$), rat($3,4$)     \\ %detrat($0.0060503328,0.0312600528$)
		4  &  2MN  &   1   &   0.205   &     detrat($0.001,0.006$), rat($5,6$)     \\ %detrat($0.0010083888,0.0060503328$)
		5  &  2MN  &  17   &   0.210   &          detrat($0,0.001$), rat($7,9$)           \\ %detrat($0,0.0010083888$)
		\hline
	\end{tabular}
	\end{adjustbox}
\end{minipage}
\begin{minipage}[t]{0.59\textwidth}
	\begin{adjustbox}{width=\textwidth,center,valign=t}
    \begin{tabular}{||c|c|c|c|c||}
		\hline
		Id & Type  & $N_s$ & $\lambda$ &                          Monomials                           \\ \hline\hline
		\multicolumn{5}{||c||}{cB211.25.24/32, 4 timescales, $\tau=1.5$}                                             \\ \hline
		0  & 2MNFG &   1   &   0.167   &                       gau($\beta,c_1$)                       \\
		1  & 2MNFG &   1   &   0.167   &                    det($0.3$), rat($0,3$)                    \\
		2  & 2MNFG &   1   &   0.167   & detrat($0.045,0.3$), detrat($0.0045,0.045$), rat($4,5$)  \\ %
		3  &  2MN  &  13   &   0.193   &               detrat($0,0.045$), rat($6,9$)                \\ \hline\hline
		\multicolumn{5}{||c||}{cB211.25.48, 5 timescales, $\tau=1.0$}                                             \\ \hline
		0  &  2MN  &   1   &   0.193   &                       gau($\beta,c_1$)                       \\
		1  &  2MN  &   1   &   0.195   &                      det($0.24$)                      \\ %det($0.238419657$)
		2  &  2MN  &   1   &   0.197   &       detrat($0.033,0.24$), rat($0,5$)        \\ %detrat($0.033462408,0.238419657$)
		3  &  2MN  &   1   &   0.200   &       detrat($0.004,0.033$), rat($6,7$)        \\ %detrat($0.004182801,0.033462408$)
		4  &  2MN  &  15   &   0.205   &            detrat($0,0.004$), rat($8,9$)             \\ %detrat($0,0.004182801$)
		\hline\hline
		\multicolumn{5}{||c||}{cB211.14.64, 4 timescales, $\tau=1.5$}                                             \\ \hline
		0  & 2MNFG &   1   &   0.167   &                       gau($\beta,c_1$)                       \\
		1  & 2MNFG &   1   &   0.167   &                   det($0.2$), rat($0,3$)                    \\ %det($0.19$)
		2  & 2MNFG &   1   &   0.167   & detrat($0.02,0.2$), detrat($0.002,0.02$), rat($4,5$)  \\ %detrat($0.019,0.19$), detrat($0.0019,0.019$)
		3  &  2MN  &  23   &   0.193   &               detrat($0,0.002$), rat($6,9$)               \\ %detrat($0,0.0019$)
		\hline\hline
		\multicolumn{5}{||c||}{cB211.072.64, 6 timescales, $\tau=1.0$}                                            \\ \hline
		0  &  2MN  &   1   &   0.185   &                       gau($\beta,c_1$)                       \\
		1  &  2MN  &   1   &   0.190   &                          det($0.1$)                          \\
		2  &  2MN  &   1   &   0.195   &               detrat($0.01,0.1$), rat($0,2$)               \\
		3  &  2MN  &   1   &   0.200   &             detrat($0.0012,0.01$), rat($3,5$)              \\
		4  &  2MN  &   1   &   0.205   &            detrat($0.0003,0.0012$), rat($6,7$)             \\
		5  &  2MN  &  12   &   0.205   &               detrat($0,0.0003$), rat($8,9$)               \\ \hline\hline
		\multicolumn{5}{||c||}{cC211.06.80, 4 timescales, $\tau=1.0$}                                             \\ \hline
		0  & 2MNFG &   1   &   0.167   &                       gau($\beta,c_1$)                       \\
		1  & 2MNFG &   1   &   0.167   &                   det($0.12$), rat($0,3$)                    \\
		2  & 2MNFG &   1   &   0.167   & detrat($0.012,0.12$), detrat($0.0012,0.012$), rat($4,6$) \\
		3  &  2MN  &  14   &   0.193   &               detrat($0,0.0012$), rat($7,9$)               \\ \hline
	\end{tabular}
    \end{adjustbox}
\end{minipage}

\caption{Integrators setup used for the ensembles analyzed in this study. The number of time scales and trajectory length, $\tau$, used for each ensemble are indicated in the respective headers.}
\label{tab:simsetup}
\end{table}

\subsection{Software details}

The simulations presented in this study have been generated using the tmLQCD~\cite{Jansen:2009xp,Deuzeman:2013xaa,Abdel-Rehim:2013wba} software suite, which provides all the necessary components to perform $N_f=2+1+1$ simulations of twisted mass clover fermions, including implementations of the polynomial and rational HMC algorithms for the non-degenerate determinant.
To enable multigrid solvers to be used in simulations~\cite{Bacchio:2017pcp}, tmLQCD provides an interface to DD$\alpha$AMG~\cite{Bacchio:2016bwn}, a multigrid solver library optimized for twisted mass (clover) fermions~\cite{Alexandrou:2016izb}.
The force calculation of some monomials in the light quark sector is accelerated by a 3-level multi-grid approach.
Moreover, we extended the DD$\alpha$AMG method for the mass non-degenerate twisted mass operator.
The multi-grid solver used in the rational approximation~\cite{Alexandrou:2018wiv} is particularly helpful for the lowest terms of the rational approximation, as well as for the rational approximation corrections in the acceptance step, where it yields a speed up of two over the standard multi-mass shifted conjugate gradient (MMS-CG) solver on traditional distributed-memory machines based on  Intel Skylake or AMD EPYC architectures.

On the other hand, especially on machines based on Intel's Knight's Landing (KNL) architecture, only the most poorly-conditioned monomials benefit from the usage of DD$\alpha$AMG, to the point where (on KNL) the inversion of the non-degenerate operator does not benefit at all.
To improve efficiency, tmLQCD also provides an interface to the QPhiX~\cite{QPhiX} lattice QCD library, which we have refactored and  extended~\cite{QPhiX-github} to support twisted mass clover fermions, including the non-degenerate doublet.
For solves related to the degenerate determinant and determinant ratios, this allows us to efficiently and flexibly combine mixed-precision CG and SIMD vector lengths of 8 or 16 as required by AVX512.
On KNL, single-precision QPhiX kernels are up to a factor of 5 more efficient than their tmLQCD-native equivalents.
Also in the MMS-CG solves in the non-degenerate sector, the double-precision kernels in QPhiX are up to a factor of 2 more efficient than the tmLQCD-native equivalents on KNL.
Combined, these efficiency improvements lead to overall speedup factors of 2-3 in the HMC on this architecture with smaller overall gains on Skylake and EPYC.

%% file: app_ode_procedure.tex
%%%%%%%%%%%%%%%%%%%%%%%%%%%%%%%%%%%%%%%%%%%%%%%%%%
\section{Extraction of $a M_\pi$ and $a f_\pi$ using the ODE procedure}
\label{sec:appB}
%%%%%%%%%%%%%%%%%%%%%%%%%%%%%%%%%%%%%%%%%%%%%%%%%%

The spectral decomposition of the pion correlation function (\ref{eq:P5P5}) can be investigated adopting the ODE procedure of Ref.~\cite{Romiti:2019qim}. 
This method is able to extract exponential signals from the temporal dependence of a lattice correlator without any {\it a priori} knowledge of the multiplicity of each signal and it does not require any starting input for the masses and the amplitudes of the signals.

The ODE approach is sensitive to the noise of the correlator, so that pure oscillating signals (conjugate pairs of imaginary masses) may typically appear in the ODE spectral decomposition.
Therefore, we adopt an improved version of the ODE procedure, in which a subsequent $\chi^2$-minimization procedure is applied to the non-noisy part of the ODE spectral decomposition~\cite{Romiti:2019qim}. 
In this way the accuracy of the {\it physical} (i.e.~non-noisy) part of the ODE spectral decomposition is improved.

The time intervals $[t_{min}, t_{max}]$ adopted for the analysis and the extracted values of the pion mass and decay constant in lattice units are collected in Table~\ref{tab:ODE}.
\begin{table}[!hbt]
\renewcommand{\arraystretch}{1.2}	 
\begin{center}	
{\small
\begin{tabular}{||c|c||c|c||c|c||}
\hline
ensemble& $\beta$ & $V / a^4$ & $[t_{\rm min}/a, \, t_{\rm max}/a]$ & $a M_\pi$ & $a f_\pi$\\
\hline
cA211.53.24& $1.726$ & $24^3 \times ~48$ & $[5, \, 24]$   & $0.16621~(40)$ & $0.07106~(36)$\\
cA211.40.24&               & $24^3 \times ~48$ & $[5, \, 24]$   & $0.14473~(76)$ & $0.06809~(30)$ \\
cA211.30.32&               & $32^3 \times ~64$ & $[6, \, 32]$   & $0.12523~(18)$ & $0.06675~(15)$ \\
cA211.12.48&               & $48^3 \times ~96$ & $[6, \, 48]$   & $0.08000~(28)$ & $0.06139~(34)$ \\
\hline
cB211.25.24& $1.778$ & $24^3 \times ~48$ & $[6, \, 24]$   & $0.10750~(189)$ & $0.05351~(48)$ \\
cB211.25.32&               & $32^3 \times ~64$ & $[6, \, 32]$   & $0.10454~(43)$ & $0.05656~(37)$ \\
cB211.25.48&               & $48^3 \times ~96$ & $[6, \, 48]$   & $0.10454~(13)$ & $0.05727~(11)$ \\
cB211.14.64&               & $64^3 \times 128$ & $[7, \, 64]$ & $0.07845~~(8)$ & $0.05476~(12)$ \\
cB211.072.64&               & $64^3 \times 128$ & $[7, \, 64]$ & $0.05659~~(8)$ & $0.05266~(15)$ \\
\hline
cC211.06.80& $1.836$ & $80^3 \times 160$ & $[7, \, 80]$ & $0.04721~~(7)$ & $0.04504~(10)$ \\
\hline
\end{tabular}
}
\end{center}
\renewcommand{\arraystretch}{1.0}
\caption{\it The time intervals $[t_{min}, t_{max}]$ adopted for the extraction of the pion mass and decay constant in lattice units obtained by applying the ODE method to the pion correlation function (\ref{eq:P5P5}).}
\label{tab:ODE}
\end{table} 

Within the ODE procedure we searched for 8 exponential signals in the time intervals of Table~\ref{tab:ODE} and in all cases at least two physical (non-noisy) exponential signals were found.
Then, a $\chi^2$-minimization procedure was applied using the physical ODE solution as the starting point.
The minimized values of the $\chi^2$ variable turned out to be always less than $1$.

The extracted values as well as their statistical errors of the ground-state mass and decay constant, collected in Table~\ref{tab:ODE}, are nicely consistent with the corresponding ones obtained by the direct single exponential fit (\ref{eq:larget}) shown in Table~\ref{tab:plateaux}.

Using the above pion data for $f_\pi$ the NLO analysis of Section~\ref{sec:fPi}, including the discretization term proportional to $a^2 M_\pi^2$, yields for the GF scale $w_0$ the value
\be
     w_0 = 0.1740 ~ (16) ~ {\rm fm} ~ 
    \label{eq:w0_fPi_ODE}
\ee
in agreement with the result~(\ref{eq:w0_NLO}).
Analogously, the use of the data for $X_\pi$ and of the NLO fit (\ref{eq:XPi_fit}) with $A_2^\prime = F_{FVE} = 0$ (see Section~\ref{sec:XPi}) leads to
\be
     w_0 = 0.17389 ~ (61) ~ {\rm fm} ~
     \label{eq:w0_XPi_ODE}
\ee
in agreement with the corresponding result shown in the second row of Table~\ref{tab:XPi}.

%% file: app_max_twist_corrections.tex
%%%%%%%%%%%%%%%%%%%%%%%%%%%%%%%%%%%%%%%%%%%%%%%%%%
\section{Maximal twist corrections for masses and decay constants}
\label{sec:appC}
%%%%%%%%%%%%%%%%%%%%%%%%%%%%%%%%%%%%%%%%%%%%%%%%%%

We follow the general approach of Ref.~\cite{Frezzotti:2004wz} to the mixed
action formulation of twisted mass lattice QCD, which ensures an unitary continuum
limit (provided sea and valence renormalized quark masses are matched). Here however
we allow for small deviations (due e.g.~to numerical errors) from the maximal twist
case, i.e.~for $m_0 \neq m_{cr}$.

\subsection{$N_f$=2+1+1 isosymmetric QCD with twisted clover Wilson quarks}
\label{sec:appC_sect1}

The lattice action can be conveniently written in terms of gauge, sea quark
and valence quark plus valence ghost fields. If the sea quarks are arranged
in two-flavour fields $\chi_\ell$ and $\chi_h$ and the valence quarks are
described by one-flavour fields $\chi_f$, with $f=u,d,...$, we have
\bea S & = & S_{g}[U] + S^{\ell}_{\rm tm}[\chi_\ell,\bar\chi_\ell,U;\mu_\ell,0;m_0]
+ S^{h}_{\rm tm}[\chi_h,\bar\chi_h,U;\mu_\sigma,\mu_\delta;m_0] \nonumber \\[2mm]
 & + & S_{val}[\{ \chi_f,\bar\chi_f \}, U; \{ \mu_f \}, m_0]  \; ,
\label{S211_tot}
\eea
with the valence sector given by
\bea
S_{val} & = & \bar\chi_u [D_{Wclov} + m_0 + i\mu_\ell \gamma_5]\chi_u +
\bar\chi_d [D_{Wclov} + m_0 - i\mu_\ell \gamma_5]\chi_d \nonumber \\[2mm]
& + & \bar\chi_s [D_{Wclov} + m_0 - i\mu_s \gamma_5]\chi_s
+ \bar\chi_c [D_{Wclov} + m_0 + i\mu_c \gamma_5]\chi_c \; \nonumber \\[2mm] 
& + & \ldots \; + \; {\rm ghost}\;{\rm terms} \; , \quad
\label{S211_val}
\eea
where ellipses stand for possible replica ($\chi'_f$) of the valence quarks with
$\mu'_f = -\mu_f$ and the ghost terms exactly cancel the valence
fermion contributions to the effective action. Here we find it convenient to express
all fermion fields in the canonical quark basis for untwisted Wilson fermions and
denote by $D_{Wclov}$ the well-known clover improved (gauge covariant) Dirac matrix:
$D_{Wclov} = D_{Wclov}[U] = \gamma \cdot \tilde\nabla[U] - (a/2) [\nabla^* \cdot \nabla][U]
+ i(c/4) \sigma \cdot F_{clover}[U]$.

We start by discussing the light valence quark sector, the extension to heavier flavours
is straightforward. Following Refs.~\cite{Frezzotti:2000nk,Frezzotti:2003ni},
we define (as customary) the twist angle $\omega_\ell$ in terms of the {\em bare} mass parameters of
the {\em light valence quark} $(u,d)$ doublet $X_\ell = (\chi_u,\chi_d)^T$, viz.\
%bare mass parameters that are renormalized by multiplication with $Z_P^{-1}$, viz.\
\be
\sin\omega_\ell = \frac{\mu_\ell}{\sqrt{ Z_A^2m_{PCAC}^2 + \mu_\ell^2} } \; , \qquad
\cos\omega_\ell = \frac{Z_A m_{PCAC}}{\sqrt{ Z_A^2m_{PCAC}^2 + \mu_\ell^2} } \;
\ee
with $Z_A$ the renormalization constant of $\bar X_\ell \gamma_\mu \gamma_5 (\tau^{1,2,3}/2) X_\ell$, which, being
independent of quark mass parameters, is defined in the chiral limit $\mu_f \to 0$, $m_0 \to m_{cr}$. 
Maximal twist corresponds to $|\omega_\ell| = \pi/2$, i.e.\ to 
angle $\theta_\ell \equiv \pi/2 - \omega_\ell$ equal to zero or $\pi$. We thus have
%, which is defined by
\bea
\cos\theta_\ell & = & \sin\omega_\ell = \frac{1}{\sqrt{1 + (Z_A^2m_{PCAC}^2)/\mu_\ell^2 } } \; , \nonumber \\[2mm]
\sin\theta_\ell & = & \cos\omega_\ell = \frac{1}{\sqrt{1 + \mu_\ell^2/(Z_A^2m_{PCAC}^2) } } \; .
\eea
Here $\mu_\ell$ is the bare twisted mass parameters for the $(u,d)$ doublet and $m_{PCAC}$
denotes the untwisted bare quark mass of the $(u,d)$ doublet as obtained from the
non-singlet WTI's -- hence a function of $m_0$ plus the other bare parameters.
We recall that $m_{PCAC} \propto m_0 - m_{cr}$. The {\em renormalized}
twisted and untwisted quark mass parameters that appear in the chiral WTI's read
(up to discretization effects)
\be
\mu_\ell^R = \mu_\ell \frac{1}{Z_P}  \, , \qquad
m^R = Z_A m_{PCAC} \frac{1}{Z_P} = (m_0 - m_{cr}) \frac{1}{Z_{S^0}} \; ,
\label{renmass}
\ee
where $Z_P$ and $Z_{S^0}$ are the renormalization constants of the pseudoscalar non-singlet
and the scalar singlet  densities (in the canonical basis for untwisted Wilson quarks).

Defining the twist angle and hence formulating the maximal twist condition in terms of $m_{PCAC}$, as
measured on the ensembles with 2+1+1 dynamical flavours, effectively takes care of (compensates for)
all the UV cutoff effects related to the breaking of chiral symmetry, including those coming from
the 2+1+1 sea quark flavours.

\subsection{Pion mass and decay constant}
\label{sec:appC_sect2}

We argue here that in the case of small enough numerical deviations from maximal twist the
lattice charged pion quantities
\be
 M_{\pi} |_L  \; , \qquad  [2 \mu_\ell \langle \pi^1({\bf 0}) | P^1 | 0 \rangle \; / \; 
 (M_\pi^2 \cos\theta_\ell)]|_L \; ,   \label{mpifpi-meas}
\ee
with $P^1 = \bar X_\ell \gamma_5 (\tau^1/2) X_\ell $ and $X_\ell = (\chi_u, \chi_d)^T$, approach
$M_\pi$ and $f_\pi$ as $a \to 0$ with lattice artifacts having numerically small, and
(we shall see) within errors immaterial, differences as compared to the O($a^2$) cutoff
effects occurring at maximal twist. Of course these values of $M_\pi$ and $f_\pi$ correspond
to the light quark renormalized mass $\; M_\ell^R \, = \, \sqrt{(m^R)^2 + (\mu_\ell^R)^2}\; $.

The numerical information on $M_\pi$ and $f_\pi$ comes from the simple correlator
$C_{PP}^{11}(x_0) = a^3\sum_{\bf x} \langle P^1(x) P^1(0) \rangle$ (and $C_{PP}^{22}(x_0)$).
The large-$x_0$ behaviour of $C_{PP}^{11}(x_0)$ determines $M_\pi$ and an
exact lattice WTI relates the operator $P^1$ to the four-divergence of a {\em conserved lattice
(backward one-point split) current}, which we denote by $\hat V_{\chi,\mu}^2$, v.i.z.\
\be
\partial_\mu \hat V_{\chi,\mu}^2(x) = 2 \mu_\ell P^1(x) = 2 \mu_\ell^R P^1_R(x) \; ,
\label{exaWTI} \ee
implying that the pion--to--vacuum matrix element of $\hat V_{\chi,\mu}^2 \,$
gives information on $f_\pi$ (barring the case of $\cos\theta_\ell =0$) . 
In Eq.~(\ref{exaWTI}) $P^1_{R} = Z_P P^1$ and the
equalities hold at operator level for finite lattice spacing ($a>0$). Hence the l.h.s.
of Eq.~(\ref{exaWTI}) is a renormalized operator and information on the approach of its matrix
elements to the continuum limit can be obtained by studying the behaviour as $a \to 0$ of
the corresponding matrix elements of $2 \mu_\ell^R P^1_R$.

Taking the matrix element of Eq.~(\ref{exaWTI}) between the vacuum and a one-$\pi^1$ state
of zero three-momentum and noting that in the continuum limit
\be \hat V_{\chi,\mu}^2 \; \stackrel{a \to 0}{\to} \; \left( \bar X_\ell \gamma_\mu (\tau^2/2) 
X_\ell \right)^R \; = \; \sin\theta_\ell \left(\bar\psi \gamma_\mu \frac{\tau^2}{2} \psi\right)^R
+ \cos\theta_\ell \left(\bar\psi \gamma_\mu \gamma_5 \frac{\tau^1}{2} \psi\right)^R  \label{currcont} \; ,
\ee
where $\psi = (u,d)^T$ obeys the (continuum) e.o.m.~$(\gamma \cdot D + M_\ell^R)\psi = 0$, 
for $a >0$ one obtains
\bea
2 [\mu_\ell \langle \pi^1({\bf 0}) | P^1 | 0 \rangle]|_L & = & 
[\cos\theta_\ell M_\pi^2 f_\pi \, + \, \sin\theta_\ell \langle \pi^1({\bf 0}) | \partial_0
(\bar\psi \gamma_0 \frac{\tau^2}{2} \psi)^R | 0 \rangle]|_L \nonumber \\[2mm]
& = & [\cos\theta_\ell M_\pi^2 f_\pi]|_L + {\rm O}(a) \, .
\label{fpicont}
\eea
%\bea
%&& 2 [\mu_\ell \langle \pi^1({\bf 0}) | P^1 | 0 \rangle]|_L =
%[\cos\theta_\ell M_\pi^2 f_\pi]|_L + [\sin\theta_\ell \langle \pi^1({\bf 0}) | \partial_0
%(\bar\psi \gamma_0 \frac{\tau^2}{2} \psi)^R | 0 \rangle]|_L \nonumber \\
%&& = [\cos\theta_\ell M_\pi^2 f_\pi]|_L + {\rm O}(am_\pi \sin\theta_\ell |c-c_{SW}|, \dots)
%\label{fpicont}
%\eea
%where the last step anticipates the result of our analysis \`a la Symanzik of the residual
%discretization errors occurring in our lattice setup, ellipses stand for numerically negligible
%lattice artifacts and $c-c_{SW}$ indicates the difference
This relation implies that as $a \to 0$ the ratio
\be
 [2 \mu_\ell \langle \pi^1({\bf 0}) | P^1 | 0 \rangle \; / \; (M_\pi^2 \cos\theta_\ell)]|_L \; \to \; f_\pi \; ,
\label{fpiestim}  \ee
at generic $\theta_\ell \neq \pm \pi/2$. Hence at $a >0$ the ratio~(\ref{fpiestim}) represents
a {\em bona fide} lattice estimator of $f_\pi$, while its discretization errors depend on the lattice 
artifacts in $M_\pi^2$, $\theta_\ell$ (or equivalently $m^R$, $\mu_\ell^R$) and the renormalized 
quantity $2\mu_\ell G_{\pi^1} = 2\mu_\ell \langle \pi^1({\bf 0}) | P^1 | 0 \rangle$.  

We are interested here in situations where $Z_A m_{PCAC}/\mu_\ell < 1$ but not negligibly small, 
say slightly above $0.1$. This situation indeed occurs in our gauge configuration ensemble cA211.12.48, at $a \sim 0.095$~fm and $a\mu_\ell = 0.0012$, where we find $Z_A m_{PCAC}/\mu_\ell \sim - 0.15$. 
In this case an analysis \`a la Symanzik of 
$M_\pi^2$, $m_\ell^R$ and $2\mu_\ell G_{\pi^1}$ shows (see below) that the change in the 
lattice artifacts of our lattice estimator of $f_\pi$, with respect to those purely O($a^{2n}$)
(with $n$ integer) that occur at maximal twist, is smaller than $0.001 f_\pi$, therefore
numerically immaterial within statistical errors that are typically of order $0.005 f_\pi$. 
%In particular, from our experience about simulations with two dynamical flavours in a lattice 
%setup where $c_{SW}$ is known one expects $|c-c_{SW}| < 0.15$, which makes the O($am_\pi$) 
%artifact in Eq.~(\ref{fpicont}) of relative order $<0.3\%$.

\subsubsection{The change in the lattice artifacts for $M_\pi^2$ and $f_\pi$} \label{sec:appC_sect2_1}
If $|am_{PCAC}|$ is non--zero, though definitely smaller than $|a\mu_\ell|$, the
same holds for $|a(m_0-m_{cr})|$ and one expects that the appropriate lattice estimators of $M_\pi$,
$f_\pi$ and any other physical quantity will be altered already at O($a$) as compared to their 
counterparts at maximal twist. Correcting analytically the lattice estimators for the deviation 
from maximal twist at order $a^0$ 
is hence not enough and one must also check that the out-of-maximal-twist modifications in 
order $a$ and order $a^2$ lattice artifacts are numerically negligible within statistical errors. 
Otherwise, analysing data corrected for deviations from maximal twist on some gauge ensembles 
together with data evaluated at maximal twist on other gauge ensembles might lead to
a systematic bias in the continuum extrapolation, where one typically assumes uniform O($a^2$) 
artifacts -- as expected if all data are obtained at maximal twist.

\subsubsection{Structure of the Symanzik effective Lagrangian} \label{sec:appC_sect2_1_1}
Let us analyse \`a la Symanzik the $N_f=2+1+1$ lattice QCD theory~(\ref{S211_tot}) 
out-of-maximal-twist and focus here 
on the light valence sector. We assume the reader is familiar with the basic literature on 
this topic such 
as~\cite{Luscher:1996sc,Frezzotti:2001ea,Frezzotti:2003ni} and references therein. 
The Symanzik local effective Lagrangian (LEL) to be used in our analysis of O($a$) artifacts
then reads 
\bea
L_{Syma} & = & L_4 + a L_5 + a^2 L_6 + O(a^3)  \; , \nonumber \\[2mm] 
L_4 & = & \frac{1}{4}(F \!\cdot\! F)  
%L_4 = \frac{1}{2}{\rm tr}(F \cdot F) + 
+ \bar\chi_\ell \left(\gamma\cdot D + m^R + i \gamma_5 \tau^3 \mu_\ell^R \right) \chi_\ell 
\nonumber \\[2mm]
& + & \bar\chi_h \left(\gamma\cdot D + m^R + i \gamma_5 \tau^3 \mu_h + \tau^1 \epsilon_h \right) \chi_h \nonumber \\[2mm] 
& + & \bar X_\ell^{val} \left(\gamma\cdot D + m^R + i \gamma_5 \tau^3 \mu_\ell^R \right) X_\ell^{val}
+ \ldots \; , \label{SymaLEL} 
\eea 
where $X_\ell^{val} = (\chi_u, \chi_d)^T$ describes the valence light quarks in the same basis as 
in Eq.~(\ref{S211_val}) while ellipses stands for $d \leq 4$ terms involving heavier valence quarks
and ghost terms. Upon taking the continuum limit in isosymmetric QCD with $2+1+1$ dynamical flavours,
we must have coinciding sea and valence renormalized masses for each flavour and keep constant as
$a \to 0$ the renormalized parameters
$\; g_R^2 \; , \; \mu_\ell^R \; , \; \mu_h^R \; , \; \epsilon_h^R \; $.
%For isosymmetric continuum QCD with $2+1+1$ dynamical flavours 
%(for each quark flavour the sea and the valence renormalized mass of course coincide) 
%the renormalized parameters that must be kept constant, if the limit 
%$a\to 0$ has to be approached with an O($a^2$) rate, are specified by
%$$ g_R^2 \; , \quad \mu_\ell^R \; , \quad \mu_h^R \; , \quad \epsilon_h^R \; . $$

The LEL terms $L_n$ are suitable linear combinations of the $d=n >4$ operator terms 
allowed by the symmetries of the lattice theory~(\ref{S211_tot}). In particular it turns out that
\bea
L_5 & = & \ldots + (c-c_{SW}) \frac{i}{4} \bar X_\ell^{val} \sigma \!\cdot\! F X_\ell^{val} 
- b_g m \frac{1}{4} F \!\cdot\! F \nonumber \\
& - & (b_m m^2 + \tilde{b}_m \mu_\ell^2) \, \bar X_\ell^{val} X_\ell^{val} 
- b_\mu m \mu_\ell \, \bar X_\ell^{val} i \gamma_5 \tau^3 X_\ell^{val} + \ldots 
\; , \label{SymaL5}
\eea
where $m \equiv Z_A m_{PCAC} \propto m_0 - m_{cr}$, while ellipses stand here for terms involving
only heavier valence quark operators as well as sea quark and ghost terms (all of them are
omitted since they are immaterial for this Section). With $c-c_{SW}$ we indicate the difference
between the 1-loop tadpole improved estimate ($c$) employed in our simulation and the exact
value ($c_{SW}$) of the coefficient of the clover term~\footnote{
From our experience in simulations with two dynamical flavours in a lattice setup
where $c_{SW}$ is known, we expect that $|c-c_{SW}|< 0.15$, which suppresses
O($a$) lattice artifact by nearly one order of magnitude and, provided $|am/a\mu_\ell| <1$ is 
small enough, makes undesired O($a$) numerically negligible in most observables.}.

Concerning $L_6$, it will be enough to focus on its $m$-dependent sector and to note 
the structure
\be
L_6(m;\mu_\ell,\mu_h,\epsilon_h, \mu_s, \dots) = L_6(0;\mu_\ell,\mu_h,\epsilon_h, \mu_s, \dots) 
+ m O_5 + m^2 O_4 \; ,
\label{L6form}
\ee
where $O_5$ ($O_4$) is a linear combination of the $m$--independent terms allowed in $L_5$ ($L_4$).
Among the latter, since we employ in our correlators flavour diagonal OS valence quark fields
$\chi_f^{val}$ with twisted mass $\mu_f>0$, or $\chi_f^{\prime \,val}$ fields with twisted mass
$\mu'_f = -\mu_f$, for the purposes of this Section only the terms bilinear in the $X_\ell^{val}$ 
and $\bar X_\ell^{val}$, or $X_\ell^{' val}$ and $\bar X_\ell^{' val}$, fields are relevant, which
we may call $m O_{5,\ell}^{val}$ and $m^2 O_{4,\ell}^{val}$. 

We note that exact or spurionic lattice symmetries rule out~\footnote{To this goal it is enough 
to consider charge combination, $\tilde{P} \times (\mu_{f,\ell,h} \to -\mu_{f,\ell,h})$ and
$(X_f \to i\tau^2 X_f) \times (\bar X_f \to -i \bar X_f \tau^2) \times (\mu_{f,\ell,h} \to -\mu_{f,\ell,h})$ invariances, with $\tilde P$ meaning parity transformation on gauge fields combined with
$(X_f(x) \to \gamma_0X_f(x_P)) \times (\bar X_f(x) \to \bar X_f(x_P)\gamma_0)$ and 
$x_P = (x_0,-\vec{x})$.} the $L_5$ terms of the form 
\[
\mu_f i \widetilde{F} \!\cdot\! F \; , \quad \mu_f \bar X_f \tau^{0,1,2,3} i \gamma_5 \gamma \!\cdot\! D  X_f \; , \quad \mu_f \bar X_f \tau^{0,1,2,3} \gamma \!\cdot\! D  X_f \; ,
\]
as well as the analogous $L_6$ terms of the form
\[
m \mu_f i \widetilde{F} \!\cdot\! F \; , \quad m \mu_f \bar X_f \tau^{0,1,2,3} i \gamma_5 \gamma \!\cdot\! D  X_f \; , \quad m \mu_f \bar X_f \tau^{0,1,2,3} \gamma \!\cdot\! D  X_f \; .
\]

%the $L_6$-terms of the form $m \mu_f \widetilde{F} \!\cdot\! F $ or
%$m \mu_f (\bar\chi_f^{val} i \gamma_5 \gamma \!\cdot\! D \chi_f^{val})$ are ruled out by discrete 
%lattice symmetries~\footnote{Under 
%$\tilde{P} \times \mu_{f,\ell,h} \to -\mu_{f,\ell,h}$ the lattice action~(\ref{S211_tot}) is 
%invariant, while the excluded terms change sign.}, for all quark flavours $f$.  

% O(a) discretization effects on $m^R$ and $M_\pi^2$ ...

\subsubsection{The discretization effects on $M_\pi |L$} \label{sec:appC_sect2_1_2}
In the case of $|am| \sim 0.0002 < |a\mu_\ell|=0.0012$, for the 
quantity $M_\pi^2 |_L$ the O($a$) deviation from its
continuum limit value $\, M_\pi^2 = 2B^R \sqrt{ (m^R)^2 + (\mu_\ell^R)^2 } + 
O\left ((m^R)^2+(\mu_\ell^R)^2 \right) \sim 2B^R \mu_\ell^R$ is given by
\be
\delta_1 M_\pi^2 = \delta_{1A} M_\pi^2 + \delta_{1B} M_\pi^2 \; ,
\ee
where, writing operators in terms of the physical basis fermion doublet fields 
$$\psi_\ell = e^{i\omega_\ell \gamma_5\tau^3/2}
X_\ell \; , \qquad \bar \psi_\ell = \bar X_\ell e^{i\omega_\ell \gamma_5\tau^3/2} \;  $$
and exploiting parity and isospin symmetries of continuum $N_f=2+1+1$ QCD, one has
\bea
\delta_{1A} M_\pi^2 & = & a \sin\theta_\ell \langle \pi^{1,2}({\bf 0}) |  (c-c_{SW})
 \bar \psi_\ell^{val} \frac{i}{4} \sigma \!\cdot\! F \psi_\ell^{val}  - 
 (b_m m^2 + \tilde{b}_m \mu_\ell^2) \bar\psi_\ell^{val} \psi_\ell^{val} |\pi^{1,2}({\bf 0}) \rangle
 \nonumber \\
& \lesssim & \; 0.003 \frac{\alpha_s(\Lambda_{QCD})}{4\pi} \Lambda_{QCD}^2 \sim 0.001 M_\pi^2
\label{D1AMpi2}
\eea
and (approximating $\cos\theta_\ell$ with $1$, since $\sin\theta_\ell \simeq 0.2$)
\bea
\delta_{1B} M_\pi^2 & = & am \langle \pi^{1,2}({\bf 0}) | (\, b_\mu \mu_\ell \bar\psi_\ell^{val} 
\psi_\ell^{val} \,+\, b_g \frac{1}{4} F \!\cdot\! F \,) | \pi^{1,2}({\bf 0}) \rangle 
\; \nonumber \\[2mm]
& \lesssim & \; 2B^R \mu_\ell^R \; 0.0005 \; .   \label{D1BMpi2}
\eea
The numerical estimate in Eq.~(\ref{D1AMpi2}) results from $|c-c_{SW}| \lesssim 0.15$,
$\alpha_s(\Lambda_{QCD}) \sim 1$, $a \Lambda_{QCD} \simeq 0.1$ and $\Lambda_{QCD}/M_\pi \simeq 2$,
while $b_m = -1/2 + {\rm O}(g_0^2) $ and (making the choice advocated in Ref.~\cite{Frezzotti:2001ea}) 
$\tilde{b}_\mu = -1/2$. \\
The numerical estimate in Eq.~(\ref{D1BMpi2}) follows from
$| \langle \pi^{1,2}({\bf 0}) | \mu_\ell \bar\psi_\ell^{val}
\psi_\ell^{val} | \pi^{1,2}({\bf 0}) \rangle | \sim |2 B^R \mu_\ell^R|$
and $|b_\mu| = {\rm O}(g_0^2) < 1$. 
In fact soft pion theorems (i.e.\ spontaneously broken continuum chiral symmetry) imply 
that the contribution of $| b_g \langle \pi^{1,2}({\bf 0}) | 
\frac{1}{4} F \!\cdot\! F \,) | \pi^{1,2}({\bf 0}) \rangle |$, with $b_g = {\rm O}(g_0^2 N_f)$, 
is O($\mu_\ell$) and thus of the same order of magnitude as 
$| b_\mu \langle \pi^{1,2}({\bf 0}) | \mu_\ell \bar\psi_\ell^{val} \psi_\ell^{val} | \pi^{1,2}({\bf 0}) \rangle |$. \\
The undesired O($a$) modification in $M_\pi^2$ is estimated to be smaller than $0.001\, M_\pi^2$, 
hence immaterial within our statistical errors of a few permil. The O($a^2$) change in $M_\pi^2$ due to the non-zero value of $am$ is also of order $0.001\, M_\pi^2$ or smaller, because of the 
form~(\ref{L6form}) of $L_6$, with $|am| \simeq 0.0002$ and 
$a^2\Lambda_{QCD}^2 \sin\theta_\ell \simeq 0.001$. \\ 

% Symanzik description of < P^1 P^1 >_latt ....

\subsubsection{The discretization effects on $f_\pi |L$} \label{sec:appC_sect2_1_3}
Based on Eq.~(\ref{fpiestim}) the lattice artifacts on $f_\pi$ can be estimated in terms of
the cutoff effects in $(\mu_\ell/\cos\theta_\ell) \langle \pi^1({\bf 0}) | P^1 | 0 \rangle|_L$
and in $M_\pi^2|_L$. We discussed above the lattice artifacts of $M_\pi^2|_L$. 
Concerning $(\mu_\ell/\cos\theta_\ell) \langle \pi^1({\bf 0}) | P^1 | 0 \rangle|_L$, we can 
see it as the product of the renormalized quantities
$Z_P^{-1} (\mu_\ell / \cos\theta_\ell) = M_\ell^R$ and 
$Z_P \langle \pi^1({\bf 0}) | P^1 | 0 \rangle]|_L = G_{\pi^1}^R$, and then
discuss separately the lattice artifacts in each of these two factors~\footnote{Of course
we do not worry about possible cutoff effects on $Z_P$, which cancel in the product.}. 

As for $Z_P^{-1} (\mu_\ell / \cos\theta_\ell) = \sqrt{(m^R)^2 + (\mu_\ell^R)^2}$, the form of 
the $m$-dependent terms in $L_5$, i.e.\ those with coefficients $b_m$ and $b_\mu$ (which are all in modulus $\lesssim 1$), implies that the O($a$) corrections to 
$m^R$ and $\mu_\ell^R$, and hence to $M_\ell^R$ are only of relative size $< |am| \simeq 0.0002$. 
Even smaller are the corrections to $M_\ell^R$ of order $a^2m$.

Let us then consider the out-of-maximal-twist induced cutoff artifacts in the matrix
element $Z_P \langle \pi^1({\bf 0}) | P^1 | 0 \rangle]|_L = G_{\pi^1}^R$. They clearly
arise from from the lattice two-point correlator $C_P^{11}(x_0)$. At order $a$, since 
for the operator  
%arise from from the lattice two-point correlator $C_P^{11}(x_0) = a^3 \sum_{\vec{x}} 
%\langle P^1(x) P^1(0) \rangle |_L$. At order $a$, since for the operator 
$P^1 = \bar\psi_\ell \gamma_5 (\tau^1/2) \psi_\ell$  it is known that $c_P=0$,
$\tilde{b}_P=0$ and $b_P = 1 + {\rm O}(g_0^2)$, implying $|b_P am| \sim 0.0002$,
the numerically dominant lattice artifacts stem from the term
%\[ 
$a \int d^4 y \langle P^1(x) P^1(0) L_5(y) \rangle |_{\rm cont} $ 
%\]
in the Symanzik description of $C_{11}(x_0)$. Inserting intermediate states and considering 
the possible $y_0$-orderings one checks that, owing to the structure of $L_5$ and $|am| \simeq 
0.0002$, the numerically leading cutoff effects in $G_{\pi^1}$ are proportional to 
$$ 
(c-c_{SW}) \sin\theta_\ell a \left[ \langle \pi^1 | \bar \psi_\ell^{val} \frac{i}{4} 
\sigma \!\cdot\! F \psi_\ell^{val}  | \pi^1 \rangle G_{\pi^1} +
\sum_{n} \langle 0 | \bar \psi_\ell^{val} \frac{i}{4} 
\sigma \!\cdot\! F \psi_\ell^{val} | n \rangle \langle n | P^1 | \pi^1 \rangle  \right]
$$
and are hence of relative order $|\sin\theta_\ell (c-c_{SW})| a\Lambda_{QCD} \alpha_s/(4\pi) $,
which, if $|am| \simeq 0.0002$ and $|c-c_{SW}|<0.15$, noting that $a\Lambda_{QCD} \alpha_s/(4\pi) 
< 0.1$, turns out to be $\lesssim 0.0002$. Compared to this the O($a^2 m$) artifacts in 
$G_{\pi^1}$ are expected to be smaller by at least one order of magnitude.

Hence our lattice estimator of $f_\pi$ is estimated to be affected by the small deviation
from maximal twist observed on the ensemble cA211.A12.48, where $|a m| \sim 0.0002$, only 
at a level of $\lesssim 0.0004\, f_\pi$, which is negligible as compared to current statistical errors.

%%Analysis of $[2 \mu_\ell \langle \pi^1({\bf 0}) | P^1 | 0 \rangle / (M_\pi^2 \cos\theta_\ell)]|_L$ ... 

\subsubsection{Analysis of $M_\pi^2$ and $f_\pi$ in terms of the renormalized light quark mass}

The discussion above is valid also in case the observables $M_\pi^2$ and $f_\pi$ are 
analysed as functions of the renormalized light quark mass
\be 
 M_\ell^R  = \sqrt{(m^R)^2 + (\mu_\ell^R)^2} =  Z_P^{-1} (\mu_\ell / \cos\theta_\ell) \; .
\ee
As already noted in Section~\ref{sec:appC_sect2_1_3}, for the ensemble cA211.12.48 the observed
small deviation from maximal twist leads to an undesired O($am$) relative change in $M_\ell^R$
that is only of order $|am| \simeq 0.0002$ and thus fully negligible in comparison to other errors.

A rather obvious but practically important and general caveat follows in the case one
insists in analysing the lattice data, e.g.~for $M_\pi$ or $f_\pi$, obtained on gauge ensembles
with non-zero $am$-values in terms of $\mu_\ell^R$ rather than $M_\ell^R$. Since for $am \neq 0$
a generic observable $Q_{obs}$ actually refers to $M_\ell^R > \mu_\ell^R$, one should, besides possibly 
applying an analytic correction to the datum for $Q_{obs}$ (as requested e.g.~for $f_\pi$, but not 
for $M_\pi$), also shift the value of the observable itself according to
\be
Q_{obs}(\mu_\ell^R) = Q_{obs}(M_\ell^R) +  
\left. \frac{\partial Q_{obs}}{\partial M_\ell^R}\right|_{M_\ell^R} (\mu_\ell^R - M_\ell^R)
+ {\rm O}( (\mu_\ell^R - M_\ell^R)^2) \; ,  
\ee
where in practice, since typically $|am| < 0.001$, only the terms linear in 
$\mu_\ell^R - M_\ell^R$ are numerically important.

%Analogous to above, but using for the ensemble out-of-max-twist as renormalized quark mass
%$$      M_\ell^R = \sqrt{(m^R)^2 + (\mu_\ell^R)^2} 
%$$
%rather than $\mu_\ell^R = Z_P^{-1}\mu_\ell$. Recall Eq.~(\ref{renmass}). 
%
%Alternatively: if one insists on using $\mu_\ell^R$ the observables $Q_{obs}$ must be
%shifted from $M_\ell^R$ to $\mu_\ell^R$:  
%$$ Q_{obs}(\mu_\ell^R) \simeq Q_{obs}(M_\ell^R) - 
%\frac{\partial Q_{obs}}{\partial M_\ell^R)}|_{M_\ell^R} (\mu_\ell^R - M_\ell^R) \; .  $$

%\vspace*{0.5cm}

\subsection{Mass and decay constant of the kaon and heavier PS-mesons}
\label{sec:appC_sect3}

Generalizing the arguments of Section~\ref{sec:appC_sect2} for the mass and decay constant of the pion to the 
case of the kaon or heavier pseudoscalar (PS) mesons is rather straightforward. Indeed,
within the $N_f=2+1+1$ lattice QCD framework of Section~\ref{sec:appC_sect1}, as far as the control of the effects 
of a small but non-zero value of $a m = a Z_A m_{PCAC}$ is concerned, there is not much difference
between a PS meson made out of two light valence quarks (pion, with $|\mu_u| = |\mu_d|$)
and a PS meson made out of a light (mass $\mu_\ell$) and a heavier (mass $\mu_x$) valence
quark. This is a consequence of the fact that the extraction of the PS meson mass and decay 
constant relies on general positivity properties of two-point correlation
functions (close enough to the continuum limit) and on exact chiral WTI's, which hold valid
irrespectively of the value of valence quark masses. One might thus deal in a fully analogous
way with PS mesons made out of two non-light valence quarks. For definiteness, however, we shall 
here focus on the case of the kaon, i.e.~$\mu_x = \mu_s \gg \mu_\ell $ and $\mu_d = \mu_\ell >0$.

It turns out that in the case of small enough numerical deviations from maximal twist,
say $0.1 \, a\mu_\ell <  |am| \ll a\mu_\ell \ll a\mu_s$, the lattice charged kaon quantities
\be
 M_{K} |_L  \; , \qquad  [(\mu_\ell + \mu_s) \langle K({\bf 0}) | P^{s,d} | 0 \rangle \; / \;
 (M_K^2 \cos( (\theta_\ell + \theta_s)/2) ]|_L \; ,   \label{mKfK-meas}
\ee
with $P^{s,d} = \bar X_{s,d} \gamma_5 (\tau^1/2) X_{s,d} $ and 
$X_{s,d} = (\chi_s, \chi_d)^T$, approach
$M_K$ and $f_K$ as $a \to 0$ with lattice artifacts having numerically small, and
within errors immaterial, differences as compared to the O($a^2$) cutoff
effects occurring at maximal twist. These values of $M_K$ and $f_K$ correspond
to the light ($d$) quark renormalized mass $\; M_\ell^R \, = \, \sqrt{(m^R)^2 + 
(\mu_\ell^R)^2}\; $ and to the strange ($s$) quark renormalized mass $\; M_s^R \, = \, 
\sqrt{(m^R)^2 + (\mu_s^R)^2}\; $.

In full analogy to the definitions adopted for light valence quarks (see Section~\ref{sec:appC_sect1})
we define $\; \mu_s^R = \mu_s / Z_P \;$, where $Z_P$ is (can be conveniently taken as) the 
same mass-independent renormalization constant of the PS non-singlet quark bilinear operator 
as above, and 
\bea
\cos\theta_s & = & \sin\omega_s = \frac{1}{\sqrt{1 + (Z_A^2m_{PCAC}^2)/\mu_s^2 } } \; , \nonumber \\[2mm]
\sin\theta_s & = & \cos\omega_s = \frac{1}{\sqrt{1 + \mu_s^2/(Z_A^2m_{PCAC}^2) } } \; .
\eea

\subsubsection{On the lattice estimators~(\ref{mKfK-meas}) of $M_K$ and $f_K$ at O($a^0$) }

The numerical information on $M_K$ and $f_K$ comes in fact from the simple correlator
$C_{KK}^{s,d}(x_0) = a^3\sum_{\bf x} \langle P^{s,d}(x) P^{d,s}(0) \rangle$.
The large-$x_0$ behaviour of $C_{KK}^{s,d}(x_0)$ determines $M_K$ as the kaon mass that,
owing to renormalizability (and unitarity in the continuum limit) of the lattice
theory of Section~\ref{sec:appC_sect1}, corresponds to the light and strange quark masses 
$M_\ell^R$ and $M_s^R$.
An exact lattice WTI relates the operator $P^{s,d}$ to the four-divergence of a {\em conserved lattice
(backward one-point split) current}, which we denote by $\hat V_{\chi,\mu}^{s,d}$, viz.\
\be
\partial_\mu \hat V_{\chi,\mu}^{s,d}(x) = (\mu_\ell +\mu_s) P^{s,d}(x) = (\mu_\ell^R
+ \mu_s^R) P^{s,d}_R(x) \; ,
\label{exaWTI_sd} \ee
implying that the kaon--to--vacuum matrix element of $\hat V_{\chi,\mu}^{s,d} \,$
gives information on $f_K$, barring the case of $\cos((\theta_\ell + \theta_s)/2) =0$ .
In Eq.~(\ref{exaWTI_sd}) $P^{s,d}_{R} = Z_P P^{s,d}$ and the
equalities hold at operator level for $a>0$. Hence the l.h.s.
of Eq.~(\ref{exaWTI_sd}) is a renormalized operator and information on the approach of its matrix
elements to the continuum limit can be obtained by studying the behaviour as $a \to 0$ of
the corresponding matrix elements of $ (\mu_\ell^R + \mu_s^R ) P^{S,d}_R$.

Taking the matrix element of Eq.~(\ref{exaWTI_sd}) between the vacuum and a one-$K$ state
of zero three-momentum and noting that in the continuum limit 
\bea
\hat V_{\chi,\mu}^{s,d} \; \stackrel{a \to 0}{\to} \; \left( \bar X_{s,d} \gamma_\mu (\tau^2/2) 
X_{s,d} \right)^R \; & = & \; \sin \frac{\theta_\ell+\theta_s}{2}  
\left(\bar\psi_{s,d} \gamma_\mu \frac{\tau^2}{2} \psi_{s,d}\right)^R \nonumber \\[2mm]
& + & \cos \frac{\theta_\ell+\theta_s}{2} \left(\bar\psi \gamma_\mu \gamma_5 \frac{\tau^1}{2} \psi\right)^R  
\label{currcont_sd} \; ,
\eea
where $\psi_{s,d} = (s,d)^T$ obeys the (continuum) e.o.m.~$(\gamma \cdot D + \frac{M_s^R + M_\ell^R}{2} +
\frac{M_s^R - M_\ell^R}{2} \tau^3 )\psi_{s,d} = 0$,
for $a >0$ one obtains
\be
[(\mu_\ell +\mu_s) \langle K({\bf 0}) | P^{s,d} | 0 \rangle]|_L =
%[\cos\theta_\ell M_\pi^2 f_\pi \, + \, \sin\theta_\ell \langle \pi^1({\bf 0}) | \partial_0
%(\bar\psi \gamma_0 \frac{\tau^2}{2} \psi)^R | 0 \rangle]|_L =
[\cos((\theta_\ell + \theta_s)/2) M_K^2 f_K]|_L + {\rm O}(a) \, .
\label{fKcont}
\ee
This relation implies that as $a \to 0$ the ratio
\be
 [(\mu_\ell + \mu_s) \langle K({\bf 0}) | P^{s,d} | 0 \rangle \; / \; (M_K^2 \cos((\theta_\ell + \theta_s)/2))]|_L \; \to \; f_K \; ,
\label{fKestim}  \ee
for generic $\theta_\ell + \theta_s \neq \pm \pi$. Hence at $a >0$ the ratio~(\ref{fKestim}) represents
a {\em bona fide} lattice estimator of $f_K$, while its discretization errors depend on the lattice
artifacts in $M_K^2$, $\theta_\ell$, $\theta_s$ (or equivalently $m^R$, $\mu_\ell^R$, $\mu_s^R$) and 
the renormalized quantity $(\mu_\ell + \mu_s) G_{K} = (\mu_\ell + \mu_s) \langle K({\bf 0}) | 
P^{s,d} | 0 \rangle$.  

\subsubsection{Theoretical control and numerical size of O($a$) and O($a^2$) artifacts}

Concerning the impact of a non-zero $am$ value on the lattice artifact in the kaon sector 
it is important to note that, as phenomenology dictates $\; \mu_\ell \simeq 0.037 \mu_s \;$, 
we have in general
\bea
\theta_s \ll \theta_\ell   \; , \qquad 
\cos((\theta_\ell + \theta_s)/2) & \simeq & \cos(\theta_\ell/2) [1 - {\rm O}(\theta_s^2)] -
{\rm O} (\theta_s \theta_\ell) \nonumber \\[2mm]
& \simeq & \cos(\theta_\ell/2) - {\rm O} (\theta_s \theta_\ell) \; .
\eea
In particular, for the gauge ensemble cA211.12.48, where $am = Z_A m_{PCAC}/\mu_l \simeq -0.15$,
we find
\be
\theta_\ell \simeq -0.15 \; , \quad \theta_s \simeq -0.006 \; , \quad
\cos((\theta_\ell + \theta_s)/2) \simeq \cos(\theta_\ell/2) - {\rm O}(0.0009) \; .
\ee

For the discussion of O($a$) lattice artifacts in $M_K$ and $f_K$ one should consider of
course the occurrence of flavour diagonal terms involving the valence quark fields $\chi_s$, 
$\bar\chi_s$ both in $L_5$ (see Eq.~(\ref{SymaL5})), where they take the form
\be
L_5 \; \supset \;
(c-c_{SW}) \frac{i}{4} \bar \chi_s^{val} \sigma \!\cdot\! F \chi_s^{val}
- (b_m m^2 + \tilde{b}_m \mu_s^2) \, \bar \chi_s^{val} \chi_s^{val}
-b_\mu m \, \mu_s \, \bar \chi_s^{val} i \gamma_5 \chi_s^{val} 
\; , \label{SymaL5_val_s}
\ee
and in $L_6$, for which the structure in Eq.~(\ref{L6form}) remains valid. Looking back to
Eq.~(\ref{SymaL5}) for the light valence quark and gluonic terms in $L_5$, one finds that
the discussion for the kaon case closely follows the one for the pion (see Section~\ref{sec:appC_sect2_1}),
provided one replaces
the valence quark field pair $X_\ell^{val} = (\chi_u,\chi_d)^T$, used for the latter, with 
the valence quark field pair $X_{s,d}^{val} = (\chi_s,\chi_d)^T$ relevant for the kaon, as well
as $\theta_\ell$ with $(\theta_s+\theta_\ell)/2$. This implies that for $M_K^2$ and $f_K$ the 
numerically dominant changes in the lattice artifacts as compared to the maximal twist case, which 
for $M_\pi^2$ and $f_\pi$ were proportional to $\sin\theta_\ell$, turn out to be proportional to
\be
\sin ((\theta_\ell + \theta_s)/2) \simeq \sin(\theta_\ell /2) - {\rm O}(0.006) 
\simeq 0.5 \sin\theta_\ell \; ,
\ee
thereby getting reduced by a factor of about two with respect to the pion case.

In conclusion, if $|am| \sim 0.0002$ (as it happens for our ensemble cA211.12.48), we estimate
a lattice artifact modification, with respect to the case of maximal twist, that does not exceed $0.0005\, M_K^2$ for $M_K^2$ and $0.0002\, f_K$ for $f_K$ and is hence safely negligible 
as compared to our current statistical errors. 

From the arguments above it should also be clear that the same quantitative estimates hold also 
for the lattice artifact changes induced by $am \neq 0$ in the mass and the decay constant of 
heavy-light PS mesons with a charm or even heavier non-light valence quark having a mass 
$\mu_x \gg \mu_s$. In fact the heavier the valence quark, the smaller $|\theta_x| \simeq |m/\mu_x|$ 
and the more numerically irrelevant the effect of a small non-zero value of $am$.

%% file: app_determination_of_gf_scales.tex
%%%%%%%%%%%%%%%%%%%%%%%%%%%%%%%%%%%%%%%%%%%%%%%%%%
\section{Determination of the GF scales $\sqrt{t_0}$ and $t_0 / w_0$}
\label{sec:appD}
%%%%%%%%%%%%%%%%%%%%%%%%%%%%%%%%%%%%%%%%%%%%%%%%%%

In this Appendix we describe the calculations of the relative GF scales $w_0 / a$, $\sqrt{t_0} / a$ and $t_0 / (w_0 a)$ at the physical pion point and we summarize the determination of the absolute scales $\sqrt{t_0}$ and $t_0 / w_0$ using the SU(2) ChPT analysis of the data for $X_\pi$, carried out in Section~\ref{sec:XPi} in the case of the GF scale $w_0$.

\subsection{Determination of the relative GF scales}
\label{sec:appD_sect1}

\input{extrapolations_tmCloverNf211_GF_scales.tex}

\subsection{Determination of the GF scales $\sqrt{t_0}$ and $t_0 / w_0$}
\label{sec:appD_sect2}

The SU(2) ChPT analysis of the data for $X_\pi$, carried out in Section~\ref{sec:XPi} adopting the GF scale $w_0$, can be repeated in the case of the scales $\sqrt{t_0}$ and $t_0 / w_0$.
The values of the relative GF scales $w_0 /a $, $\sqrt{t_0} / a$ and $t_0 / (w_0 a)$ have been calculated at the physical pion point in the previous Section~\ref{sec:appD_sect1} and, for sake of clarity, we recollect them in Table~\ref{tab:relGF}.
\begin{table}[!h]
\renewcommand{\arraystretch}{1.2}	 
\begin{center}
\begin{tabular}{||c||c|c|c||}
\hline
$\beta$ & $w_0 / a$ & $\sqrt{t_0} / a$ & $t_0 / (w_0 a)$\\ \hline
\hline
~1.726~ & ~1.8352~(35)~ & ~1.5660~~(22) & ~1.3359~~(12)~\\ \hline
~1.778~ & ~2.1299~(16)~ & ~1.80396~(68) & ~1.52789~(33)~\\ \hline
~1.836~ & ~2.5045~(17)~ & ~2.1094~~~(8) & ~1.77670~(37)~\\ \hline
\hline
\end{tabular}
\end{center}
\renewcommand{\arraystretch}{1.0}
\caption{\it Values of the relative GF scales $w_0 /a $, $\sqrt{t_0} / a$ and $w_0 / (w_0 a)$ obtained at the physical pion point in Section~\ref{sec:appD_sect1}.}
\label{tab:relGF}
\end{table}

Adopting fitting functions similar to the ones given by Eq.~(\ref{eq:XPi_fit}) used in the case of the scale $w_0$, we obtain
\bea
     \label{eq:w0_appD}
     w_0 & = & 0.17383 ~ (57)_{\rm stat+fit} ~ (26)_{\rm syst} ~ [63] ~ \mbox{fm} ~ , ~ \\[2mm]
     \label{eq:t0s_appD}
     \sqrt{t_0} & = & 0.14436 ~ (54)_{\rm stat+fit} ~ (30)_{\rm syst} ~ [61] ~ \mbox{fm} ~ , ~ \\[2mm]
     \label{eq:t0w0_appD}
     \frac{t_0}{w_0} & = & 0.11969 ~ (52)_{\rm stat+fit} ~ (33)_{\rm syst} ~ [62] ~ \mbox{fm} ~ . ~
\eea

\begin{figure}[htb!]
\begin{center}
\includegraphics[scale=0.80]{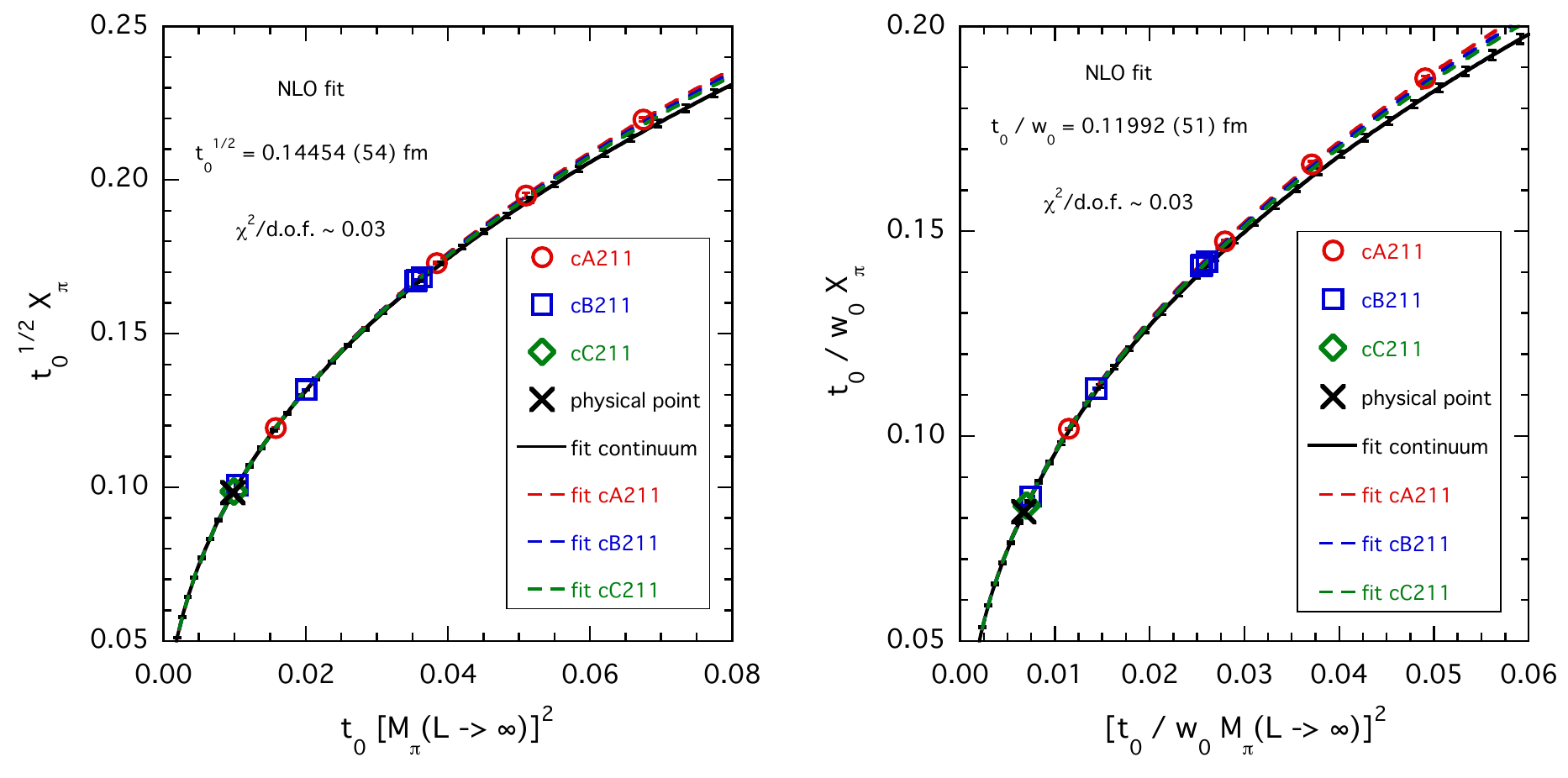}
\end{center}
\vspace{-0.75cm}
\caption{\it \small The same as in Fig.~\ref{fig:XPi_NLO}, but in the case of the GF scales $\sqrt{t_0}$ (left panel) and $t_0 / w_0$ (right panel).}
\label{fig:XPi_GF}
\end{figure}

The quality of the fitting procedure in the case of the GF scales $\sqrt{t_0}$ and $t_0 / w_0$ is illustrated in Fig.~\ref{fig:XPi_GF} and it should be compared with the one shown in Fig.~\ref{fig:XPi_NLO} in the case of the GF scale $w_0$.

Some values obtained for the continuum-limit fitting parameters $f$ and $\bar{\ell}_4^{phys}$ and for the discretization parameters $D_0^\prime$ and $D_1^\prime$ are collected in Table~\ref{tab:GFscales}.
\begin{table}[htb!]
\renewcommand{\arraystretch}{1.2}	 
\begin{center}
\begin{tabular}{||c||c|c||c|c||}
\hline
GF scale & $f$ (MeV) & $\bar{\ell}_4^{phys}$ & $D_0^\prime$ & $D_1^\prime$\\ \hline
\hline
$w_0$         & ~124.4~(1.2)~ & ~3.24~(29)~ & ~-0.167 (52)~ & ~8.0~(2.3)~\\ \hline
$\sqrt{t_0}$ & ~124.5~(1.2)~ & ~3.16~(27)~ & ~-0.065 (40)~ & ~6.3~(1.6)~\\ \hline
$t_0 / w_0$ & ~124.7~(1.3)~ & ~3.08~(27)~ & ~-0.003 (32)~ & ~4.9~(1.2)~\\ \hline
\hline
\end{tabular}
\end{center}
\renewcommand{\arraystretch}{1.0}
\vspace{-0.25cm}
\caption{\it Values of the fitting parameters $f$ and $\bar{\ell}_4^{phys}$ and of the discretization parameters $D_0^\prime$ and $D_1^\prime$ obtained in the case of the three GF scales $w_0$, $\sqrt{t_0}$ and $t_0 / w_0$ using the data on $X_\pi$ and adopting fitting functions similar to Eq.~(\ref{eq:XPi_fit}) with $A_2^\prime = F_{FVE} = 0$.}
\label{tab:GFscales}
\end{table} 
It can clearly be seen that the pion mass dependence of $X_\pi$ in the continuum limit is stable against the choice of the specific GF scale, while the values of the discretization parameters $D_0^\prime$ and $D_1^\prime$ depend on the above choice.
The discretization effects on $X_\pi$ appear to be smaller in the case of the GF scale $t_0 / w_0$.

\begin{figure}[!htb]
\begin{center}
\includegraphics[scale=0.750]{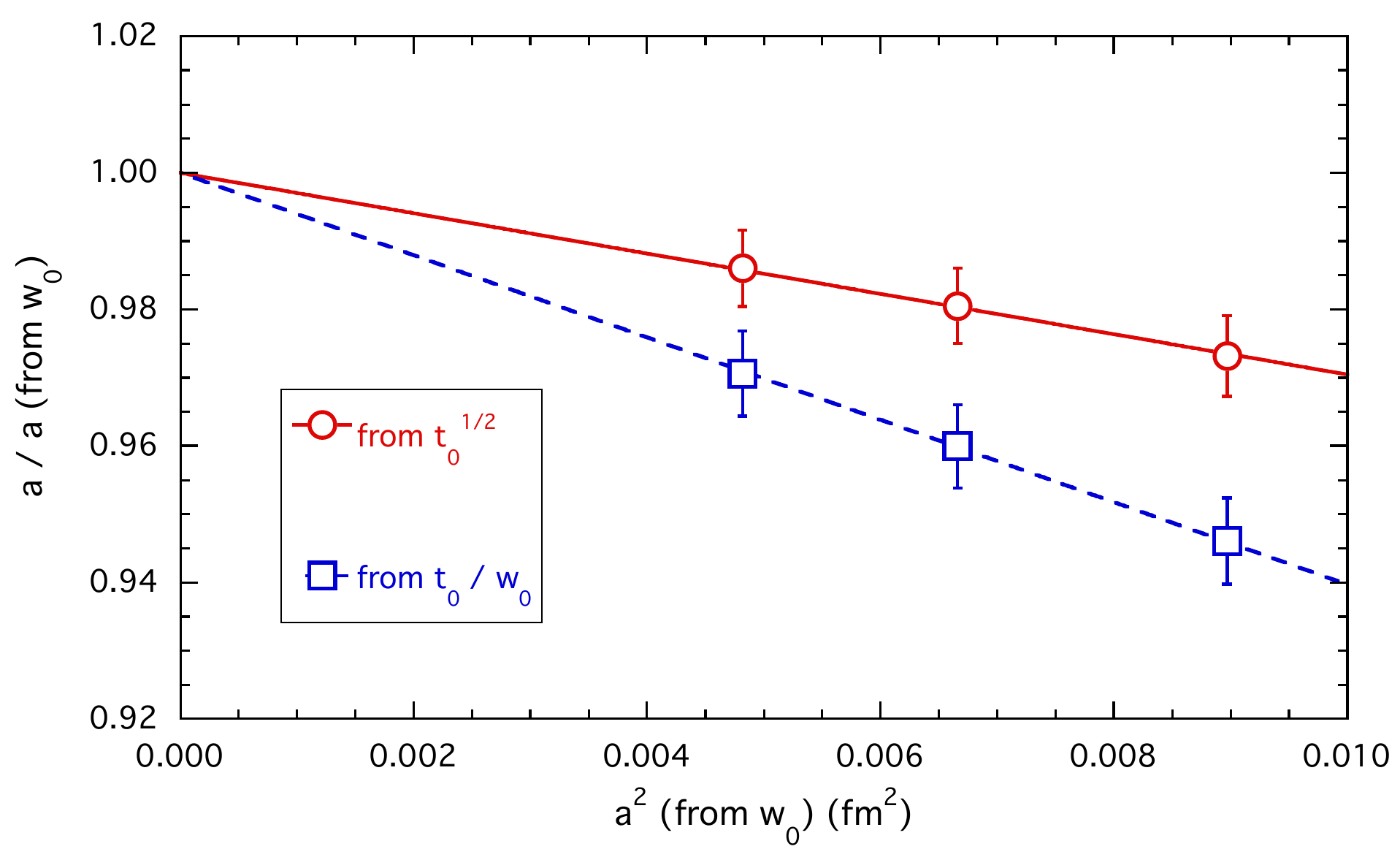}
\end{center}
\vspace{-0.90cm}
\caption{\it \small Ratio of the lattice spacing $a$ obtained from the GF scales $\sqrt{t_0}$ (red circles) and $t_0 / w_0$ (blue squares) with the one determined from the GF scale $w_0$ (see Table~\ref{tab:spacings}). The solid and dashed lines are linear fits.}
\label{fig:a_GF}
\end{figure}

Finally, the values of the lattice spacing $a$ corresponding to the three GF scales (\ref{eq:w0_appD}-\ref{eq:t0w0_appD}) and to the relative scales given in Table~\ref{tab:relGF} are shown in Table~\ref{tab:spacings}.
\begin{table}[htb!]
\renewcommand{\arraystretch}{1.2}	 
\begin{center}
{\small
\begin{tabular}{||c||c|c|c||}
\hline
GF scale & $a(\beta=1.726)$ (fm) & $a(\beta=1.778)$ (fm) & $a(\beta=1.836)$ (fm)\\ \hline
\hline
$w_0$         & 0.09471  (39) & 0.08161 (30) & 0.06941 (26)\\ \hline
$\sqrt{t_0}$ & 0.09217  (41) & 0.08002 (34) & 0.06844 (29)\\ \hline
$t_0 / w_0$ & 0.08960  (47) & 0.07834 (41) & 0.06737 (35)\\ \hline
\hline
\end{tabular}
}
\end{center}
\renewcommand{\arraystretch}{1.0}
\vspace{-0.25cm}
\caption{\it Values of the lattice spacing $a$ corresponding to the three GF scales $w_0$, $\sqrt{t_0}$, $t_0 / w_0$ and to the corresponding relative scales given in Table~\ref{tab:relGF}.}
\label{tab:spacings}
\end{table} 
The three determinations of $a$ differ by ${\cal{O}}(a^2)$ effects, as shown in Fig.~\ref{fig:a_GF}.
In particular, we get: $a(\sqrt{t_0}) / a(w_0) \simeq 1 -  0.09\,(2) ~ a^2(w_0) / w_0^2$ and $a(t_0 / w_0) / a(w_0) \simeq 1 - 0.18\,(2) ~ a^2(w_0) / w_0^2$.

%% file: extrapolations_tmCloverNf211_GF_scales.tex
In this section we provide the details of the determinations of the
gradient flow (GF) scales at the physical point. Our analysis is based on
the values of the gradient flow scales in Table \ref{tab:GF_scales} as obtained on the ensembles in Table \ref{tab:simudetails}. We calculate the scales following the definitions in \cite{Luscher:2010iy,Luscher:2011bx,Borsanyi:2012zs} using the standard Wilson action for the gradient flow evolution and the symmetrized discretization of the action density as described in \cite{Luscher:2010iy}. Apart from the usual scales
$s_0/a \equiv \sqrt{t_0}/a$ and $w_0/a$, we also consider the derived
scale $(t_0/w_0)/a$ as well as the dimensionless ratio
$(s_0/w_0)$. The former is interesting, because it
exhibits very mild quark-mass dependence and reduced autocorrelations, while the dimensionless
ratio $(s_0/w_0)$ can be used for assessing lattice artefacts and crosschecking the consistency of the various
analysis procedures and determinations of the scales in physical units.

\begin{table}[h!]
\begin{adjustbox}{width=\textwidth}
%% BaKo: replaced tabularx with tabular, which works well with adjustbox
%\begin{tabularx}{\textwidth}{@{\extracolsep{\fill}}||l|r| llll|llll ||}
\begin{tabular}{||l|r| llll|llll ||}
 \hline
  ensemble & $N_{\text{meas}}$ & $s_0/a$ & $w_0/a$ & $(t_0/w_0)/a$ & $s_0/w_0$ &
                                                             $\tau_\textrm{int}^{s_0}$ &  $\tau_\textrm{int}^{w_0}$ &  $\tau_\textrm{int}^{t_0/w_0}$ &  $\tau_\textrm{int}^{s_0/w_0}$     \\
\hline
  cA211.53.24 & 1122 & 1.5306(21) & 1.7597(43) & 1.33139(89)   & 0.86982(100)  & 23(6) & 25(7)   &   7(1)   &  18(4)  \\ 
  cA211.40.24 & 1219 & 1.5384(18) & 1.7766(33) & 1.33213(96)   & 0.86592(64)   & 20(5) & 18(4)   &   7(1)   &   9(2)  \\
  cA211.30.32 & 2559 & 1.5460( 9) & 1.7928(17) & 1.33314(47)   & 0.86233(32)   & 22(5) & 21(4)   &   9(1)   &  10(2)  \\
  cA211.12.48 & 326 & 1.5614(22) & 1.8249(33) & 1.33590(155)  & 0.85559(29)   & 69(30) & 63(27) & 59(25)   &  16(5)  \\
  cB211.25.24 & 1145 & 1.7937(22) & 2.0992(46) & 1.53260(108)  & 0.85445(77)   & 21(5) & 25(6)   &   5(1)   &   12(2)  \\
  cB211.25.32 & 990 & 1.7922(19) & 2.0991(47) & 1.53018(72)   & 0.85380(91)   & 35(10) & 45(14)   &   6(1)   &  28(7)  \\
  cB211.25.48 & 1175 & 1.7915( 8) & 2.0982(19) & 1.52966(41)   & 0.85384(38)   & 28(8) & 31(9)   &   9(2)   &  20(5)  \\
  cB211.14.64 & 619 & 1.7992( 5) & 2.1175(11) & 1.52875(23)   & 0.84968(23)   & 30(8) & 32(9)   &   8(1)   &  23(6)  \\
  cB211.072.64 & 191 & 1.8028( 8) & 2.1272(19) & 1.52784(42)   & 0.84750(41)   & 45(18) & 52(22) &  16(5)   &  41(16) \\
  cC211.06.80 & 785 & 2.1094( 8) & 2.5045(17) & 1.77670(37)   & 0.84226(27)   & 46(17) & 42(16) &  14(3)   &  26(8)  \\
\hline
\end{tabular}
%\end{tabularx}
\end{adjustbox}
\caption{\label{tab:GF_scales}GF scales and corresponding integrated autocorrelation times in units of trajectories of length $\tau=1.0$ from the symmetrized action density. The $N_{\text{meas}}$ measurements on different ensembles were performed using different separations and the autocorrelation times were scaled appropriately. Similarly, for the cB211.25.24, cB211.25.32 and cB211.14.64 ensembles, the autocorrelation times were scaled to take into account the $\tau=1.5$ trajectory lengths used there.}
\end{table}

The errors are calculated by taking into account
the autocorrelations of the various quantities,  which are also listed in Table
\ref{tab:GF_scales}. It is well known that the autocorrelations can be
sizable for the GF scales. In particular on coarse lattices it is
known that they can 
be larger than the ones for the topological charge
\cite{Deuzeman:2012jw}. More importantly, they are
also expected to grow towards the chiral limit. The data shown in \Cref{tab:GF_scales} confirms this behaviour in terms of the lattice spacing and the pion mass. The large autocorrelations observed on the ensemble cA211.12.48, corresponding to the coarsest lattice spacing and the smallest pion mass, could be also related to metastability effects known for this ensemble. 
Nonetheless, the 
data in Table \ref{tab:GF_scales} also indicates that for our simulations the autocorrelations
are well under control even at the physical point. One interesting point to
note is the fact that the autocorrelations of the scale $(t_0/w_0)/a$
are reduced by roughly a factor 3 with respect to the ones of the usual GF scales
$\sqrt{t_0}/a$ and $w_0/a$. This is also reflected in the very small
statistical error for $(t_0/w_0)/a$, roughly a factor 2-3 smaller than for
$\sqrt{t_0}/a$ and $w_0/a$ at the physical point.

\begin{figure}[h!]
  \begin{center}
    \includegraphics[scale=0.3]{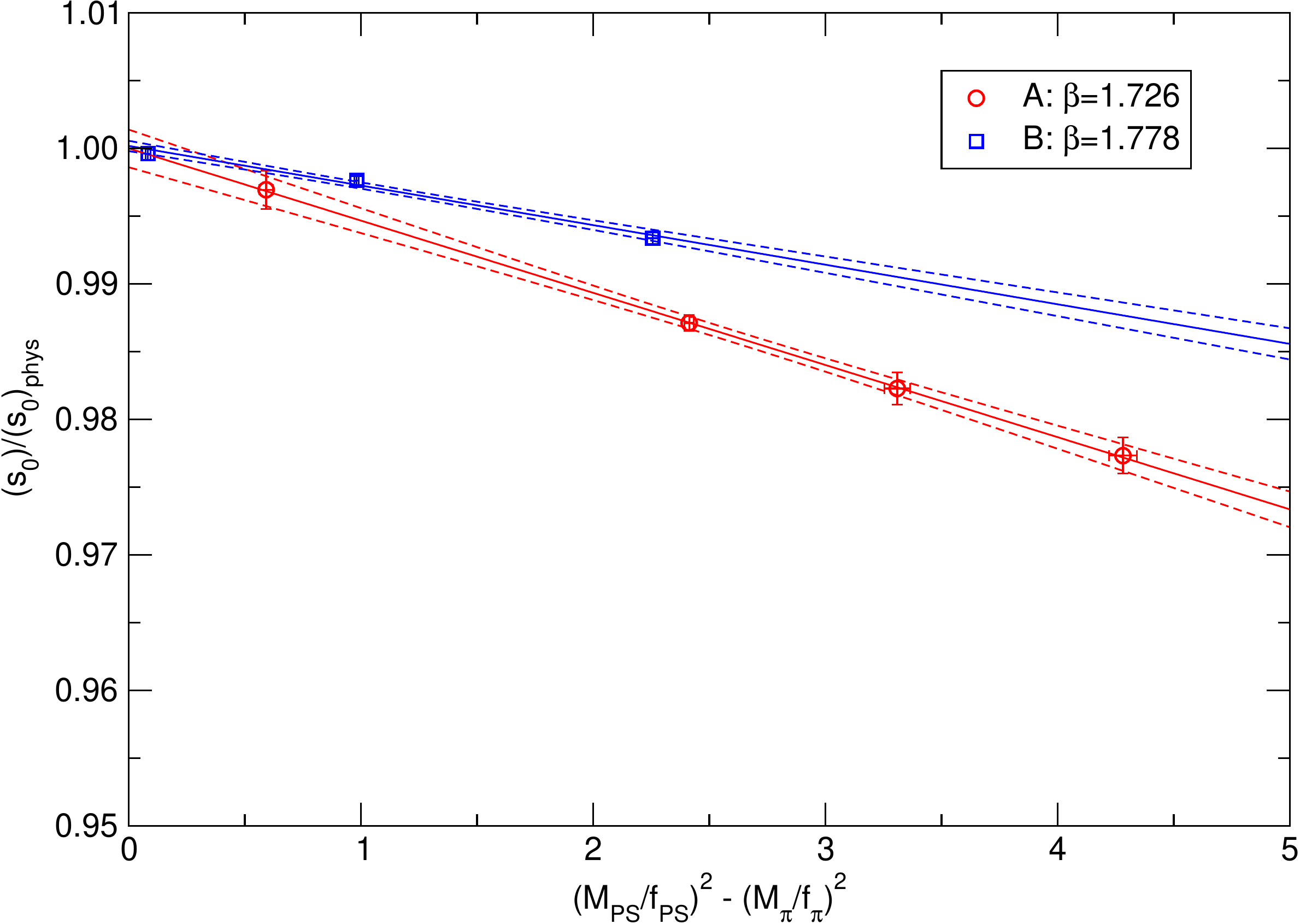}\hfill
    \includegraphics[scale=0.3]{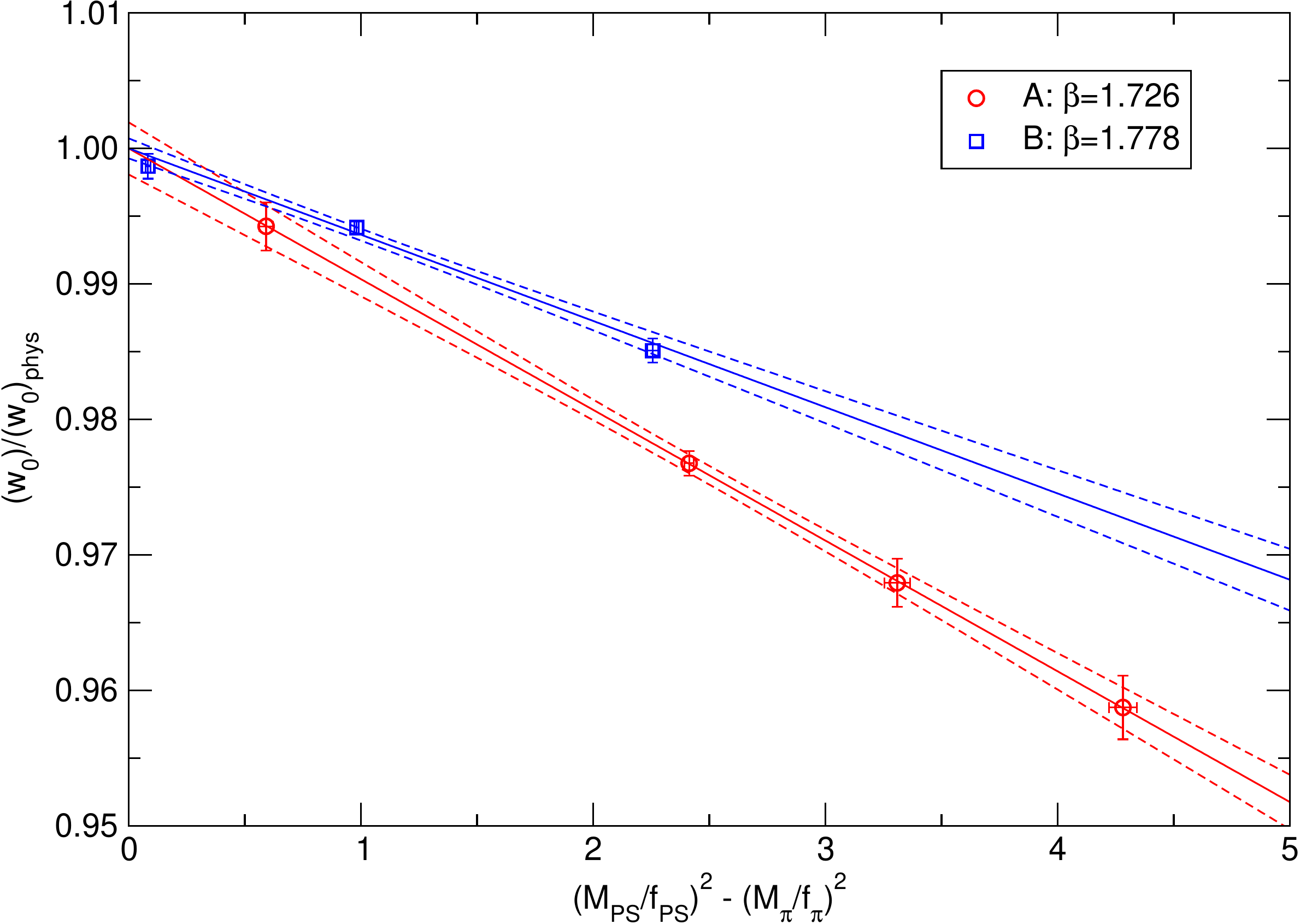}\\
    \vspace*{0.5cm}
    \includegraphics[scale=0.3]{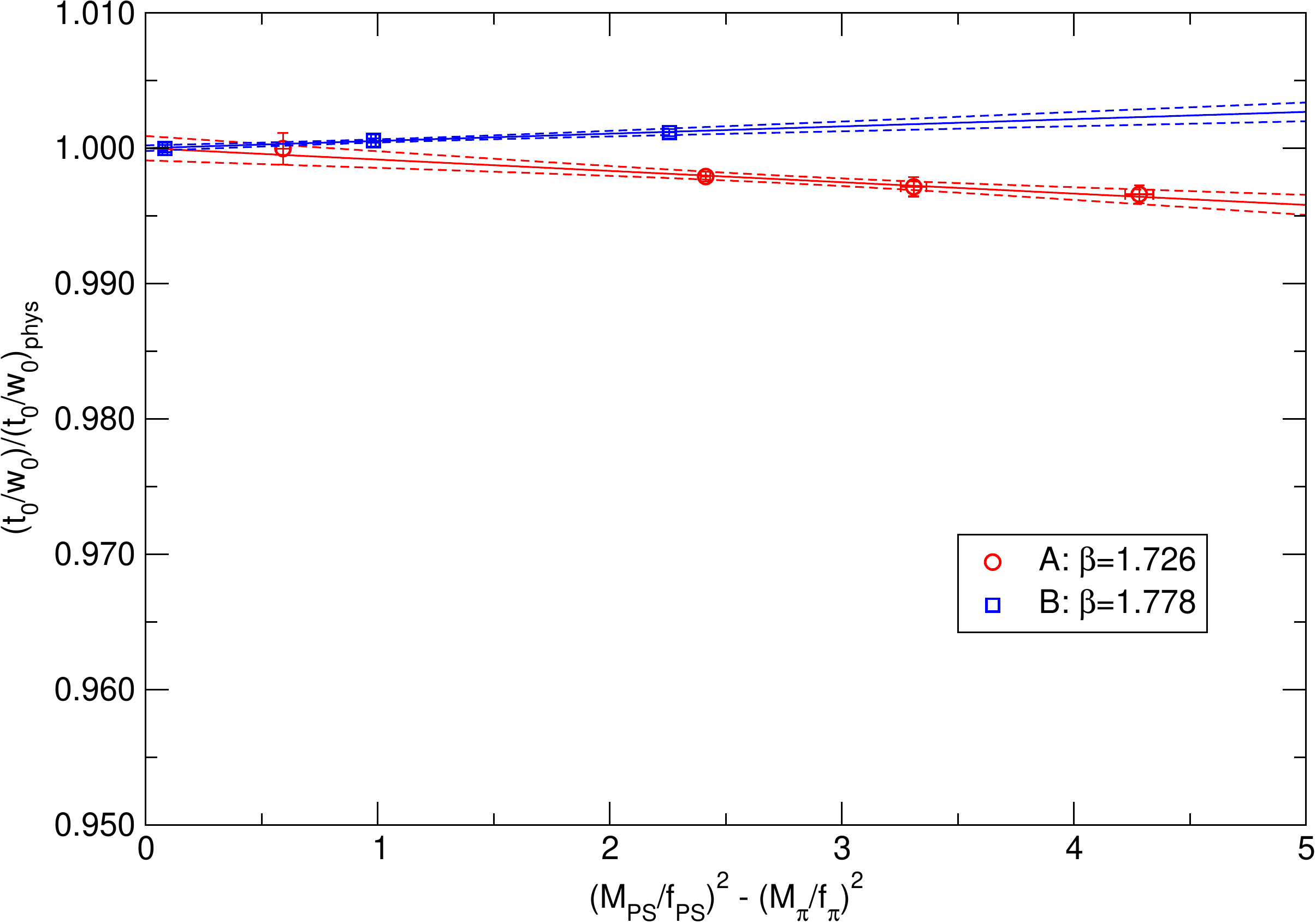}
  \end{center}
  \vspace*{-0.75cm}
\caption{\label{fig:GFscales_extrapolation} Light quark-mass dependence and extrapolations to the physical point of the GF
  scales $s_0 \equiv \sqrt{t_0}$, $w_0$, $t_0/w_0$, normalised by their values at the physical point for ease of comparing the results at different lattice spacings.}
 % , and of the dimensionless ratio $s_0/w_0$}.
\end{figure}

In order to use the scales in the analysis of the light meson sector,
we need the values at the physical pion-mass point. To achieve this,   
we perform an extrapolation for the two sets of ensembles cA211 and cB211\footnote{The results for the ensembles cB211.25.24 and cB211.25.32 shown in Table~\ref{tab:GF_scales} are not used for the determination of the relative GF scales at the physical pion point, since finite-volume effects are found to be negligible with respect to the other uncertainties for $M_\pi L \gtrsim 3.5$ (see also Table~\ref{tab:simudetails}).} 
to the physical point in terms of 
$\Delta^2  \equiv (M_\pi/f_\pi)^2-(M_\pi/f_\pi)^2_\text{phys.}$, such that the physical
point is reached when this quantity is zero, while for the ensemble
cC211.06.80 we directly use the value at the physical point as given in
Table \ref{tab:GF_scales}. We note that for this ensemble $\Delta^2 =
0.025$ such that the potential corrections would be tiny and in fact smaller
than the statistical errors.
The details of the extrapolations are
summarized in Table \ref{tab:GF_scales_extrapolation}. In order
to illustrate the extrapolations and compare them at different lattice spacings, in
Fig.~\ref{fig:GFscales_extrapolation} we show the scales normalized by their
values at the physical point as a function of $\Delta^2$.

\begin{table}[h!]
\begin{center}
\small{
\begin{tabularx}{\textwidth}{@{\extracolsep{\fill}}||l||ccc||ccc||}
 \hline
          & $(s_0/a)_\text{phys.}$ & $c$ & $\chi^2/\text{d.o.f.}$  &  $(w_0/a)_\text{phys.}$ & $c$ & $\chi^2/\text{d.o.f.}$  \\
  \hline
  cA211 & 1.5660(22)  & -0.0082(8) & 0.02 & 1.8352(35) & -0.0174(13)  & 0.00  \\
  \hline
  cB211 & 1.80396(68) & -0.0053(5) & 2.14 & 2.1299(16) & -0.0136(12)  & 1.84 \\
  \hline
\end{tabularx}
}

\vspace*{0.5cm}

%  \footnotesize{
\begin{tabularx}{\textwidth}{@{\extracolsep{\fill}}||l||ccc||ccc||}
 \hline
          & $((t_0/w_0)/a)_\text{phys.}$ & $c$  &$\chi^2/\text{d.o.f.}$ &  $(s_0/w_0)_\text{phys.}$ & $c$ & $\chi^2/\text{d.o.f.}$ \\
  \hline
  cA211 & 1.3359(12) & -0.0011(4)            & 0.15 & 0.8531(10) & 0.0038(4) & 0.05 \\
  \hline
  cB211 & 1.52789(33) & \phantom{-}0.0008(3) & 0.18 & 0.84697(37)& 0.0030(3) & 1.26 \\
  \hline
\end{tabularx}
%}
\end{center}

\caption{\label{tab:GF_scales_extrapolation}Results for the extrapolations of the GF
  scales to the physical point in terms of
  $\Delta^2 = (M_\pi/f_\pi)^2-(M_\pi/f_\pi)^2_\text{phys.}$, i.e., $X(M_\pi)/a = (X/a)_\text{phys.} + c \cdot \Delta^2$. Note that for the
  ensembles cA211 we have $N_\text{d.o.f.}=2$, while for cB211 $N_\text{d.o.f.}=1$.}  
\end{table}
From the plots and the data in the table it is obvious that the
quark-mass dependence of the scale $(t_0/w_0)/a$ is very small, i.e.,
the corrections are 
less than 0.5\% at our largest pion mass ensemble cA211.53.24, to be compared to 2.5\% for $\sqrt{t_0}/a$ and 4.5\% for $w_0/a$. We note that the quark-mass dependence exhibits clearly visible lattice artefacts, but the dependence seems to become weaker towards the continuum limit. One peculiar feature is the fact that for the scale $(t_0/w_0)/a$ the slope of the quark-mass dependence changes sign when going from the coarser to the finer lattice spacing. This certainly warrants further investigation, once more data is available, however, one should keep in mind that for this quantity the slope is consistent with zero within less than 3$\sigma$.

\begin{figure}[h!]
  \begin{center}
    \includegraphics[scale=0.5]{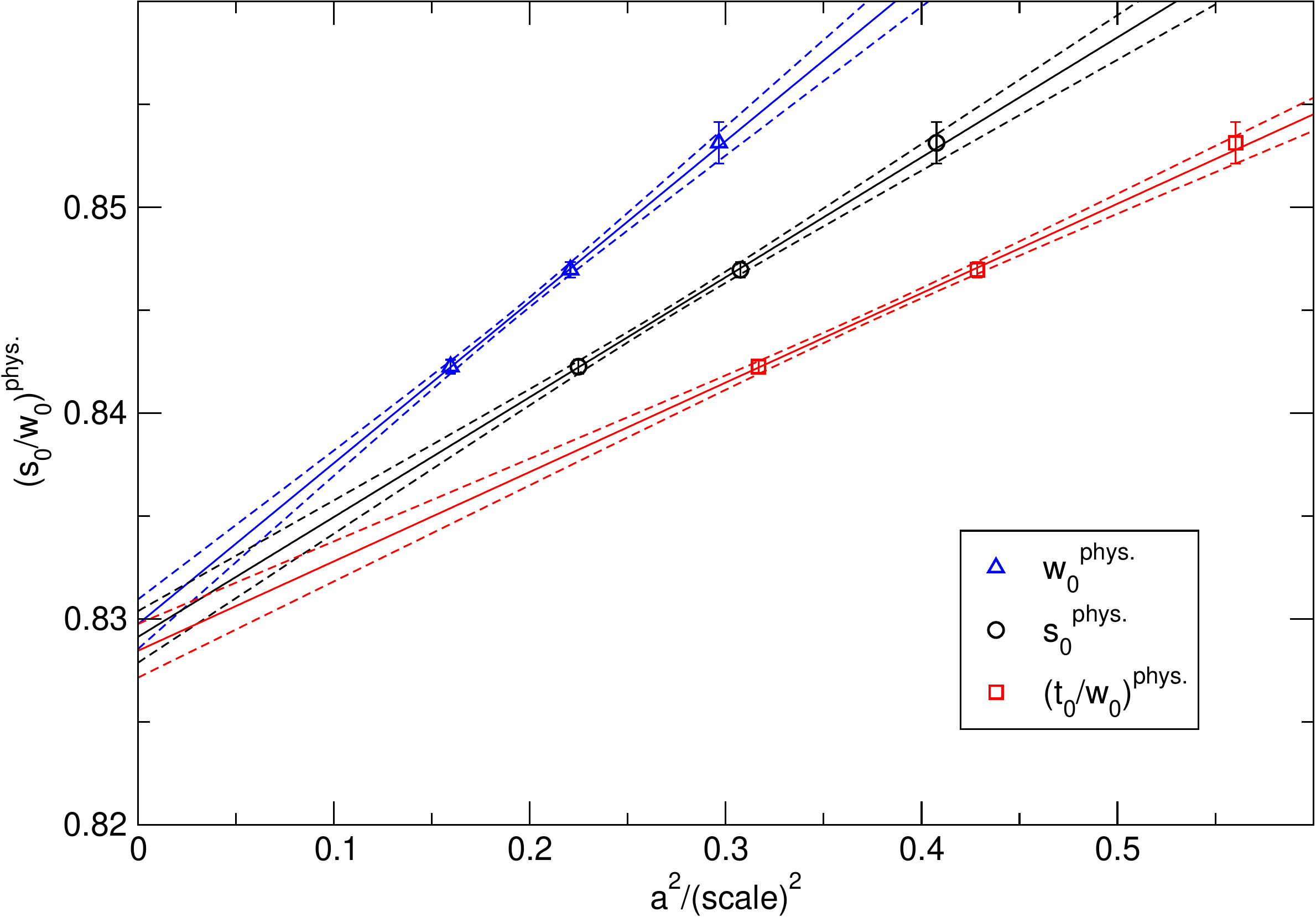}
  \end{center}
    \vspace*{-0.75cm}
  \caption{\label{fig:s0overw0_vs_a2} Continuum extrapolations of the dimensionless ratio of GF scales $s_0/w_0$ at the physical point in terms of the lattice spacing in units of the GF scales $s_0^\text{phys.}, w_0^\text{phys.}$ and $t_0^\text{phys.}/w_0^\text{phys.}$ at the physical point.}
\end{figure}

In order to examine the lattice artefacts of the GF scales further, we now turn to the dimensionless ratio $s_0/w_0$. In Figure \ref{fig:s0overw0_vs_a2} we show the ratio at the physical point as a function of the lattice spacing expressed in units of the three GF scales $s_0^\text{phys.}, w_0^\text{phys.}$ and $t_0^\text{phys.}/w_0^\text{phys.}$ at the physical point. Note that for the lattice spacing cC211 this corresponds to the value obtained on the ensemble cC211.06.680. The data show a precise ${\cal O}(a^2)$-scaling towards the continuum and allow continuum extrapolations in terms of $a^2$. The continuum extrapolations using in turn $a^2/(t_0/w_0)^2,  a^2/t_0$ and $a^2/w_0^2$ yield $(s_0/w_0)^\text{phys.} = 0.8285(13), 0.8291(13)$ and $0.8298(12)$, respectively, with $\chi^2/\text{d.o.f.}=0.20, 0.12$ and $0.06$. The values in the continuum are perfectly consistent with each other and averaging them in the usual way gives $(s_0/w_0)^\text{phys.} = 0.8291(13)(5)[14]$, where the second error reflects the spread of the results while the error in the square bracket is the combined one. These results provide a nontrivial check of the expected scaling behaviour with quantities determined with an accuracy of between 1 - 2 permille for the scales and sub-permille for the ratio, and they nicely confirm the automatic ${\cal O}(a)$-improvement in place for TM Wilson fermions at maximal twist.

\begin{figure}[h!]
  \begin{center}
    \includegraphics[scale=0.5]{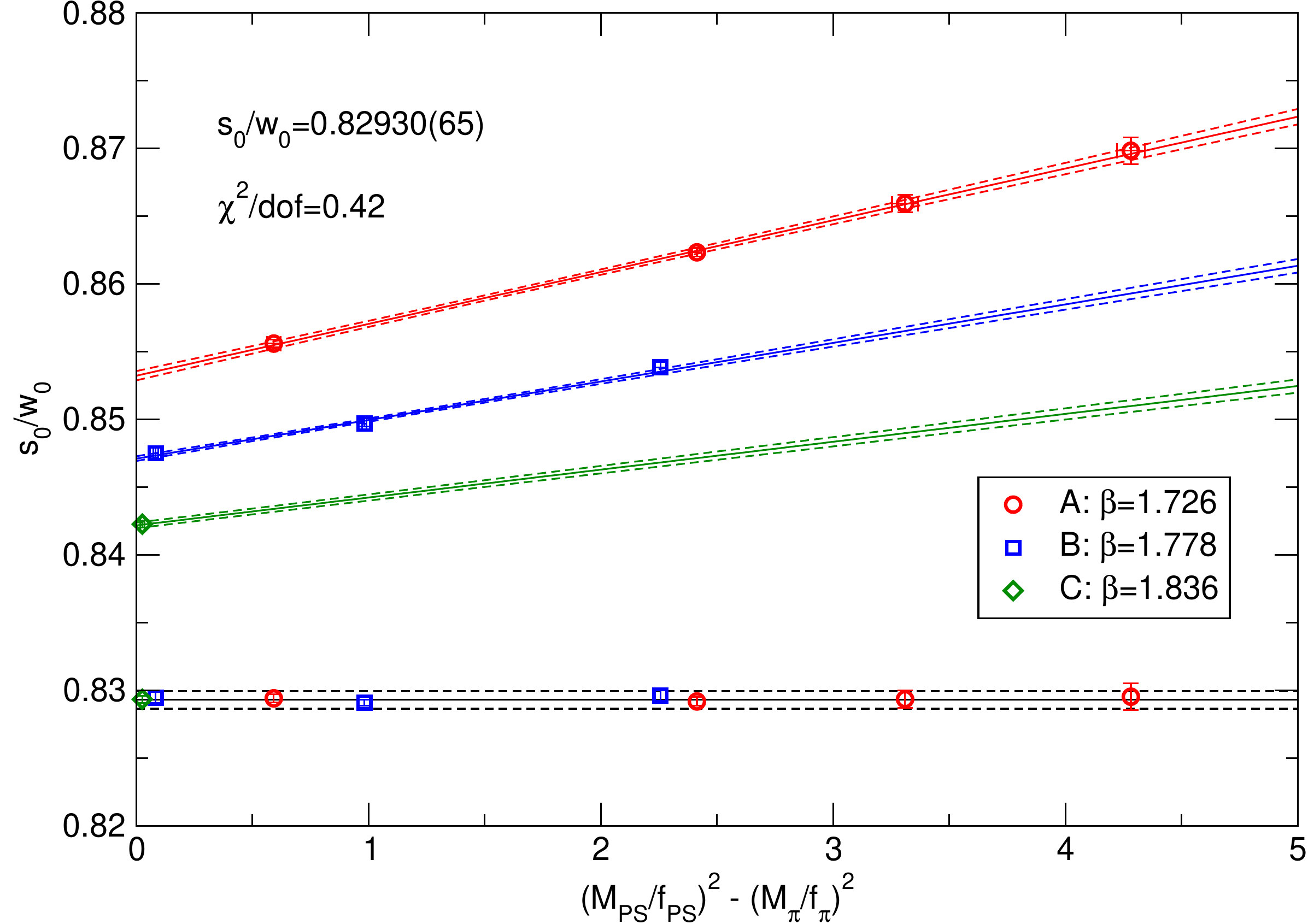}
  \end{center}
    \vspace*{-0.75cm}
  \caption{\label{fig:s0overw0_sym_globalFit} Global fit of the lattice spacing and light quark-mass dependence of the dimensionless ratio $s_0/w_0\equiv\sqrt{t_0}/w_0$. The black line with the error band at the bottom shows the fit result in the continuum, while the data points on this line represent the data corrected by the lattice artefacts as described by the global fit.}
\end{figure}

Given the fact that the ratio at the physical point shows a very nice ${\cal O}(a^2)$-scaling, we may attempt a global fit in order to extrapolate simultaneously to the physical pion mass and to the continuum limit, using
\begin{equation}
\frac{s_0}{w_0}\left(\frac{M_\text{PS}}{f_\text{PS}}, \left(\frac{a}{w_0}\right)^2 \right) = \left(\frac{s_0}{w_0}\right)^\text{phys.}_\text{cont.} + A_1 \cdot \left(\frac{a}{w_0}\right)^2 + (B_0 + B_1 \cdot \frac{a^2}{w_0^2}) \left(\frac{M_\text{PS}^2}{f_\text{PS}^2} - \frac{M_\pi^2}{f_\pi^2} \right) \, ,
\end{equation}
which includes a light quark-mass dependence proportional to $\Delta^2$ described by $B_0$ and lattice artefacts proportional to $a^2/w_0^2$ described by $A_1$ and $B_1$. The latter coefficient describes the lattice artefacts on the quark-mass dependence. The global fit suggests that $B_0$, describing the quark-mass dependence in the continuum, is well consistent with zero, i.e., $B_0=0.001(7)$. That is, in the continuum the ratio $\sqrt{t_0}/w_0$ appears to have no dependence on the pion mass at all and the observed pion-mass dependence at finite lattice spacings is apparently just a lattice artefact. However, given the fact that we do not have data for the ratio at the lattice spacing cC211 away from the physical point, and hence no information on the quark-mass dependence at the finest lattice spacing, it is not clear how solid this conclusion is. Nevertheless, we may attempt to fit our data with $B_0=0$ fixed, and in figure \ref{fig:s0overw0_sym_globalFit} we show the results for this global fit.
The coloured lines with error bands show the light quark-mass dependence of the ratio and the extrapolations to the physical point for each lattice spacing, while the black line with the error band at the bottom shows the fit result in the continuum. The data points on this line represent our data corrected by the lattice artefacts as described by the global fit.

For the ratio at the physical point and in the continuum the fit yields
\begin{equation}
\left(\frac{s_0}{w_0}\right)^\text{phys.}_\text{cont.} = 0.82930(65)
\end{equation}
with $\chi^2/d.o.f.=0.42$, $N_\text{d.o.f.}=5$ and $A_1= 0.0806(30), B_1= 0.0129(5)$.

We note that the ratio is determined with a precision in the sub-permille region, i.e., better than 0.8 permille. As such, it provides an interesting consistency crosscheck on any other, independent determination of the scales, e.g., through hadronic quantities.